\documentclass[onecolumn,authoryear]{els-mrw} 

\usepackage{amsmath,amssymb,amsfonts,amsthm,makeidx,graphicx}
\usepackage{txfonts}
\usepackage{helvet}

\usepackage{subcaption}

\usepackage{amsmath}
\usepackage{amssymb}
\usepackage{float}
\usepackage{bm}
\usepackage{color,graphicx}
\usepackage{slashed}
\usepackage{comment}
\usepackage{xspace}
\usepackage{multirow}
\usepackage{cancel}
\newcommand{\Tr}{\mathrm{Tr}\,}

\begin{document}

\chapter{Generalized Parton Distributions: Phenomenology, Extraction, and Hadron Imaging}\label{chap1}

\author[1]{Simonetta Liuti}%
\author[1]{Zaki Panjsheeri}%

\author[1,2]{Kemal Tezgin}%

\address[1]{\orgname{University of Virginia}, \orgdiv{Physics Department}, \orgaddress{382 McCormick Rd., Charlottesville, VA 22904}}
\address[2]{\orgname{Virginia Tech}, \orgdiv{Department of Physics}, \orgaddress{Blacksburg, VA 24061, USA}}

\articletag{Chapter Article tagline: update of previous edition,, reprint..}

\maketitle
%\begin{glossary}[Glossary]
%\term{Europe} the model is a coherent view of capital markets data that allows users to interact with the content in a consistent manner.
%\term{Primates} regardless of the source. Essentially, of sources. Properly deployed.
%\end{glossary}

\begin{glossary}[Nomenclature]

\begin{tabular}{@{}lp{34pc}@{}}
GPD & Generalized Parton Distribution\\
CFF & Compton Form Factor \\
PDF & Parton Distribution Function\\
EMT & Energy-Momentum Tensor \\
DVCS & Deeply Virtual Compton Scattering\\
DVMP & Deeply Virtual Meson Production\\
DDVCS  & Double DVCS \\
\end{tabular}
\end{glossary}

\begin{abstract}[Abstract]
Generalized Parton Distributions (GPDs) provide a framework for investigating the correlated momentum and spatial structure of quarks and gluons in hadrons and for accessing fundamental properties such as angular momentum and the QCD energy-momentum tensor. In this review, we discuss the present status of GPD phenomenology, emphasizing the challenges involved in connecting deeply virtual exclusive measurements to the underlying partonic structure. We organize this problem in terms of two successive inverse problems: the extraction of Compton Form Factors (CFFs) from measured observables and the reconstruction of GPDs from the convolution integrals defining the CFFs. We review the theoretical description and phenomenological parametrizations of GPDs, current strategies for CFF and GPD extraction, and the  role of lattice QCD, Bayesian inference, uncertainty quantification, and artificial intelligence, including neural networks and interpretable machine-learning approaches. We discuss the limitations of present determinations and the opportunities offered by the Jefferson Lab program, complementary exclusive processes, and the future Electron-Ion Collider. Finally, we consider how increasingly precise and multidimensional information, together with new statistical and AI methodologies, will transform GPD phenomenology in the emerging era of precision hadron imaging.
\end{abstract}

%
%%%%%%%%%%%%%%%%%%%%%%%%%%%%%%%%%%%%%%%%%%%%%%%%%%
%% First section: Introduction to GPDs and DVCS %%
%%%%%%%%%%%%%%%%%%%%%%%%%%%%%%%%%%%%%%%%%%%%%%%%%%

\section{Introduction and Physics Motivation}
\label{sec:introduction}
%%%%
Determining how quarks and gluons are distributed in both momentum and position inside the proton is one of the central goals of contemporary strong-interaction physics. 
In the infinite-momentum frame, parton distribution functions (PDFs) describe their momentum distributions along the proton direction, while form factors (FFs) encode their spatial distributions as they can be related, through a Fourier transform, to transverse densities. 
Generalized Parton Distributions (GPDs) connect and extend these two complementary descriptions by correlating the longitudinal momentum of partons with the momentum transferred to the proton. 
They reduce to ordinary PDFs in the forward, zero-momentum-transfer limit, while their integrals over the parton momentum yield the proton FFs. 
Furthermore, the GPDs second Mellin moments give access to the total angular momentum carried by quarks and gluons and, in combination with information on their spin contributions, to their orbital angular momentum. More generally, these moments are related to form factors of the QCD energy--momentum tensor, providing access to the mechanical properties of the proton, including the distribution of its internal forces .

The phenomenological challenge is that none of this information is directly observable. GPDs enter deeply virtual exclusive scattering (DVES) processes through convolution integrals with perturbative QCD kernels, giving rise to quantities such as Compton form factors (CFFs). Experiments, in turn, measure cross sections and polarization asymmetries constructed from amplitudes containing the CFFs. Reconstructing partonic structure from experiment, therefore, requires solving a sequence of interconnected inverse problems,
\\

\text{\qquad\qquad\qquad Experimental observables}
$\longrightarrow$
\text{CFFs}
$\longrightarrow$
\text{GPDs}
$\longrightarrow$
\text{spatial and dynamical properties of the nucleon}.
\\

\noindent The limited kinematic coverage of existing measurements, correlations among the contributing GPDs, and the integral nature of their relation to observables make the reconstruction underconstrained, giving rise to an ill-posed inverse problem.
GPD phenomenology is now entering a qualitatively new stage. The Electron-Ion Collider (EIC) will greatly enlarge the kinematic domain and precision of exclusive measurements, particularly for gluons and sea quarks and in the transition toward the small-\(x\) regime. At the same time, lattice QCD is beginning to provide increasingly detailed constraints on the same underlying hadronic structure. The problem is consequently evolving from extracting a limited number of quantities from relatively sparse measurements toward high-dimensional inference combining increasingly large and strongly correlated sources of experimental and theoretical information.
This development coincides with rapid advances in artificial intelligence (AI) and modern statistical inference. Flexible neural-network representations, Bayesian methods, generative models, and interpretable approaches such as symbolic regression offer new ways to address the dimensionality and non-uniqueness of the GPD inverse problem. Their role is potentially more profound than providing more flexible fitting functions: they make it possible to construct inference frameworks in which experimental measurements, lattice-QCD information, and exact or approximate theoretical constraints can be incorporated simultaneously. These methods can incorporate theoretical constraints from QCD directly into the learning and inference process, reducing the space of physically admissible solutions.

The convergence of new experimental capabilities at the EIC, advances in lattice QCD, and AI-based inference therefore opens the possibility of moving beyond predominantly model-driven extractions toward quantitative, uncertainty-controlled determinations of the quark and gluon structure of the nucleon. The central challenge for the coming era will not simply be to generate increasingly flexible descriptions of GPDs, but to determine which features of the reconstructed nucleon are actually required by the data and QCD, with what uncertainty, and at what level of spatial and dynamical resolution.

\subsection{From Exclusive Measurements to GPDs}

\begin{figure}[H]
    \includegraphics[width=1.0\linewidth]{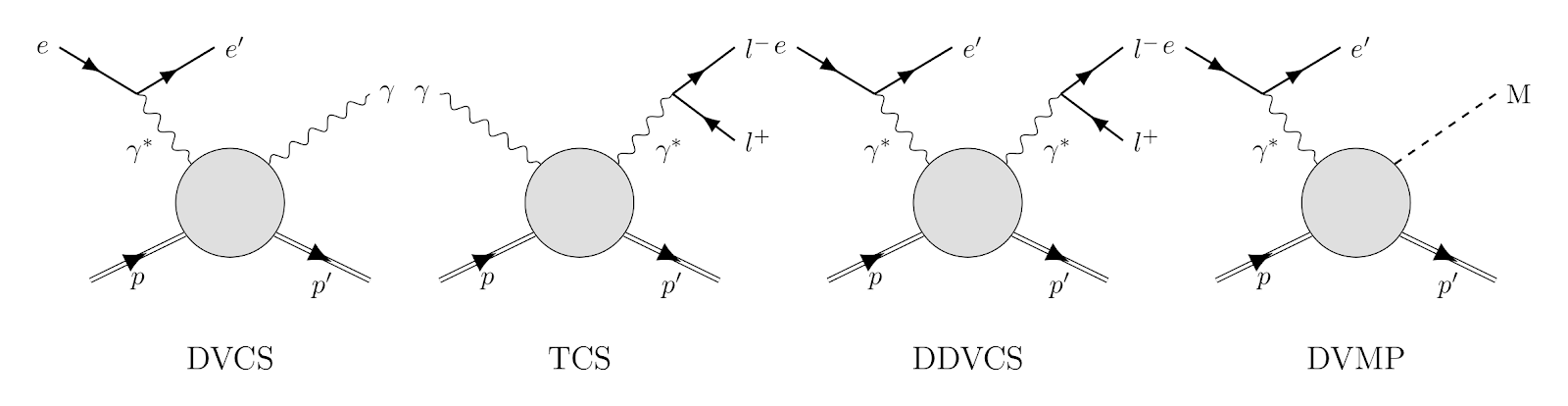}
    \caption{Four important processes facilitating access to GPDs: deeply
  virtual Compton scattering (DVCS), timelike Compton scattering (TCS), double deeply virtual Compton scattering (DDVCS), and deeply virtual meson production (DVMP).}
    \label{fig:dvcstcsdvmp}
\end{figure}

Among deeply virtual exclusive scattering (DVES) processes, Deeply Virtual Compton Scattering (DVCS) and Deeply Virtual Meson Production (DVMP), shown in Fig.~\ref{fig:dvcstcsdvmp}, have been the most extensively studied channels for accessing GPDs.
In both cases, however, the connection between the measured observables and the underlying GPDs is indirect. In DVCS, GPDs enter through Compton Form Factors (CFFs), defined as convolution integrals of the GPDs with perturbatively calculable hard-scattering kernels. DVMP involves analogous convolutions, with the factorized amplitude additionally depending on the distribution amplitude of the produced meson. Consequently, the extraction of GPDs requires the inversion of these convolution relations after the corresponding amplitudes have first been constrained from experimental observables.

The experimental study of DVCS has progressed considerably since the first observations of process-sensitive asymmetries and cross sections at HERMES, HERA, and Jefferson Lab (JLab) established the feasibility of accessing GPDs through hard exclusive reactions   \cite{HERMES:2001bob,H1:2001nez,CLAS:2001wjj}. These pioneering measurements initiated an extensive experimental program at HERA, HERMES, JLab, and subsequently COMPASS, providing increasingly precise measurements of unpolarized and polarized cross sections and asymmetries over complementary kinematic regions \cite{ZEUS:2003pwh,H1:2005gdw,Chen:2006na,HERMES:2006pre,MunozCamacho:2006hx,JeffersonLabHallA:2007jdm,H1:2007vrx,CLAS:2007clm,Airapetian:2008aa,ZEUS:2008hcd,CLAS:2008ahu,H1:2009wnw,HERMES:2009cqe,HERMES:2010hnl,HERMES:2010dsx,HERMES:2011bou,HERMES:2012gbh,CLAS:2015bqi,CLAS:2015uuo,JeffersonLabHallA:2015dwe}.
%The experimental program has also expanded well beyond DVCS on the proton. 
Closely related channels such as Timelike Compton Scattering (TCS) \cite{Berger:2001xd,Boer:2015hma} and Double Deeply Virtual Compton Scattering (DDVCS), \cite{Guidal:2002kt,Belitsky:2002tf,Belitsky:2003fj} provide complementary access to GPDs through different hard-scattering kernels and kinematic configurations. Deeply virtual meson production offers additional sensitivity to flavor, polarized, gluon, and chiral-odd GPDs. Hadron-induced reactions provide another emerging direction: exclusive Drell--Yan processes and, more generally, single-diffractive hard exclusive reactions have been proposed as complementary probes of GPDs, with the additional hard scales and final-state kinematics providing access to combinations of partonic variables that are difficult to resolve in conventional DVCS, \cite{Qiu:2022bpq}. A qualitatively different opportunity for accessing the gluon sector is provided by ultra-peripheral collisions (UPCs) at RHIC and the LHC, where the strong electromagnetic fields of relativistic hadrons and nuclei act as sources of quasi-real photons, enabling photon--hadron interactions at very high center-of-mass energies. 
The next major expansion of the experimental landscape beyond the JLab 12-GeV program will come from the EIC. Whereas JLab provides high-luminosity precision measurements predominantly in the valence region, the EIC will extend exclusive measurements over a much broader range in $x_{Bj}$ and$Q^2$, with particular sensitivity to sea-quark and gluon GPDs. 
%DVCS, exclusive vector-meson production, and related exclusive channels will provide access to the spatial and momentum structure of the nucleon in regions that are only weakly constrained by present measurements. 
The extended $Q^2$ lever arm will also provide new sensitivity to QCD evolution and to the separation of quark and gluon contributions.

%Deeply virtual exclusive measurements have also been extended to nuclear targets. Early HERMES measurements explored the nuclear dependence of DVCS observables over a range of nuclei \cite{HERMES:2009xsg}, while subsequent exclusive measurements at JLab established coherent and incoherent DVCS on \(^{4}\mathrm{He}\), providing sensitivity to both nuclear GPDs and the partonic structure of bound nucleons \cite{CLAS:2017udk,CLAS:2018ddh,CLAS:2021ovm,Fucini:2021psq}. This program is now being substantially expanded with CLAS12 and ALERT through recoil and spectator-tagged measurements on light nuclei. Looking further ahead, the EIC will extend nuclear GPD studies to a broad range of nuclei and into the sea-quark and gluon dominated regimes.

In parallel, the theoretical description of DVCS has evolved toward increasingly precise formulations of both its perturbative QCD structure and the connection between GPDs, CFFs, helicity amplitudes, and measured observables. QCD factorization separates the short-distance part of the reaction, described by perturbatively calculable hard-scattering coefficient functions, from the long-distance hadronic structure encoded in the GPDs. The resulting CFFs are convolution integrals of these two components, with the hard coefficients calculated order by order in the strong coupling $\alpha_s$. This factorized structure provides the basis for systematically improving the theoretical accuracy of the GPD-to-CFF mapping through higher-order perturbative calculations and QCD evolution, while the subsequent CFF-to-observable relation is determined by the helicity and kinematic structure of the reaction.
%%%
The early comprehensive treatments established the structure of the DVCS and the amplitudes for its Bethe-Heitler (BH) background process, including twist-three contributions and their characteristic azimuthal dependence \cite{Belitsky:2001ns,Belitsky:2008bz,Belitsky:2010jw,Belitsky:2012ch}. Electromagnetic gauge invariance and the role of power-suppressed contributions were investigated in ~\cite{Anikin:2000em,Radyushkin:2000ap,Kivel:2000fg,Belitsky:2000vx}, while systematic treatments of target-mass and finite $t$ corrections have clarified the importance of finite $Q^2$ effects in the kinematic range of present experiments \cite{Braun:2012bg,Braun:2012hq}. 
More recently, the DVCS cross section has been reformulated directly in terms of helicity amplitudes, providing an alternative organization of the reaction that makes the connection between nucleon and photon helicity transitions, CFFs, and experimentally measured observables explicit, \cite{Kriesten:2019jep,Kriesten:2020apm,Kriesten:2021sqc}. This framework provides the basis for generalized Rosenbluth-type separations, for both DVCS and related DVES processes, in which the distinct kinematic dependence of the helicity amplitudes can be exploited to enhance sensitivity to CFF combinations of different twist.
The perturbative description of the GPD-to-CFF mapping has continued to improve. Hard-scattering coefficient functions have been calculated beyond leading order,  \cite{Ji:1997nk,Belitsky:1997rh,Mankiewicz:1997bk,Ji:1998xh,Belitsky:1999sg,Freund:2001hm,Freund:2001rk,Freund:2001hd,Pire:2011st,Moutarde:2013qs}, together with studies of soft-collinear resummation \cite{Altinoluk:2012fb,Altinoluk:2012nt,Bertone:2022frx}. More recently, substantial progress has been made toward a complete next-to-next-to-leading-order (NNLO) description of DVCS. Two-loop coefficient functions have been derived for the vector and axial-vector sectors, and their phenomenological impact on the dominant CFFs has been investigated \cite{Braun:2020yib,Braun:2021grd,Braun:2022bpn}. In particular, the new NNLO analysis in \cite{Braun:2022bpn} showed that higher-order corrections can be sizable in phenomenologically relevant kinematics, in part because of cancellations between quark and gluon contributions. The calculation has subsequently been completed for the flavor-singlet axial-vector and gluon-transversity coefficient functions, \cite{Ji:2023xzk}, providing the full set of NNLO coefficient functions required for DVCS phenomenology. Most recently, conformal moments of the two-loop coefficient functions have been obtained \cite{Braun:2025noa}, enabling their implementation in Mellin--Barnes representations and opening the way to GPD extractions consistently performed at NNLO accuracy. These developments will be particularly important for matching the theoretical precision to that of the forthcoming high-statistics measurements at JLab and the EIC.
The increasingly rich experimental information now within reach, together with continued theoretical developments, calls for correspondingly sophisticated phenomenological analyses capable of consistently extracting the underlying GPDs and quantifying their uncertainties.
%%
%Starting with the first quantitative fits to DVCS observables \cite{Kumericki:2007sa,Guidal:2008ie,Kumericki:2009uq,Moutarde:2009fg,Guidal:2009aa,Guidal:2010ig,Guidal:2010de,Goldstein:2010gu,Kumericki:2011zc,Kumericki:2011rz,GonzalezHernandez:2012jv,Kumericki:2013br,Boer:2014kya}, 
%and progress has been made in extracting CFFs and constraining GPD parametrizations. 
As discussed in the following sections, current efforts focus on the statistical and inverse nature of the extraction, including correlations and degeneracies among CFFs, parametrization dependence, uncertainty quantification, Bayesian and Monte Carlo methods, and flexible machine-learning representations.

This review is organized as follows. We first introduce the basic properties of GPDs and the theoretical constraints relevant to their phenomenological determination (Section \ref{sec:definitions}). In Section \ref{sec:EMT} we define the connetion of GPDs with the QCD energy momentum tensor and its form factors parametrization.  Experimental access through deeply virtual exclusive processes and the connection between measured observables and  CFFs in Section \ref{sec:experiment}. The GPD inverse problem and extraction are is reviewed in Section \ref{sec:inverse_prob}, with particular attention to uncertainty quantification, and the emerging role of AI and modern statistical methods. Finally, in Section \ref{sec:Limitations} we discuss the current limitations and open challenges in GPD phenomenology and the prospects for overcoming them with future measurements and new analysis methodologies in the EIC era.

%We next discuss the complementary information provided by lattice QCD and its incorporation into phenomenological analyses. 
% 
%Finally, this review is mostly dedicated to discussing GPDs and their extraction from DVCS on a nucleon
%target. For other related processes, we refer the reader to the reviews in \refcite{Favart:2015umi} (DVMP),
%\refcite{Dupre:2015jha} (nuclear DVCS), and to 
%\refcite{Amrath:2008vx} (DVCS from a pion). 

\section{Definitions and Fundamental Constraints}
\label{sec:definitions}

Below, we define the kinematic variables relevant to GPD physics, introduce the GPDs through their underlying QCD correlation functions, and summarize their basic properties.
\begin{enumerate}

\item Kinematic variables \(x,\xi,t,Q^2\)

GPDs are written in terms of lightcone variables, where four-vectors are defined $v = (v^{+}, v^{-}, v_{\perp}) $ where $v^{\pm} = 1/\sqrt{2} (v^{0} \pm v^{3})$. GPDs are functions of four variables, the momentum fraction $x$, the skewness $\xi$, the proton momentum transfer $t$, and the hard scale $Q^{2}$. The momentum fraction and skewness are typically written in one of two systems, by Ji with the ``symmetric" variables, $x, \xi$, \cite{Ji:1996nm}, and Radyushkin with the ``asymmetric" variables $X, \zeta$, \cite{Radyushkin:1997ki}. The proton momentum transfer is defined as $t = (p' - p)^{2}$, where $p,p'$ are the incoming and outgoing proton momenta.
We adhere to the ``symmetric" system of variables in what follows. To write the parameterization of QCD correlation functions in terms of GPDs, it is useful to define $P = (p + p')/2$, $\Delta = p' - p$, the average and relative proton momenta. 
The four variables $x, \xi, t, Q^{2}$ are not truly independent, in that, \textit{e.g.}, the $\xi$ and $t$ variables are coupled through a $t_{\mathrm{min}} = - 4 \xi^{2} M^{2} / (1-\xi^{2})$, or equivalent a $\zeta_{\mathrm{max}}$, where we have for simplicity neglected target mass corrections. Therefore, the limit $\xi \neq 0, t = 0$ is unphysical.

\item Quark and gluon GPDs \(H,E,\widetilde H,\widetilde E\)

The QCD quark-quark correlation functions $W_{\Lambda \Lambda'}^{[\Gamma]}$ that GPDs parameterize have the form,
\begin{eqnarray}
{W}_{\Lambda \Lambda'}^{[\Gamma]} = \frac{1}{2} \int \frac{dz^{-}}{2\pi} e^{i x P^{+} z^{-}} \langle p', \Lambda'| \bar{\psi}(-\frac{1}{2} z) \Gamma \mathcal{W} \psi{(\frac{1}{2} z)}| p, \Lambda \rangle |_{z^{+} = 0, z_{\perp} = 0},
\end{eqnarray}
where $\mathcal{W}$ represents the gauge link connecting a path between $-z/2$ and $z/2$. The gluon-gluon correlation functions are defined analogously, 
\begin{eqnarray}
    W_{\Lambda \Lambda'}^{\mu \nu; \rho \sigma} = \frac{1}{x P^{+}} \int \frac{d z^{-}}{2 \pi} e^{i x P^{+}z^{-}} \langle p', \Lambda' | 2 \: \Tr [G^{\mu \nu}(-\frac{z}{2}) \mathcal{W} G^{\rho \sigma}(\frac{z}{2}) \mathcal{W'}] \: |p, \Lambda \rangle_{z^{+} = 0, z_{\perp} = 0}. 
\end{eqnarray}
Important subtleties in the theory lie in the choice of this gauge link, typically taken as staple or straight, where the latter corresponds to the GPD limit of the generalized transverse momentum-dependent distribution (GTMD). A full classification of what is briefly defined here can be found in  \cite{Meissner:2009ww, Lorce:2013pza}. 

The four GPDs that enter at twist-two and correspond to initial and final quark states that do not flip their spin, or ``chiral even," are denoted $H, E$ for the vector case and $\tilde{H}, \tilde{E}$ for the axial vector case. They are defined through the vector and axial-vector twist-two quark-quark QCD correlation functions as, 
\begin{eqnarray}
    W_{\Lambda \Lambda'}^{[\gamma^{+}]}
    &=& \frac{1}{2 P^{+}} \bar{U}(p', \Lambda') \Big(\gamma^{+} H^{q}  +  \frac{i \sigma^{+\Delta}}{2M} E^{q} \Big) U(p, \Lambda) \\
    %= H^{q} \delta_{\Lambda \Lambda'} + \frac{\Lambda \Delta^1 + i \Delta^{2}}{2M} E^{q} \delta_{-\Lambda \Lambda'}\\
    W_{\Lambda \Lambda'}^{[\gamma^{+} \gamma_{5}]} &=&  \frac{1}{2P^{+}} \bar{U}(p', \Lambda') \Big(\gamma^{+} \gamma_{5} \tilde{H}^{q} + \frac{\Delta^{+}\gamma_{5}}{2M} \tilde{E}^{q}\Big) U(p, \Lambda)
    %= \Lambda \tilde{H}^{q} \delta_{\Lambda \Lambda'} + \frac{\Delta^{1} + i \Lambda \Delta^{2}}{2 M } \xi \tilde{E}^{q} \delta_{-\Lambda \Lambda'}, 
\end{eqnarray}
and for the vector and axial-vector twist-two gluon-gluon QCD correlation function, 
\begin{eqnarray}
    \delta_{\perp}^{ij} W_{\Lambda \Lambda'}^{+i; +j} &=& \frac{1}{2P^{+}} \bar{U}(p', \Lambda') \Big(\gamma^{+} H^{g} + \frac{i \sigma^{+ \Delta}}{2 M } E^{g}\Big) U(p, \Lambda) \\
    -i \epsilon_{\perp}^{ij} W_{\Lambda \Lambda'}^{+i; +j} &=& \frac{1}{2P^{+}} \bar{U}(p', \Lambda') \Big(\gamma^{+} \gamma_{5} \tilde{H}^{g} + \frac{\Delta^{+} \gamma_{5}}{2 M } \tilde{E}^{g} \Big) U(p, \Lambda). 
\end{eqnarray}
%The $H, \tilde{H}$ GPDs enter in a spin configuration in which the proton does not flip its spin, while the $E, \tilde{E}$ GPDs enter in the case in which the proton flips its spin. This explains exactly why there are no $E, \tilde{E}$ analogues to the PDFs, as the proton cannot flip its spin in a forward scattering amplitude.  

\noindent {\it Twist three}

\noindent Beyond the leading-twist description, twist-three GPDs encode additional quark--gluon correlations and transverse partonic dynamics, and enter exclusive observables through power-suppressed contributions of order $1/Q$ (Table \ref{tab:GPD}). They also provide important information on quark orbital angular momentum and on the distinction between genuine quark--gluon correlations and Wandzura--Wilczek-type contributions (we refer the reader to \cite{Meissner:2009ww} for the explicit definition of the correlation function.

\begin{table}[htp]
\centering
\begin{tabular}{|c|c|c|}
\hline
  $P_q P_p$ & TMD & GPD  \\
\hline 
%%% tw 3
  UU & $f^\perp$ &  $ 2\widetilde{H}_{2T} + E_{2T} $ \\
\hline 
  LL & $g_L^\perp$ & $2\widetilde{H}_{2T}' + E_{2T}' $ \\
\hline
 UL & $f_L^{\perp \, {\bf (*)}}$ & $ \widetilde{E}_{2T} - \xi E_{2T}$ \\
\hline
  LU & $g^{\perp \, {\bf (*)}}$ &  $ \widetilde{E}_{2T}' - \xi E_{2T}'$\\
\hline
 UT & $f_T^{\bf (*)}$ & $ H_{2T} + \, \tau \widetilde{H}_{2T} $
\\
\hline
  LT & $g_T'$ & $ H_{2T}' + \, \tau \widetilde{H}_{2T}'$  \\
 \hline
\end{tabular}
\caption{Twist-three GPDs and their helicity content. The first column shows the polarizations for the quark $P_q$, and proton, $P_p$; the second column shows the quark-proton polarization configurations in the transverse momentum distributions (TMD) sector, in the third column shows the the GPDs in the notation of \cite{Meissner:2009ww} are dispalyed. The asterisk denotes naive T-odd twist-three TMDs (a similar table appears in  \cite{Kriesten:2020wcx}).} 
\label{tab:GPD}
\end{table}

\item Chiral-even and chiral-odd sectors

The $H, E, \tilde{H}, \tilde{E}$ GPDs defined above are chiral-even as they correspond to correlation functions that contain operators that do not flip the quark helicity. However, QCD processes certainly allow for quark states that flip the helicity of the quark, and these correspond to ``chiral-odd" GPDs. At twist-two, the $H_{T}, E_{T}, \tilde{H}_{T}, \tilde{E}_{T}$ quark chiral-odd GPDs are defined through, 
\begin{eqnarray}
    W_{\Lambda \Lambda'}^{[i \sigma^{i+} \gamma_{5}]} &=& \frac{i \epsilon^{ij}}{2 P ^{+}} \bar{U}(p', \Lambda') \Big(i \sigma^{+j} H_{T} + \frac{\gamma^{+} \Delta^{j} - \Delta^{+} \gamma^{j}}{2 M } E_{T} + \frac{P^{+} \Delta^{j}}{M^{2}} \tilde{H}_{T} - \frac{P^{+} \gamma^{j}}{M} \tilde{E}_{T}\Big) U(p, \Lambda)
    %\\&=& \Big(\frac{i \epsilon^{ij} \Delta^{j}}{2M} (E_{T} + 2 \tilde{H}_{T}) + \frac{\Lambda \Delta^{i}}{2M} (\tilde{E}_{T} - \xi E_{T} ) \Big) \delta_{\Lambda \Lambda'} + \Big( (\delta_{i1} + i \Lambda \delta_{i2}) H_{T} - \frac{i \epsilon^{ij} \Delta^{j} (\Lambda \Delta^{1} + i \Delta^{2})}{2M^2} \Big) \delta_{-\Lambda \Lambda'}. \nonumber 
\end{eqnarray}

\item Forward limits and connections to PDFs

The GPDs $H, \tilde{H}, H_{T}$ reduce to the PDFs in the forward limit, in which $\xi, t = 0$. Explicitly, $H^{q}(x, 0, 0) = q(x)$, $H^{g}(x, 0, 0) = x g(x)$, $\tilde{H}^{q}(x, 0, 0) = \Delta q(x) $, $H^{g}(x, 0, 0) = x \Delta g(x)$, $H_{T}^{q}(x, 0, 0) = \delta q(x)$. 

\item Electromagnetic and energy-momentum tensor form factors

Integration over the momentum fraction $x$ shows crucial correspondences of the GPDs with the electromagnetic form factors and elements of the energy momentum tensor. The first Mellin moment of the GPDs $H, E$ are given by the Dirac $F_{1}$ and Pauli $F_{2}$ form factors, 
\begin{eqnarray}
    \int_{-1}^{1} dx H^{q}(x, \xi, t, Q^{2}) &=& F_{1}^{q}(t) \\
    \int_{-1}^{1} dx E^{q}(x, \xi, t, Q^{2}) &=& F_{2}^{q}(t)
\end{eqnarray}
The second Mellin moment of the GPDs is related to the $A, B, C$ generalized form factors (GFFs) of the energy-momentum tensor, 
\begin{eqnarray}
    \int_{-1}^{1} dx x H(x, \xi, t, Q^{2}) &=& A(t, Q^{2}) + (2\xi)^{2} C(t, Q^{2}) \\
    \int_{-1}^{1} dx x E(x, \xi, t, Q^{2}) &=& B(t, Q^{2}) - (2\xi)^{2} C(t, 
    Q^{2}). 
\end{eqnarray}
The sum of the two formulae above gives exactly the Ji sum rule further discussed in Section \ref{sec:EMT}. 

\item Polynomiality, positivity, symmetry properties, 
and QCD evolution

The first and second Mellin moment properties above are special cases of a general property of the Mellin moments of GPDs called polynomiality: as a consequence of Lorentz invariance, the $x^{n}-$weighted integrals of GPDs are polynomials in the skewness $\xi$, the coefficients of which are given by functions of only $t$ and $Q^{2}$ that are GFFs. Explicitly, 
the Mellin moments are defined as, 
\begin{equation}
    M_n^q(\xi, t) \equiv \int_{-1}^{1} dx x^{n-1} F_q(x, \xi, t), \quad\quad F=H,E, 
\end{equation}
and in terms of GFFs, 
\begin{eqnarray}
\label{eq:H_moment}
    \int_{-1}^{1} dx x^{n-1} H_q(x, \xi, t) & = & \sum_{i=0,even}^{n-1} (2\xi)^i A^q_{n,i}(t) 
    + \mod(n-1,2) (2\xi)^{n} C_{n}^q(t) \\
\label{eq:E_moment}
    \int_{-1}^{1} dx x^{n-1} E_q(x, \xi, t) & = & \sum_{i=0,even}^{n-1} (2\xi)^i B^q_{n,i}(t) 
    - \mod(n-1,2) (2\xi)^{n} C_{n}^q(t) .
\end{eqnarray}
The GPDs are also bounded by the PDFs, following a property called positivity. For the GPDs $H, E$, for example, this is realized as, 
\begin{eqnarray}
    (1-\xi^2)\Big(H^{q} -\frac{\xi^2}{1-\xi^2}E^{q}\Big)^2 + \frac{t_{\mathrm{min}} - t}{4 M^2} ( E^{q})^2 \leq q(x_{\mathrm{in}}) q(x_{\mathrm{out}}), 
\end{eqnarray}
where definitions for $x_{\mathrm{in}}, x_{\mathrm{out}}$ are given in, e.g., \cite{Diehl:2003ny} as functions of $x$ and $\xi$.
Furthermore, similarly to the PDFs, GPDs obey symmetry conditions in their $x$ dependence. In particular, for the GPD $H$, $H^{q}(x) = -H^{\bar{q}}(-x)$, and, therefore, the ``$+$" component, $H^{+}_{q} = H^{q} + H^{\bar{q}}$, is antisymmetric with respect to $x$, and the ``$-$" component, $H^{-}_{q} = H_{q} - H_{\bar{q}}$, is symmetric with respect to $x$. The gluon distribution $H^{g}$ possesses the same symmetry, but in the case of the polarized distributions $\tilde{H}$, the symmetry is instead, $\tilde{H}^{q}(x) = H^{\bar{q}}(-x)$, 
and so the ``$+$" and ``$-$" components have the opposite symmetry. 

A final property that we mention here is the QCD evolution of the GPDs that governs their dependence on $Q^{2}$. As GPDs are collinear objects, their evolution follows a similar structure to that of PDFs, with the caveat that instead of one active variable in the $x$ space like in the case of PDFs, the GPDs have an additional dependence on $\xi$ in evolution. In other words, the perturbatively calculated splitting functions have both an $x$ and $\xi$ dependence. Additionally, the GPDs possess two different modes of $Q^{2}$ evolution, the ``DGLAP" region and the ``ERBL" region that correspond to the evolution in different ranges of $x$, where for quarks, $x > \xi$ obeys DGLAP and and $-\xi < x < \xi$ obeys ERBL evolution, while for antiquarks, $-1 < x < -\xi$ obeys DGLAP and $- \xi < x< \xi$ obeys ERBL evolution. We refer the reader to the literature for details on the evolution equations and splitting functions, {\it e.g.} \cite{Musatov:1997pu,Golec-Biernat:1998zbo,Bertone:2023jeh,Braun:2022bpn,Panjsheeri:2025vpa}.

\item {Impact-parameter distributions}
%\label{sec:impact}
\noindent GPDs are powerful tools for imaging the internal structure of the proton. Without invoking the full GPD formalism, one can gain significant insight regarding the proton's spatial structure through the electromagnetic form factors probed in elastic electron-nucleon scattering. The Dirac $F_{1}$ and Pauli $F_{2}$ form factors are related to the proton's electric charge $\rho(b_{\perp})$ and magnetic moment $\rho_{M}(b_\perp)$ density distributions, respectively, upon Fourier transformation in transverse momentum transfer $\Delta_{\perp}$,
where $b_{\perp}$ is the impact parameter in the transverse plane. 
%The significance of defining these densities in the transverse plane is described in Sec. \ref{sec:2D_3D}. 
A rich literature exists for the investigation of these densities, \cite{Miller:2007uy, Venkat:2010by}, where much emphasis has been put on analyzing the consequences of the flavor separation of these form factors in experiment \cite{Cates:2011pz, Qattan:2012zf, Qattan:2015qxa}.

The special role played by GPDs in imaging the proton's spatial structure lies in the correlations between their \(x\) and \(t\) dependence. In a light-front description, these correlations connect the fraction of the proton's longitudinal light-cone momentum carried by a parton with its transverse spatial distribution, the latter being obtained, at zero skewness, $\xi$, through a two-dimensional Fourier transform with respect to the transverse momentum transfer,
%At zero skewness, (\xi=0), where (t=-\boldsymbol{\Delta}\perp^2), the latter is obtained through a 
\begin{equation}
q(x,\mathbf{b}_\perp)=
\int \frac{d^2\boldsymbol{\Delta}_\perp}{(2\pi)^2}\,
e^{-i\mathbf{b}\perp\cdot\boldsymbol{\Delta}_\perp}
H(x,0,-\boldsymbol{\Delta}_\perp^2)\, \qquad \quad q_{X}(x, {\bf{b}_{\perp}}) = \int \frac{d^2\boldsymbol{\Delta}_\perp}{(2\pi)^2}, 
e^{-i\mathbf{b}\perp\cdot\boldsymbol{\Delta}_\perp}
\Big( H(x,0,-\boldsymbol{\Delta}_\perp^2) + i \frac{\Delta_{y}}{2M} E(x,0,-\boldsymbol{\Delta}_\perp^2) \Big)
\end{equation}
where $\mathbf{b}_\perp$ denotes the transverse position of an unpolarized parton relative to the proton's transverse center of momentum, and $q(x, {\bf{b}_{\perp}})$ and $q_{X}(x, {\bf{b}_{\perp}})$ are the spatial densities for an unpolarized and transversely polarized proton, respectively, \cite{Soper:1976jc, Burkardt:2000za, Diehl:2002he}.
Studying nontrivial $x, b_{\perp}$ correlations is at the heart of proton imaging. See Fig. \ref{fig: FT} for an illustration of these Fourier transformations. Average partonic radii $\langle b_{T}(x)\rangle^{1/2}$, defined as expectation values in $b_{\perp}$ of the density $\rho(x, b_{\perp})$ that follow a decreasing behavior as a function of $x$ would indicate the presence of small size configurations and serves as a probe of color transparency, \cite{Jain:1995dd, Belitsky:2003nz, Liuti:2004hd, Dutta:2012ii, Brodsky:2022bum}.
%In the above discussion on spatial imaging, the Fourier transformations of the electromagnetic form factors and the GPDs into coordinate space were defined for the transverse components $b_{\perp}$ of the impact parameter. 
%The transverse plane in lightcone variables has the distinct advantage in that it is isomorphic to the Galilean group and therefore avoids the trouble of relativistic corrections, Refs. \cite{Kogut:1969xa, Soper:1976jc}. 
%However, the proton is after all a three-dimensional object, 
%so appealing only to a two-dimensional picture would leave the quest for imaging the proton fundamentally incomplete. 
%Additionally, studies such as Ref. \cite{Miller:2007uy} indicate that there certainly is a difference in whether one works in 2D or 3D, considering the fact that the Breit frame and the transverse plane Fourier transformations give completely opposite images for the charge density of the neutron. 
The difficulty in defining a three-dimensional spatial density for a relativistic system is more fundamental than the presence of relativistic corrections. It is related to the well-known limitation on localizing a relativistic particle below its Compton wavelength. At distances of order \(1/M\), the energy required for localization becomes sufficiently large that particle creation can no longer be neglected, and a description in terms of a fixed number of constituents loses its meaning. For the proton, its Compton wavelength is not negligible compared with its spatial extent, making a conventional three-dimensional density interpretation intrinsically problematic. This difficulty motivates the use of the light-front formulation, where transverse spatial distributions can be defined without requiring localization in the longitudinal direction.
%As Jaffe strongly emphasized, the Compton wavelength must be much smaller than the intrinsic size of the system in order to localize the nucleon and for the three-dimensional Fourier transformation to be free of relativistic corrections, \cite{Jaffe:2020ebz}. This condition is met for nuclei, Jaffe demonstrates, as their masses are large and therefore their Compton wavelengths small compared to their intrinsic size, but for a nucleon the two are of the same order of magnitude.
As emphasized in \cite{Jaffe:2020ebz}, a spatial density requires the system to be localized in a wave packet whose extent is small compared with the intrinsic size being resolved. 
%Relativity, however, prevents arbitrarily sharp localization, with the Compton wavelength setting the relevant scale. 
Only when the intrinsic size of the system is parametrically larger than its Compton wavelength can one choose a localization scale that resolves the internal structure while keeping relativistic effects under control, thereby recovering an essentially wave-packet-independent three-dimensional density from the Fourier transform of the corresponding form factor. This hierarchy is well satisfied for nuclei, but not for the nucleon, whose Compton wavelength and intrinsic size are not sufficiently separated. 
%Consequently, the conventional three-dimensional Fourier-transform interpretation does not define an unambiguous intrinsic spatial density of the nucleon.
Attempts to reconcile the 2D and 3D pictures include the Abel transformation, \cite{Moiseeva:2008qd, Rajan:2018zzy, Panteleeva:2021iip}, refinements in the choice of wavepackets, e.g., \cite{Epelbaum:2022fjc}, and Wigner rotations, \cite{Lorce:2020onh}, but there always persists the impasse of the Compton wavelength $\lambda \sim \frac{1}{M}$, where $M$ is the hadron mass, leading some to suggest that the goal of 3D imaging is unachievable, \cite{Miller:2025zte}. 
Advantages intrinsic to the lattice appear to provide a path forward, \cite{Liu:2026pbf}, but an experimental probe that resolves the problem of the Compton wavelength is an open question for further investigation.

\begin{figure}[h!]
\includegraphics[width=15cm]{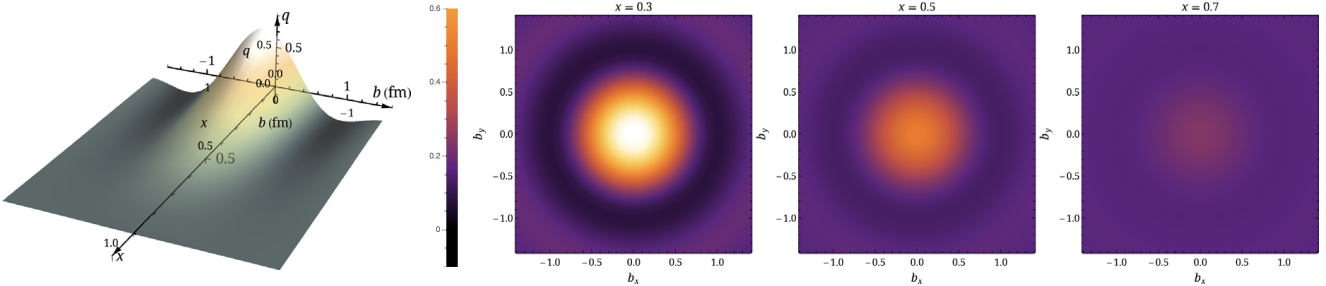}
\centering
\caption{Adapted from \cite{Lin:2020rxa}. Fourier transformation of lattice QCD calculations of the GPD $H$ in $x, t$ space at the physical pion mass within the LaMET framework.}
\label{fig: FT}
\end{figure}

Alongside the ongoing effort to image the proton through GPDs, several other exciting and complementary approaches to hadron imaging are being developed. Mantysaari and Schenke have performed significant studies on coherent and incoherent $J/\Psi$ production to model the distribution of quarks and gluons, which they refer to as ``gluon hotspots," \cite{Mantysaari:2016jaz, Mantysaari:2016ykx}. Contrary to this point of view, meanwhile, there is yet another approach to imaging the proton emphasizing the transport of baryon number. Rossi and Veneziano have posited that the only gauge invariant configuration in SU(3) is a 3-junction, in which the valence quarks are three prongs connected by a Y-shaped junction of gluons, a ``baryon junction," and Kharzeev proposed an experimental avenue to probe this through baryon stopping, \cite{Rossi:1977cy, Kharzeev:1996sq}. Intriguing recent experimental advances point to progress in this direction, e.g. \cite{STAR:2024lvy}. Unifying the gluon hotspots, baryon junctions, and GPD-based pictures would provide a comprehensive account of the proton's complicated spatial structure. Interfacing baryon junctions with GDPs, \cite{Panjsheeri:2025zrm}, which has analogues in PDF studies, e.g. \cite{Liu:2012ch, Hou:2022ajg}, and of testing the gluon hotspot model with GPDs, \cite{Panjsheeri:2024ysh, Panjsheeri:2024gmw}, are matters of ongoing research.

\end{enumerate}

\section{The energy-momentum tensor}
\label{sec:EMT}

The energy-momentum tensor (EMT) of QCD encodes fundamental information of hadrons. Properties such as mass, spin, and the $D$-term, as well as their associated spatial distributions, can be obtained by deciphering the information contained in the EMT. In this section, we focus on the spin-1/2 EMT. 

For a spin-1/2 hadron, a symmetric EMT (Belinfante-improved EMT) can be parametrized by four EMT form factors, \cite{Ji:1996ek}, as
\begin{equation}
    \left\langle p'\left|\hat T_a^{\mu\nu}(0)\right|p\right\rangle = \bar u(p') \left[ A_a(t)\frac{P^\mu P^\nu}{M} + J_a(t)\frac{i\left(P^\mu\sigma^{\nu\rho}+P^\nu\sigma^{\mu\rho}\right)\Delta_\rho}{2M} + D_a(t)\frac{\Delta^\mu\Delta^\nu-g^{\mu\nu}\Delta^2}{4M} + M\bar c_a(t)g^{\mu\nu} \right] u(p)
\end{equation}
where $P=(p_1+p_2)/2$, $\Delta = p' - p$, and $t=\Delta^2$, and $a$ denotes all partonic species, $a=u, d, \dots , g$. Although individual form factors depend on the renormalization scale and do not need to be conserved at the partonic level, the conservation of the total EMT, $\partial_\mu T^{\mu\nu} = 0$, constrains the form factor $\sum_a \bar{c}_a (t) = 0$.

Moreover, the total form factors $A$ and $J$ are constrained in the forward limit at $t=0$ as $A(0)=1, J(0)=\frac12$, which reflects that the total longitudinal and angular momentum are carried by the constituents of the particle, respectively, \cite{Ji:1996ek, Cotogno:2019xcl}. In contrast, the form factor $D(0)$, referred to as the $D$-term, has no constraint and must be determined experimentally, which has led to a series of studies to determine the $D$-term of the proton. 

One obstacle to measuring EMT form factors is that EMT couples extremely weakly to gravity which makes it practically impossible to measure them directly. GPDs come to the rescue here by providing an indirect route to this difficulty. The second Mellin moments of unpolarized GPDs, $H^q$ and $E^q$, are related to the EMT form factors as follows 
\begin{align}
    \int_{-1}^{1}dx\,x\,H^q(x,\xi,t) &= A_q(t)+ \xi^2D_q(t) \, , \\
    \int_{-1}^{1}dx\,x\,E^q(x,\xi,t) &= 2J_q(t) - A_q(t) - \xi^2D_q(t) \, .
\end{align}
leading to the Ji's angular momentum sum rule in the forward limit
\begin{equation}
\label{eq:Ji_sumrule}
    J_q(0) = \frac12 \int_{-1}^{1}dx\,x \left[ H^q(x,0,0) + E^q(x,0,0) \right] \, .
\end{equation}
So, knowing the unpolarized GPDs can also fully determine the EMT form factors, $A, J,$ and $D$, hence all fundamental properties of the proton. The notation in Sec. \ref{sec:definitions} corresponds to the EMT form factors here by $D = 4 C, J = \frac{1}{2}(A + B)$, which a common difference in notation in the literature.

But the next question is how to relate these form factors to the physical distributions inside the nucleon. Similar to the charge distributions obtained from the electromagnetic form factors, as argued in \cite{Polyakov:2002yz, Polyakov:2018zvc}, one can reveal the 3D spatial content of the EMT form factors, modulo the relativistic corrections (see Sec.~\ref{sec:definitions}), by taking the inverse Fourier transform with respect to the momentum transfer in the Breit frame, where the momentum transfer is purely spatial, i.e., $\Delta^0 = 0$
\begin{equation}
    T^{\mu\nu}(\vec r, \vec s) = \int \frac{d^3\Delta}{2E(2\pi)^3}\, e^{-i\vec{\Delta}\cdot\vec r} \left\langle p', s' \left| \hat T^{\mu\nu}(0) \right|p, s\right\rangle \, .
\end{equation}
Energy, angular momentum, pressure, and shear force distributions are related to the specific components of this static tensor. Specifically
\begin{equation}
    \varepsilon(r) \equiv  T^{00}(\vec r), \qquad s^i J_{\mathrm{mono}}(r) + s^j \left( \frac{r^i r^j}{r^2} -\frac{\delta^{ij}}{3} \right)  J_{\mathrm{quad}}(r) \equiv \epsilon^{ijk}r^jT^{0k}(\vec r, \vec s), \qquad \left(\frac{r^i r^j}{r^2} -\frac{\delta^{ij}}3\right)s(r) +\delta^{ij}p(r) \equiv T^{ij}(\vec r) \, .
\end{equation} 
It was shown in \cite{Lorce:2017wkb} that the angular momentum distribution can be decomposed in terms of monopole, $J_{\mathrm{mono}}$, and quadrupole, $J_{\mathrm{quad}}$, contributions, but they are linearly related to each other, as shown in \cite{Schweitzer:2019kkd}, by
\begin{equation}
    J^a_{\mathrm{quad}}(r)=-\frac{3}{2} J^a_{\mathrm{mono}}(r) \, .
\end{equation}
On the other hand, the spatial components of the static EMT can be decomposed in terms of a trace part, pressure ($p$), and a traceless part, shear force ($s$), and they are related to each other by, due to the total EMT conservation, see \cite{Goeke:2007fp, Polyakov:2018zvc} for details, by the first-order differential equation
\begin{equation}
    \frac{dp(r)}{dr} +\frac{2}{3}\frac{ds(r)}{dr} +\frac{2}{r}s(r)=0 \, .
\end{equation}
Therefore, GPDs play a fundamental role in revealing energy, angular momentum, pressure, and shear force distributions through their connection to EMT form factors. For an analogy between the EMT, superconductivity, and the cosmological constant, see \cite{Liu:2023cse}. 
\\

One notices here that so far the particular distribution that corresponds to the angular momentum orbital component (OAM) has not been identified. 
%but rather only the combination of distributions that give the total angular momentum. 
In Ji's description, one can take $L = J - S$, where $J$ is given by Eq.\eqref{eq:Ji_sumrule}, and $S$ is identified with $\Delta \Sigma_{q}$. The relation to the Jaffe--Manohar definition \cite{Jaffe:1989jz} can be understood through the gauge-link structure: while a straight gauge link corresponds to Ji OAM, a staple-shaped light-cone gauge link yields the Jaffe--Manohar OAM , with the difference between the two associated with the effect of the color field on the struck quark \cite{Burkardt:2012sd}.
Later developments by Lorc\'e and Pasquini singled out an example of a distribution more general than GPDs, called a generalized transverse momentum-dependent distribution (GTMD), namely $F_{14}$, to be responsible for OAM, Ref. \cite{Lorce:2011kd}, although it has remained very much a matter of speculation whether GTMDs can be directly observed \cite{Echevarria:2022ztg,Bhattacharya:2023yvo}. On the other hand, Kiptily and Polyakov demonstrated that a combination of a particular twist-three GPD, $\tilde{E}_{2T}$, and the twist-two GPDs $H, E$ also correspond to OAM,  \cite{Kiptily:2002nx}. A proof of the correspondence between the GTMD $F_{14}$ and the GPD $\tilde{E}_{2T}$ using Lorentz invariance relations and QCD equation of motion relations was given in \cite{Rajan:2016tlg,Rajan:2017cpx}. The correspondence of these sum rules within the theory has been verified in lattice calculations, \cite{Engelhardt:2017miy, Engelhardt:2024kcf}.
The above discussion applies for the longitudinal component of angular momentum. Sum rules that also exist for the transverse spin case, \cite{Ji:2020hii, Bakker:2004ib, Leader:2011cr, Ji:2012vj, Hatta:2012jm, Leader:2012ar, Lorce:2018zpf, Alkasassbeh:2024aws}, as well as for spin-1 systems like the deuteron, e.g., \cite{Taneja:2011sy}.

\section{Experimental Access} 
\label{sec:experiment}

%\begin{enumerate}
%\item Introduction: Experiments at JLab, HERA, COMPASS, Future EIC and their sensitivity to 
%unpolarized and polarized observables: beam-spin, target-spin, and charge asymmetries
%\item DVCS and Bethe–Heitler interference
%\item DVMP
%\item Timelike Compton scattering
%\item Double DVCS/Other exclusive channels/Nuclei
%\item UPCs
%\end{enumerate}

Accessing GPDs is widely accepted to be a challenging task because they are not directly observable, but appear in a convoluted form in the expression of observables. Moreover, the channels that are sensitive to GPDs may interfere with other processes. Deeply virtual Compton scattering (DVCS), for instance, is known as the golden channel for accessing leading-twist, chiral-even GPDs. The DVCS signal, however, is often obscured by the Bethe-Heitler process, a purely electromagnetic subprocess that can be expressed in terms of the Dirac and Pauli form factors. In a wide range of kinematics, the BH process introduces significant background to access GPDs through the DVCS process, as illustrated for instance in \cite{Vanderhaeghen:1999xj, Aschenauer:2013hhw, Aschenauer:2025cdq}. Although the BH contribution highly dominates the cross section in certain kinematics, thanks to the rather accurate knowledge of Dirac and Pauli form factors, it is possible to constrain the DVCS amplitude through its interference with DVCS. But even so, DVCS measurements alone are not sufficient for complete flavor separation of GPDs, and available data have limited kinematic coverage, see Fig~\ref{fig:EICimpact}. For this reason, the extraction of GPDs requires a coordinated effort across the hadron-physics community, including experimental data collection from complementary processes, such as TCS, DDVCS, and DVMP (see Fig.~\ref{fig:dvcstcsdvmp}),  performing first-principles lattice-QCD calculations, and conducting phenomenological and model studies to gain insights about them.

\subsection{DVCS and BH processes}
\label{sec:DVCS-BH}

The DVCS process, $\gamma^* p \rightarrow \gamma p'$, is often regarded as the golden channel for accessing GPDs, \cite{Ji:1996nm}. Experimentally, it is accessed through an exclusive process in which all initial and final state particles are detected in the electroproduction of a photon off a nucleon target
\begin{equation}
    e + p \rightarrow e' + p' + \gamma
\end{equation}
In DVCS, the incoming lepton emits a spacelike virtual photon that interacts with the partonic structure inside the nucleon to emerge as a real photon. However, the DVCS process has a substantial background, often dominating, from the BH process, as has been illustrated in numerous studies \cite{Vanderhaeghen:1999xj, Aschenauer:2013hhw, Aschenauer:2025cdq}, which has exactly the same final state as DVCS but instead emits the outgoing photon from the incoming or outgoing lepton. 

Because the DVCS and BH processes have the same final state, the total amplitude becomes $\mathcal{T} = \mathcal{T}_{DVCS} + \mathcal{T}_{BH}$, and the differential cross-section can be written in terms of three contributions: pure DVCS, pure Bethe-Heitler, and their interference term as 
\begin{equation}
    \frac{d\sigma}{dx_{Bj}\,dQ^2\,d|t|\,d\phi\,d\phi_S}= \frac{\alpha_{em}^3}{16\pi^2 x_{Bj} (s-M^2)^2 \sqrt{1+\gamma^2}} \bigg|\frac{\mathcal{T}}{e^3}\bigg|^2
\end{equation}
where $\alpha_{em}$ is the electromagnetic fine structure constant and $|\mathcal{T}|^2$ = $|\mathcal{T}|_{DVCS}^2$ + $|\mathcal{T}|_{BH}^2$ + $\left(\mathcal{T}_{DVCS}^* \mathcal{T}_{BH} + \mathcal{T}_{BH}^* \mathcal{T}_{DVCS}\right)$. 

\begin{figure}[H]
    \centering
    \includegraphics[width=0.85\linewidth]{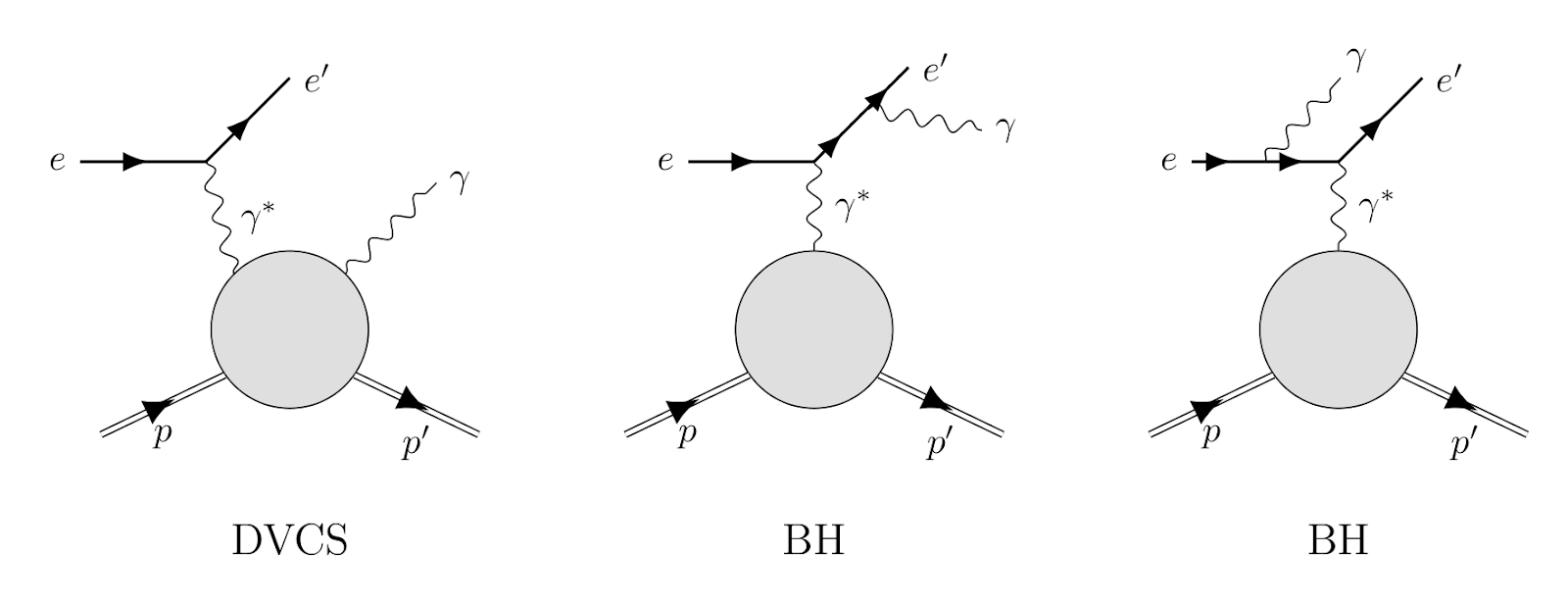}
    \caption{DVCS and BH processes. In BH processes, the photon can be emitted from either the incoming or outgoing electron.}
    \label{fig:dvcs_bh}
\end{figure}

The squared amplitudes of DVCS and BH, as well as their interference term in the cross section, was decomposed in terms of Fourier harmonics of the azimuthal angle $\phi$, with terms proportional to $\mathrm{cos(n\phi)}$ and $\mathrm{sin(n\phi)}$ in \cite{Belitsky:2001ns,Belitsky:2010jw}. If transverse target polarization is present, additional $\phi_S$ terms appear in the decomposition. At leading twist, the expression of the cross section considerably simplifies as it only involves four chiral-even Compton form factors, $\mathcal{H},\mathcal{E},\mathcal{\tilde{H}},\mathcal{\tilde{E}}$, associated with the four twist-2 chiral-even GPDs, $H, E, \tilde{H}, \tilde{E}$, respectively, whereas the BH amplitude includes only two nonperturbative inputs, the Dirac form factor $F_1(t)$, and the Pauli form factor $F_2(t)$, in addition to the known kinematic factors and leptonic propagators. 

Going beyond the leading-twist approximation introduces additional difficulties in extracting CFFs and GPDs, as the cross-section expressions become substantially more involved. Twist-3 contributions, which are suppressed by powers of $1/Q$ relative to the leading-twist amplitude, involve additional CFF combinations and Fourier harmonics. Moreover, genuine twist-3 GPDs associated with quark-gluon-quark correlators appear at this order \cite{Belitsky:2000vx}. Therefore, if higher-twist contributions are neglected at moderate values of $Q^2$, a leading-twist fit absorbs these effects into the extracted CFFs, resulting in biased estimates of their magnitudes and kinematic dependence. All these additional complexities make GPD extraction more sensitive to model dependence.
In this context, new formulations of deeply virtual exclusive processes exist that propose different organizations of the twist-two, twist-three terms, and  kinematic power corrections in the cross section which are more amenable to phenomenological analyses \cite{Braun:2011zr,Braun:2012hq,Kriesten:2019jep,Kriesten:2020wcx,Kriesten:2020apm,Qiu:2022pla}. 

In addition to the leading-twist approximation, which determines the relation between observables and CFFs, the perturbative order in the strong coupling $\alpha_s$ determines the coefficient functions as well as the GPD evolution that relate CFFs to GPDs. At leading order (LO), the DVCS hard-scattering coefficient functions couple directly to quark GPDs, whereas at next-to-leading order (NLO) one-loop corrections to the quark coefficient functions and direct gluon contributions enter the Compton form factors \cite{Ji:1998xh}. In an NLO analysis, one must consistently combine NLO coefficient functions with NLO GPD evolution using specified renormalization and factorization scales. Therefore, perturbative corrections beyond LO can substantially modify both the real and imaginary parts of CFFs, which in turn affects the GPD extraction \cite{Moutarde:2013qs}.  

Within collinear factorization for DVCS, \cite{Collins:1998be}, CFFs can generically be written in terms of GPDs in a convoluted form as 
\begin{equation}
    \mathcal F(\xi,t,Q^2) = \sum_{i} \int_{-1}^{1}dx\, C_{\mathcal F}^{i} \left( x,\xi,\frac{Q^2}{\mu_F^2}, \alpha_s(\mu_R)\right) F^i(x,\xi,t;\mu_F^2) 
\end{equation}
where $i$ runs through all partonic degrees of freedom $i = u,d,s,\ldots,g$. Here, $C_{\mathcal F}^{i}$ includes the charge factors and denotes the hard scattering coefficient function which depends on the hard scale $Q^2$, factorization scale $\mu_F$, and the renormalization scale $\mu_R$, and can be expanded perturbatively order by order in powers of $\alpha_s(\mu_R)$. Up to NLO, for instance, the coefficient function is expanded as 
\begin{equation}
    C_{\mathcal F}^{\,i} = C_{\mathcal F}^{\,i,(0)} + \frac{\alpha_s(\mu_R)}{2\pi} C_{\mathcal F}^{\,i,(1)} + \mathcal O(\alpha_s^2) \,,
\end{equation}
with the note that gluons first enter directly at NLO, i.e., $C_{\mathcal F}^{\,g,(0)}=0$. More recently, DVCS coefficient functions were also calculated at NNLO order in $\alpha_s$ in \cite{Braun:2020yib, Braun:2021grd, Braun:2022bpn, Ji:2023xzk}. 

For DVCS off a proton target, $\gamma ^* p \rightarrow \gamma p$, at the leading twist amplitude and at leading order in $\alpha_s$, both vector, ${\cal F}_q$ = (${\cal H}_q$, ${\cal E}_q$), and axial-vector, $\widetilde{\cal F }_q$= ($\widetilde{\cal H}_q$ $\widetilde{\cal E}_q$), components contribute, resulting in the following CFF expressions \cite{Vanderhaeghen:1999xj}
\begin{subequations}
\label{eq:CFFdef}
\begin{eqnarray}
\mathcal{F}_q(\xi,t)  & = &  C^+ \otimes F_q \equiv \int_{-1}^{1} dx  \,  C^+(x,\xi) F_q(x,\xi,t), 
 \\
\mathcal{\widetilde{F}}_q(\xi,t) & = & C^-  \otimes  \widetilde{F}_q  \equiv \int_{-1}^{1} dx  \,  C^-(x,\xi) \widetilde{F}_q(x,\xi,t) , 
%\widetilde{\mathcal{F}} = \int_{-1}^{1} dx  \widetilde{F}(x,\xi,t)  C^-(x,\xi) ,
\label{eq:CFF}
\end{eqnarray}
\end{subequations}
where $F_q= H_q, E_q$, $\widetilde{F}_q= \widetilde{H}_q, \widetilde{E}_q$ are the quark GPDs. The proton CFFs can then be obtained by summing over flavors weighted with their associated charge terms 
\begin{equation}
    \mathcal F^p = \sum_q e_q^2\,\mathcal F^q, \qquad \widetilde{\mathcal F}^{\,p} = \sum_q e_q^2\,\widetilde{\mathcal F}^{\,q} \, .
\end{equation}
Also, at leading order, the charge-independent coefficient functions simplify significantly and are given by 
 \begin{equation}
 \label{eq:CFFpm}
 { C^\pm(x,\xi) = \frac{1}{x-\xi + i \epsilon} \pm \frac{1}{x+\xi - i \epsilon} . } %C^\pm(x,\xi) = \frac{1}{x-\xi - i \epsilon} \pm \frac{1}{x+\xi - i \epsilon},
 \end{equation}
With these conventions, the integrals expressing the CFFs can be separated into the real and imaginary parts, with the real part given by the Cauchy principal value of the convolution integral, whereas the imaginary part comes from the poles at $x=\pm \xi$
\begin{subequations}
\label{eq:CFF}
\begin{eqnarray}
\mathcal{F}_q(\xi,t)  & = & \Re e \mathcal{F}_q + i \, \Im m \mathcal{F}_q = P.V. \int_{-1}^{1} dx  \left( \frac{1}{x-\xi} + \frac{1}{x+\xi} \right)  F_q(x,\xi,t) {- i \pi \, \big(F_q(\xi,\xi,t) - F_q(-\xi,\xi,t)\big)}  , 
 \\
\mathcal{\widetilde{F}}_q(\xi,t) & = &  \Re e \mathcal{\widetilde{F}}_q +  i \, \Im m \mathcal{\widetilde{F}}_q = P.V. \int_{-1}^{1} dx  \left( \frac{1}{x-\xi} - \frac{1}{x+\xi} \right)  \widetilde{F}_q(x,\xi,t) {- i \pi \, \big(\widetilde{F}_q(\xi,\xi,t) + \widetilde{F}_q(-\xi,\xi,t)\big)} , 
%\widetilde{\mathcal{F}} = \int_{-1}^{1} dx  \widetilde{F}(x,\xi,t)  C^-(x,\xi) ,
\end{eqnarray}
\end{subequations}
with the coefficient functions obeying the following symmetric/antisymmetric relations under $x \rightarrow -x$
\begin{subequations}
\begin{eqnarray}
C^-(-x,\xi) & = & C^-(x,\xi) \quad\quad symmetric \\
C^+(-x,\xi) & = & -C^+(x,\xi) \quad\,\, antisymmetric \, .
\end{eqnarray}
\end{subequations}

%\subsubsection{Flavor-dependent transverse imaging}
The difference in the $u$ and $d$ quarks is heavily emphasized in the literature of imaging from electromagnetic form factors, as discussed in Sec. \ref{sec:definitions}. Calculations on the lattice for the strangeness and charm form factors, \textit{e.g.} \cite{Green:2015wqa, Sufian:2020coz} show a nonzero contribution to the nucleon form factors from these flavors. Studies of the impact of the strangeness form factor on imagining, \cite{Diehl:2007uc}, and flavor dependence in the context of quantum entanglement, \cite{Hatta:2025obw}, highlight the significance of GPD flavor separation. Recent advances in the lattice calculations of the second Mellin moments, $A_{20}, B_{20}$, of the GPDs $H, E$, \cite{Hackett:2023rif}, have served as a robust constraint on GPD phenomenology, particularly the flavor separation of the antiquark $\bar{u}, \bar{d}$ GPDs,  \cite{Panjsheeri:2025vpa}. 

The imaging of the GPD $E$ emphasizes the importance of flavor separation, as one can show that the GPD $E$ induces a ``distortion" in the transverse plane which occurs in opposite directions for the $u$ and $d$ quarks due to differences in the anomolous magnetic moment, \cite{Burkardt:2002hr}. Differences in flavor impact also the distribution of the angular momentum density, \cite{Lorce:2011kd}, as well as the imaging of transversity, \cite{Diehl:2005jf, Goldstein:2014aja}. 
%\subsubsection{2D versus 3D imaging} \label{sec:2D_3D}
%\subsubsection{Gluon imaging and small-$x$ structure}

\subsection{Deeply virtual meson production}

In the deeply virtual meson production (DVMP) process, $\gamma^*p\rightarrow M p'$, a meson $M$ is produced in the final state instead of a real photon, see Fig~\ref{fig:dvcstcsdvmp}. The DVMP process provides additional information regarding GPDs because the produced meson's quantum numbers determine which GPD sectors dominate, e.g., at leading twist, longitudinal vector mesons mainly probe the unpolarized GPDs $H$ and $E$, whereas pseudoscalar mesons probe $\tilde{H}$ and $\tilde{E}$. Moreover, the flavor content of the produced meson helps to separate quark GPDs by flavor, since different mesons probe different combinations of quark GPDs. 

From the work by \cite{Collins:1996fb} it has been well known that, for longitudinally polarized virtual photons, the leading-power amplitude factorizes into convolution of a perturbatively calculable hard-scattering kernel, a nonperturbative GPD, and the leading-twist meson distribution amplitude. In the same work, it was shown that the amplitudes of transversely polarized photons are power suppressed in the asymptotic limit, and no collinear factorization theorem established for them. However, at few $\mathrm{GeV}^2$ scales, transversely polarized photons contribute to the amplitude significantly and can even become the dominant contribution in the cross section, as observed, for instance, in the electroproduction of $\pi^0$ experiments in \cite{Bedlinskiy:2012be,CLAS:2014jpc,CLAS:2016tqs,JeffersonLabHallA:2016wye,JeffersonLabHallA:2020dhq}. As first pointed out in \cite{Ahmad:2008hp}, 
%This has led to several phenomenological models of pseudoscalar meson production, 
these contributions can be accounted for if the production process involves twist-2 chiral-odd transversity GPDs coupled to twist-3 meson distribution amplitudes, at variance with what previously believed \cite{Mankiewicz:1998kg}. Several phenomenological models of pseudoscalar meson production were subsequently developed \cite{Goldstein:2013gra,Goloskokov:2011rd}. 

\subsection{Timelike Compton scattering}

Timelike Compton scattering (TCS), $\gamma p\rightarrow\gamma^*p'\rightarrow\ell^+\ell^-p'$, is the timelike analog of DVCS. In TCS, a real photon is scattered off a nucleon and produces an outgoing virtual photon with a large timelike virtuality $Q'^2=M_{\ell\ell}^2$, which decays into a lepton pair; see Fig.~\ref {fig:dvcstcsdvmp}. As shown in \cite{Berger:2001xd}, the TCS amplitude factorizes into timelike hard-scattering coefficient functions and the same universal GPDs that enter DVCS. As in DVCS, the same final state also receives a contribution from BH, but with the dominance of BH over pure TCS more pronounced in a wider kinematic region, making it more difficult to get a TCS signal \cite{Berger:2001xd, AbdulKhalek:2021gbh}. However, the TCS--BH interference term can be used to constrain the TCS amplitude. 

At leading twist and leading order in $\alpha_s$, the TCS coefficient functions are related to the complex conjugates of the corresponding DVCS coefficient functions, up to a sign depending on the vector or axial-vector sector; see \cite{Berger:2001xd}. At NLO, analytic continuation from spacelike to timelike kinematics produces additional contributions and characteristic $\pi^2$ terms as shown in \cite{Pire:2011st}. Moreover, gluon coefficient functions enter directly at NLO and can produce sizable corrections \cite{Moutarde:2013qs}. Although the TCS data is currently limited for a comprehensive extraction of GPDs, the first TCS measurement by CLAS12 was published in 2021, \cite{CLAS:2021lky}, measuring the photon beam polarization and the decay lepton angular asymmetries, thereby opening a direct experimental test of GPD universality between spacelike and timelike processes. There are also data-driven studies, \cite{Grocholski:2019pqj}, using DVCS proton data with LO and NLO spacelike-to-timelike relations to make predictions for TCS observables. 

\subsection{Double DVCS/Other exclusive channels/Nuclei}
Complementary exclusive channels can provide information on GPDs that is hard to obtain from DVCS alone. Double deeply virtual Compton scattering (DDVCS), in which the final photon is virtual and decays into a lepton pair, introduces an additional virtuality that allows the partonic momentum fraction and skewness to be varied more independently, providing enhanced sensitivity to the $x$-dependence of GPDs \cite{Guidal:2002kt,Deja:2023ahc}. 
%Other channels, including timelike Compton scattering, deeply virtual meson production, and exclusive heavy-meson production, provide complementary flavor, spin, and gluon sensitivity. 
A broader class of hard exclusive multi-particle production processes, including photon--meson and two-meson production with a large invariant mass of the produced pair, is also currently being explored as they are believed to provide an enhanced sensitivity to the $x$-dependence of the GPDs, through additional kinematic variables \cite{Cosyn:2020kfe,Duplancic:2023kwe,Crnkovic:2025man,Nabeebaccus:2026rco}.
Wide-angle Compton scattering (WACS) is an additional probe of GPDs through their $1/x$ moments at large momentum transfer \cite{Radyushkin:1998rt}. Related information can also be obtained from two-photon reactions in $e^+e^-$ collisions, where hadron-pair production accesses generalized distribution amplitudes, the crossed counterparts of GPDs.
Jefferson Lab has a substantial WACS program, including measurements of cross sections and polarization transfer, and approved 12 GeV measurements of WACS at 8 and 10 GeV photon energies and polarization observables at large $s,-t,-u$.

While this review  is largely focused on DVES off of a nucleon, nuclear targets have also been considered since the inception \cite{Polyakov:2002yz,Kirchner:2003wt}. The interpretation of nuclear scattering in terms of the partonic structure of bound nucleons remains particularly challenging, since conventional nuclear effects must be separated from genuine modifications of the underlying quark and gluon structure. For GPDs, this problem acquires an additional layer of complexity compared with ordinary PDFs, since the off-forward kinematics introduce correlations between the partonic variables and the nuclear momentum transfer, making nuclear binding, nucleon motion, and off-shell effects intrinsically intertwined with the spatial information encoded in the GPDs \cite{Liuti:2005qj,Martinez-Fernandez:2026zog,Cosyn:2026gyy}. Nuclear GPDs have also been investigated within a variety of complementary approaches, including impulse-approximation and few-body calculations for light nuclei, as well as treatments of coherent scattering and nuclear shadowing at small $x$ \cite{Scopetta:2004kj,Guzey:2008fe,Goeke:2009tu,Fucini:2018gso}. At still smaller $x$, the connection between GPDs and the high-density regime of QCD can be formulated in terms of off-forward dipole amplitudes and BK/JIMWLK evolution, providing a link between nuclear GPDs, transverse imaging, and saturation dynamics \cite{Kovchegov:2025yyl}.
Experimental measurements are being conducted by the ALERT collaboration, \cite{Armstrong:2017zqr}, at Jefferson Lab, focused on $^4\text{He}$.
The prospects for nuclear DVCS at the EIC are also beginning to be investigated through dedicated experimental and detector studies. Particular attention is being given to the far-forward detection of intact recoil nuclei and nuclear fragments \cite{Chang:2025pgi}, which is essential for separating coherent and incoherent channels as well as for fully exclusive measurements \cite{Liuti:2005gi,Martinez-Fernandez:2026web}. 
%Recent simulations of coherent DVCS on polarized $^{3}\mathrm{He}$, together with broader studies of forward nuclear detection at the EIC, demonstrate the potential for extending GPD imaging to light nuclei 

%and the spin structure of this spin-0 target allows for an especially simple formulation of the cross section in which only two CFFs enter at twist-two, unlike in DVCS off of a spin-1/2 system in which four CFFs enter at twist-two. 
%Prospects for the EIC are explored in, \textit{e.g.}, \cite{Freund:2003ix}.  

\subsection{Ultra-peripheral collisions}
Ultra-peripheral collisions (UPCs) provide a complementary avenue for accessing GPDs, particularly in the small-$x$ gluon sector, extending hadron imaging into a kinematic region largely inaccessible to fixed-target deeply virtual processes.  The quasi-real photons emitted in the UPCs enable exclusive photoproduction processes such as $\gamma  \, p(A)\to J/\psi \, p(A)$ \cite{ALICE:2021gpt,ALICE:2023gcs,ALICE:2023jgu}, and they provide, therefore, an important connection between GPD phenomenology and small-$x$ gluon dynamics \cite{Mantysaari:2017dwh,Mantysaari:2016ykx,Cepila:2017nef}. 
%The momentum-transfer dependence of exclusive vector-meson production carries information on the transverse spatial distribution of gluons, while coherent and incoherent production probe, respectively, the average gluon distribution and its event-by-event fluctuations. 

%%%%%%
%%%%%%
%%%%%%
\section{From  Measured Observables to GPDs} 
\label{sec:inverse_prob}

\subsection{The fundamental inverse problem: why extracting a GPD is much harder than measuring a cross section}
\label{subsec:inv_p1}
%%%
An important challenge in GPD phenomenology arises from the two successive inverse problems involved in their determination. GPDs are not measured directly but enter observables through convolution integrals defining the CFFs; the CFFs, in turn, must first be extracted from measured cross sections and asymmetries. The reconstruction proceeds through the sequence
\[
\text{observables}\ \longrightarrow\ \text{CFFs}\ \longrightarrow\ \text{GPDs},
\]
with a distinct inverse problem arising at each step. The first one consists in disentangling the different CFF contributions to the measured observables, while the second inverse problem requires reconstructing the underlying GPDs from their convolution integrals.
\\

%\subsection{Uncertainty quantification and propagation of experimental uncertainties into GPD uncertainties}
%\label{subsec:UQ}

A reliable GPD determination therefore requires uncertainties to be followed through the entire chain, from measured observables to CFFs and ultimately to GPDs. Experimental statistical and systematic uncertainties, correlations among observables, theoretical uncertainties, parametrization dependence, prior sensitivity, and uncertainties associated with lattice-QCD inputs can all contribute to the final result. Because the CFF-to-GPD step is an ill-posed inverse problem, uncertainties can be strongly amplified and correlated. Modern uncertainty-quantification methods therefore aim to characterize the full probability distribution of the inferred quantities rather than quoting only point estimates and symmetric errors. A central goal is to distinguish regions and features of GPDs that are genuinely constrained by the available information from those that remain model or prior dependent.

\subsubsection{First Inverse Problem: Extraction of Helicity Amplitudes and CFFs from Experiment} 
\label{sec:inverse_prob_1}

\noindent The first inverse problem concerns the determination of CFFs from experimentally measured cross sections and asymmetries.  Cross sections and spin asymmetries involve different combinations of helicity amplitudes of the exclusive reaction. In DVCS and TCS these include the interference between the deeply virtual process and the BH amplitude. 
This inversion is challenging because an individual experimental observable does not generally isolate a single helicity amplitude or CFF. Rather, different CFFs enter simultaneously through linear and bilinear combinations, with their relative sensitivity depending on the polarization observable and the kinematics. The interference with the Bethe--Heitler process provides particularly valuable sensitivity to certain linear combinations of CFFs, whereas the squared DVCS amplitude introduces additional bilinear combinations. Consequently, increasing the number and variety of polarization observables is essential for disentangling the different CFF contributions.
%
%The first inverse problem is closely tied to the representation of the exclusive reaction in terms of helicity amplitudes. 
%The experimentally measured cross sections and polarization observables are constructed from bilinear combinations of these amplitudes, which in turn are expressed in terms of the CFFs. 
A helicity-amplitude formulation therefore provides a natural framework for tracing the sensitivity of individual observables to the different vector, axial-vector, and helicity-flip CFF structures \cite{Kriesten:2019jep,Kriesten:2020apm,Kriesten:2021sqc}. Through QCD factorization (Sections \ref{sec:definitions}, \ref{sec:DVCS-BH}), the helicity amplitudes are expressed in terms of CFFs which contain the nonperturbative hadronic information. In this representation of DVCS, the reaction can be organized schematically as,
\begin{eqnarray}
\{{\cal H},{\cal E},\widetilde{\cal H},\widetilde{\cal E},\ldots\}
\longrightarrow
\{f_{\Lambda \Lambda'}^{\Lambda_{\gamma^*},\Lambda_{\gamma'}}\}
\longrightarrow
\{\sigma,A,\ldots\},
\end{eqnarray}
making explicit the intermediate amplitude structure that connects CFFs to observables (similar structures describe all the other DVES processes). 
Here, \({\cal F}={\cal H},{\cal E},\widetilde{\cal H},\widetilde{\cal E},\ldots\); 
$\Lambda$ $\Lambda'$ are the initial and final proton helicities, and $\Lambda_{\gamma^*},\Lambda_{\gamma'}$ are the initial virtual photon and final measdured photon helicities.
Organizing these effects at the amplitude level provides a systematic way of assessing their impact on the extraction of the CFFs of different twist: in DVCS, ananlogously to DIS, setting $\Lambda_{\gamma^*}= \pm 1$ gives twist two contributions, while $\Lambda_{\gamma^*}= \pm 0$, give stwist three contributions. 
The helicity-amplitude representation has motivated generalized Rosenbluth-type separations, in which the distinct kinematic dependence of different amplitude structures can be exploited to enhance the experimental separation of CFF contributions.
The existence of an explicit ``forward" map, however, does not imply that it can be uniquely inverted. Several CFFs generally contribute simultaneously to a given observable, and the available set of measurements may constrain only particular combinations of them. Consequently, the inverse mapping
\begin{eqnarray}
\{\sigma,A,\ldots\}
\longrightarrow
\{f_{\Lambda \Lambda'}^{\Lambda_{\gamma^*},\Lambda_{\gamma'}}\}
\longrightarrow
\{{\cal H},{\cal E},\widetilde{\cal H},
\widetilde{\cal E},\ldots\}
\end{eqnarray}
can contain strong correlations, weakly constrained directions, and multiple solutions. The helicity-amplitude formalism therefore helps identify where the sensitivity to the different CFFs originates, while the statistical extraction determines how much of this information can actually be resolved by the available measurements.

This distinction provides a useful separation between the reaction formalism and the inference problem. The former specifies the forward map from CFFs to observables, including the relevant helicity and power-suppressed structures; the latter asks how reliably this map can be inverted in the presence of finite and incomplete experimental information. 
%Since several CFFs can contribute simultaneously to a given observable, the available measurements generally constrain combinations of CFFs rather than determining each one independently. This gives rise to correlations, degeneracies, and potentially multiple solutions in the CFF parameter space.

The problem is further complicated by the fact that the experimentally available set of observables is generally incomplete. Even when several cross sections and asymmetries are measured at the same kinematic point, they need not provide independent constraints on all real and imaginary components of the contributing CFFs. Some directions in the multidimensional CFF space can therefore be tightly constrained while others remain weakly determined. Correlations and degeneracies naturally emerge, and different sets of CFF values may produce statistically equivalent descriptions of the measured observables.
This feature is important when interpreting a CFF extraction. A successful fit to the measured cross sections and asymmetries is not sufficient to establish that the individual CFFs have been uniquely determined. 
%The data may instead select an extended region, or several regions, of CFF parameter space. 
%The extracted result can consequently depend on which CFFs are allowed to vary, the bounds or theoretical assumptions imposed on them, and whether information from neighboring kinematic points is introduced through a common parametrization.

%The treatment of helicity amplitudes also introduces theoretical considerations into this first inverse problem. At leading twist, factorization provides a relatively direct connection between the dominant helicity amplitudes and the twist-two CFFs. At the moderate \(Q^2\) values characteristic of much of the existing fixed-target data, however, helicity-flip amplitudes, higher-twist contributions, and finite-\(Q^2\) corrections may also contribute. Neglecting these terms can transfer missing reaction-mechanism effects into the extracted leading-twist CFFs. The statistical extraction and the theoretical description of the amplitudes therefore cannot be regarded as entirely separate issues.

Local and global fits, neural-network parametrizations, replica methods, and likelihood-based Bayesian analyses described below,  represent different strategies for addressing this statistical inverse problem. %Different phenomenological strategies address the inherent underdetermination of the problem in different ways. 
Local analyses determine the CFFs independently at each kinematic point, often restricting poorly constrained CFFs through bounds or model assumptions \cite{Guidal:2004nd,Guidal:2010ig,Guidal:2008ie}. Global analyses instead introduce a common kinematic representation of the CFFs, allowing measurements at different values of $x_{Bj}$, $t$, and $Q^2$ to constrain one another. Furthermore, neural-network approaches reduce the dependence on a prescribed analytic functional form while retaining correlations across kinematics \cite{Kumericki:2007sa,Kumericki:2009uq,Kumericki:2011rz,Kumericki:2013br}.  More recently, likelihood-based Bayesian analyses have been used to examine the multidimensional CFF probability distribution locally, with the aim of identifying correlations, degeneracies, non-Gaussian behavior, and multiple solutions before imposing a common representation across kinematics \cite{Adams:2024pxw,Pandey:2026rvn}.
More recently, generative AI methods have been introduced to address explicitly the non-uniqueness of the observable-to-CFF inverse problem. The Variational Autoencoder Inverse Mapper (VAIM) learns a probabilistic inverse mapping from measured observables to the space of compatible CFF solutions, using latent variables to represent information that is not uniquely determined by the observables. Rather than returning only a single best-fit CFF configuration, the method can therefore generate multiple solutions consistent with the same measured input. This provides a complementary approach to conventional optimization and Bayesian/MCMC analyses for exploring degeneracies in CFF extraction \cite{Almaeen:2024guo,Hossen:2024qwo}.

All of these approaches answer somewhat different questions and will be addressed in more detail in Section \ref{sec:strategies}.
%a global analysis seeks a common representation capable of describing the full body of measurements, whereas a local likelihood analysis asks what the measurements at a particular kinematic point actually determine. The latter provides a useful assessment of the information content of the data themselves and can help distinguish experimentally constrained CFF combinations from constraints introduced through parametrizations, priors, or correlations across kinematics.
\\

%%%%% second inverse 
\subsubsection{Second Inverse Problem: from CFFs to GPDs}
\label{subsec:secondinvprob}
 
The connection between GPDs and experimentally accessible amplitudes is established through QCD factorization. At sufficiently large \(Q^2\), the deeply virtual Compton scattering amplitude can be separated into perturbatively calculable hard-scattering coefficient functions and nonperturbative GPDs \cite{Ji:1996ek,Radyushkin:1997ki,Collins:1998be}. 
%The resulting Compton form factors (CFFs) have the schematic form
%\begin{equation}
%\label{eq:factorization}
%{\cal F}(\xi,t,Q^2)=\sum_{i=q,g}\int_{-1}^{1} dx\,C_i(x,\xi,Q^2,\mu^2)\,F_i(x,\xi,t,\mu^2),
%\end{equation}
%where the coefficient functions \(C_i\) define the hard-scattering kernels and \(F_i\) denote the corresponding quark and gluon GPDs.
The perturbative description of this relation has been developed systematically beyond leading order \cite{Ji:1997nk}, including NLO coefficient functions and evolution \cite{Belitsky:1999hf}, conformal-moment and Mellin--Barnes representations \cite{Kumericki:2007sa}, and, more recently, NNLO corrections \cite{Braun:2020yib}. 
%These developments are essential for precision GPD phenomenology, since the perturbative order, factorization scale, and quark--gluon mixing affect the interpretation of the extracted CFFs. 
At the moderate $Q^2$ values of existing fixed-target measurements, finite $t/Q^2$, target mass, and higher-twist corrections can also modify the leading-twist relation and have been the subject of extensive theoretical studies \cite{Braun:2011zr,Braun:2012bg,Braun:2014sta}.

In a nutshell, one is faced with an inverse problem that is unique to GPDs that can be schematically represented as,
\begin{eqnarray}
\label{eq:CFF_schematic}
CFF(\xi,t,Q^2) =   \int (QCD \,kernel) \times GPD(x,\xi,t,Q^2),
\end{eqnarray}
where both the perturbatively calculable QCD kernel and the GPD depend on an extra variable, $x$ that is not directly observable in DVES measurements. Consequently, even complete knowledge of a CFF as a function of its external kinematic variables does not directly provide the full $x$-dependence of the underlying GPD (more detailed equations are given in Section \ref{sec:DVCS-BH}). 

The formalism describing DVCS observables in terms of CFFs, including their interference with the Bethe-Heitler process and their decomposition in terms of helicity and azimuthal structures was first derived in \cite{Belitsky:2001ns,Belitsky:2010jw}, and subsequently incorporated into phenomenological frameworks such as PARTONS \cite{Berthou:2015oaw} and into conformal-moment-based global analyses. 
%in has been developed into a detailed description of
Several other formulations of the DVCS amplitude and cross section have been developed, differing in the choice of reference frame and, in particular, in the momentum direction used to define the longitudinal axis and the associated momentum transfer vector. In addition to the commonly used formulations of \cite{Belitsky:2001ns,Belitsky:2010jw}, alternative conventions have been developed in \cite{Braun:2012bg} and \cite{Kriesten:2019jep}. Although these descriptions represent the same physical process and must yield equivalent results when treated consistently, the different choices of reference vectors lead to different organizations of the helicity amplitudes and kinematic corrections, an issue that becomes particularly relevant at finite $t/Q^2$.
%%
%Extensions of the factorization framework to other hard exclusive reactions have also been investigated in Refs.~\cite{Qiu:2022pla,Qiu:2023mrm,Qiu:2022bpq,Qiu:2025ksq} within the proposed framework of single-diffractive hard exclusive processes (SDHEPs). It is argued that different processes, characterized by different hard-scattering coefficient functions, can provide complementary sensitivity to the \(x\)-dependence of specific GPD projections that cannot be recovered from conventional DVCS alone.
Extensions of the factorization framework to other hard exclusive reactions have also been investigated in ~\cite{Qiu:2022pla,Qiu:2023mrm,Qiu:2022bpq,Qiu:2025ksq} within the proposed framework of single-diffractive hard exclusive processes (SDHEPs). Rather than focusing on a specific exclusive channel, the SDHEP framework identifies a broad class of $2\to3$ reactions for which the diffractive hadronic transition can be factorized from an additional hard $2\to2$ subprocess. The final state kinematics can modify the weighting of the GPD in the convolution, providing enhanced sensitivity to its $x$-dependence that cannot be recovered from the conventional GPD-to-CFF convolution.
Multi-particle exclusive reactions have been addressed in \cite{Deja:2023ahc,Nabeebaccus:2026rco,Siddikov:2025kah,Siddikov:2025orq,Crnkovic:2025man} where by exploiting additional hard scales and more differential final-state kinematics, these reactions can provide access to GPD information that is otherwise integrated over in conventional formulations (we refer the reader to Section \ref{sec:experiment} for an overview of the experimental access to GPDs).

Although these developments help define the forward theoretical map from GPDs to CFFs with increasing perturbative and kinematic accuracy, the inverse problem encountered in phenomenology consists in attempting to reconstruct the multidimensional GPDs from the $x$ convolution integrals for the various processes, 
in contrast to inclusive DIS, where the partonic momentum fraction can be directly related to an experimentally determined scaling variable ($x\equiv x_{Bj}$), at leading order.
a problem whose uniqueness is not guaranteed even when the CFFs are accurately known.
%Experimental measurements provide access to integral projections of the GPDs rather than to their full dependence on $(x,\xi,t,Q^2)$. 

The loss of information through the convolution makes reconstructing a multidimensional function intrinsically non-unique without additional information or assumptions.
%In particular, different GPDs can generate identical, or experimentally indistinguishable, CFFs. 
The mathematical nature of this ambiguity has been investigated explicitly in studies of the DVCS deconvolution problem, which demonstrated the existence of nontrivial GPD contributions, the so-called shadow GPDs, having arbitrarily small effects on the CFFs (namely contributions to a GPD that can remain invisible in a given convolution and therefore cannot be distinguished through the corresponding observable). 
%In Refs.\cite{Bertone:2021yyz}, by including NLO PQCD evolution effects {\bf ...}, it is shown that nontrivial/substantial changes in the GPD can have arbitrarily small effects on DVCS observables. Their conclusion is therefore that DVCS alone does not uniquely determine the underlying GPD, motivating multichannel analyses.
%%
%Schematically, starting from Eq.\eqref{eq:CFF_schematic}, suppose
%\begin{equation}
%    F'(x,\xi,t)= F(x,\xi,t)+S(x,\xi,t) 
%\end{equation}
%with, $ C\otimes S=0$. Then
%\begin{equation}
%C\otimes F'=C\otimes F,
%\end{equation}
%even though $ F'\neq F$. 
%%%
%Thus, even perfect knowledge of the CFF does not necessarily imply unique knowledge of the GPD. This is more fundamental than having large experimental uncertainties or an insufficiently flexible fit: it concerns the mathematical non-uniqueness of the inverse map itself. 
In particular, in \cite{Bertone:2021yyz}, by including NLO PQCD evolution effects it was shown that nontrivial/substantial changes in the GPD can have arbitrarily small effects on DVCS observables.
Studies within the QGT Collaboration have explored whether QCD evolution and measurements spanning different scales can constrain such otherwise weakly determined directions in GPD function space 
\cite{Moffat:2023svr,Guo:2025muf,Freese:2024ypk}.

These results emphasize that the difficulty is not solely a consequence of limited experimental precision or of the choice of fitting procedure, but is intrinsic to the information contained in a given convolution. Addressing this ambiguity phenomenologically requires a consistent framework in which different GPD realizations can be propagated through the convolution to CFFs and ultimately confronted with experimental observables. A systematic computational framework for carrying out these steps is provided by the PARTONS project \cite{Berthou:2015oaw}. Its modular structure implements GPD models and evolution, perturbative coefficient functions and CFF evaluation, and the calculation of deeply virtual exclusive observables. PARTONS has subsequently provided the computational framework for global phenomenological analyses and neural-network extractions of CFFs \cite{Moutarde:2018kwr,Moutarde:2019tqa}.

More recently, artificial-intelligence approaches have been developed to address different aspects of these inverse problems. These include generative inverse methods, such as the Variational Autoencoder Inverse Mapper (VAIM), designed to represent the multiplicity of CFF solutions compatible with a given set of observables \cite{Almaeen:2024guo}, as well as neural-network representations of GPDs aimed at reducing functional-form bias and exploiting correlations across kinematics \cite{Xu:2026lko,Huang:2026eai,Mezrag:2026wcf}. Interpretable approaches based on symbolic regression have also been introduced to recover compact analytic representations of GPDs from phenomenological and lattice-QCD information \cite{Dotson:2025omi}. These developments illustrate a broader shift from selecting a single parametrized solution toward exploring, constraining, and interpreting the space of solutions allowed by the available information (see also discussion in \cite{Adams:2024pxw,Pandey:2026rvn}. The statistical and AI methodologies underlying these approaches, together with their treatment of uncertainties and correlations, are discussed in detail in Sec.~\ref{sec:strategies}.

%%%%%%
%%%%%%

%\begin{enumerate}
%\item Double distributions
%
%\item Regge-inspired parameterizations
%
%\item Spectator/diquark models
%
%\item Conformal partial-wave representations
%
%\item Dispersion-relation approaches
%
%\item Flexible/global parameterizations
%
%\item Neural-network representations
%
%\item Role of theoretical constraints in reducing model dependence
%\end{enumerate}

\subsection{Extraction Strategies: Data $\rightarrow$  CFFs  $\rightarrow$ GPDs }
\label{sec:strategies}

%\textbf{Local versus global CFF extraction}

The first level of phenomenological analyses concerns the extraction of Compton form factors (CFFs) from measured observables. Traditionally, researchers have focused on either {\it local} extractions, where  the CFFs are determined independently in individual kinematic bins, with minimal assumptions about their dependence on the external variables, 
or {\it global} extractions, where, in contrast, a common parametrization or functional dependence across multiple kinematic points and datasets is implemented. 
\\

%%%
\subsubsection{Local and Global Analyses}
\label{subsec:local_global}

\noindent The local extraction method reduces model dependence but is often limited by the number and precision of observables available in each bin, while
the global methods, by extending over multiple bins, can use the available information more efficiently and provide smoother, more stable determinations. In this case, to achieve converenge, additional assumptions about the kinematic dependence of the CFFs need, however, to be introduced. 
%%%
Local analyses in DVES were pioneered by Guidal, \cite{Guidal:2010de,Guidal:2009aa,Guidal:2013rya}, using  HERMES data \cite{Airapetian:2008aa,Airapetian:2009ac,Ye:2006gza,Kopytin:2005vv}  at individual experimental kinematic points, typically allowing a subset of CFFs to vary while bounding the others relative to a reference GPD model given that the unconstrained problem did not converge (Section \ref{sec:inverse_prob_1}). In particular, the analyses implemented MINUIT/MINOS $\chi^2$ minimization, where seven of the total eight CFFs were varied locally and bounded relative to the VGG \cite{Vanderhaeghen:1999xj} parametrization values. 
More recently the local determination of all eight CFFs at a common kinematic point, was addressed in \cite{Shiells:2021xqo}, using the complete polarization structure of the observables (Section \ref{sec:experiment}) obtained with pseudodata based on the GGL parametrization, \cite{Goldstein:2010gu}. A $\chi^2$ fit was then performed using twelve constraints from all available observables, and all of the eight CFFs were
determined with finite error estimates. Although this study pointed out that using as ``many constraints as possible generally gives a better fit", no quantitative study of the statistics and correlations among the extracted parameters was discussed.

%%%%% GLOBAL
The first quantitative attempt to a global extraction was that of Kumeri\v{c}ki and M\"{u}ller (KM) \cite{Kumericki:2007sa,Kumericki:2009uq}. The analysis uses parametrized GPD/CFF representations to simultaneously describe DVCS measurements across various experimental determinations and kinematics. 
The KM analysis, for example, combines collider and fixed-target information from HERA, HERMES, CLAS, and Hall A, spanning approximately $10^{-4}\lesssim x_B\lesssim0.4$ and $1\lesssim Q^2\lesssim100~\mathrm{GeV}^2$, thereby connecting the small $x_{Bj}$ sea-quark region probed at HERA with the valence region accessed at HERMES and Jefferson Lab.
The dominant quantity constrained by the unpolarized DVCS data were the CFFs $\Re {\rm e} {\cal H}$ and $\Im {\rm m} {\cal H}$, while the remaining CFFs entered through a chosen GPD model and were adjusted as part of the global description. As a result, $E$ was poorly constrained, and $\widetilde E$ was largely controlled by theoretical assumption of a pion-pole contribution.  
In subsequent studies, \cite{Kumericki:2011rz,Kumericki:2013br,Cuic:2020iwt} the CFFs were directly extracted as parameters by replacing the conventional parametrization with a neural network. 
Assuming the dominance of the CFF $\mathcal H$, and using a procedure analogous to the Monte Carlo replica method implemented by NNPDF \cite{Ball:2008by,NNPDF:2021uiq}, an ensemble of networks trained on Monte Carlo replicas of HERMES data was used to propagate experimental uncertainties and obtain a flexible global representation of the CFF. While substantially reducing functional-form bias, the extraction largely depends on the physics assumption that contributions from the remaining CFFs can be neglected. 

A more recent important direction in global extractions was taken by Moutarde et al., \cite{Moutarde:2019tqa}, where similarly to the work in \cite{Kumericki:2011rz}, neural networks were used to eliminate a preselected analytic functional form for the CFFs, and the experimental uncertainties were propagated using the replica method. In this approach  all the CFFs were extracted using eight independent neural networks, each trained on the same kinematic points $(\xi,t,Q^2)$. The dataset was also considerably larger, spanning from HERA through fixed-target/JLab kinematics. Hall A cross sections were excluded based on the fact that they could not achieve a convergent fit.

In summary, while Kumeri\v{c}ki et al., pioneered the neural network CFF idea under the $\mathcal H$ -dominance,  Moutarde, Sznajder and Wagner  generalized it to a multi-CFF, multi-experiment global extraction.
Global analyses  demonstrated that a common GPD-based parametrization can provide a simultaneous description of DVCS measurements spanning widely separated kinematic regimes. Their interpretation, however, remains dependent on the assumptions entering the GPD parametrization and on the relative constraints provided by the different datasets. For example, the solution of \cite{Kumericki:2011rz} was strongly influenced by the Hall A cross sections and it involved a sizable contribution from $\widetilde H$, which was subsequently disfavored by the more extensive CLAS unpolarized cross-section measurements. This illustrates an important limitation of global fits: 
%a satisfactory description of the available data does not necessarily imply a unique determination of the underlying CFFs. Assessing correlations, degeneracies, and alternative solutions therefore requires a more complete exploration of the multidimensional likelihood.
%your likelihood/MCMC approach addresses a different question—what the measurements at a given kinematic point actually allow before imposing correlations across kinematics
the available experimental observables do not, in general, determine a unique set of CFFs. Several CFFs contribute simultaneously to the measured cross sections and asymmetries, often through different linear and bilinear combinations, while the number and precision of independent observables remain limited. As a consequence, different combinations of CFFs can provide comparably good descriptions of the same data, leading to strong correlations, degeneracies, and, in some cases, multiple solutions. The resulting extraction can therefore depend significantly on the assumptions, constraints, and statistical methodology adopted in the analysis.

This motivates a complementary question: what do the measurements themselves determine at a given kinematic point, before correlations across different kinematics are imposed? Local likelihood-based analyses address this question by exploring the multidimensional CFF parameter space independently in each kinematic bin. A sufficiently complete exploration of the likelihood, for example through Bayesian inference and MCMC sampling, can reveal correlated, non-Gaussian, or multimodal regions that may not be adequately characterized by a single best-fit solution and its local Hessian uncertainties. Such analyses therefore provide a direct assessment of the information content of the data and help distinguish experimentally constrained CFF combinations from constraints introduced by the global parametrization, theoretical assumptions, or priors.

A useful illustration is provided by comparing the extraction of CFFs from the high-precision Hall A unpolarized DVCS cross sections \cite{JeffersonLabHallA:2022pnx} with the likelihood analysis of ~\cite{Adams:2024pxw}. The analysis of Georges et al. reported sensitivity to all four helicity-conserving CFFs and emphasized the importance of including correlations with helicity-flip amplitudes in obtaining realistic uncertainties. A subsequent likelihood-based analysis of the same class of observables, however, showed that the unpolarized twist-two cross section fully constrains only three independent CFF combinations. By exploring the multidimensional likelihood with MCMC, rather than characterizing the solution primarily through a best fit and its local uncertainties, extended correlated and weakly constrained directions become apparent. Thus, the ability of a multi-parameter fit to return values and uncertainties for all CFFs should not necessarily be identified with each CFF being independently determined by the data. This distinction is particularly important for underconstrained inverse problems, where correlations among parameters can produce an apparently well-defined fit even when substantial degeneracies remain.
%%%

\subsubsection{Least-squares/Hessian-based vs. Likelihood-based analyses, Bayesian inference and MCMC}
\label{subsec:Hessian}

\noindent A statistical issue which is particularly relevant in this context is that the analyses discussed so far, are essentially conventional least-squares global fits. 
The original analysis in \cite{Kumericki:2011rz}, for example, did not provide a systematic study of model uncertainties or full experimental-error propagation, and the authors explicitly noted this limitation.  
Traditional CFF and GPD analyses often rely on least-squares minimization, in which a $\chi^2$ function is constructed from the differences between measured and calculated observables. Parameter uncertainties are then estimated from the curvature of the objective function around its minimum, typically through the Hessian matrix. These methods are computationally efficient and remain widely used, but the Gaussian approximation implicit in Hessian error propagation can become inadequate when the likelihood is strongly non-linear, multimodal, weakly constrained, or exhibits significant parameter degeneracies. In these situations, a good value of the global $\chi^2/\mathrm{d.o.f.}$ can provide a parameterized solution describing the selected data, but it does not establish that the underlying multidimensional CFF solution is unique.
\\

\noindent {\it }

\noindent The non-uniqueness of the CFFs solutions is particularly important when comparing different CFF extractions: agreement with the measured observables does not necessarily imply that a unique underlying CFF solution has been identified. Bayesian and likelihood-based approaches combined with Markov Chain Monte Carlo (MCMC) sampling provide a natural framework for exposing this structure by mapping the multidimensional probability distribution of the CFFs, rather than reducing the result to a single best-fit solution and its local uncertainties.

In Bayesian inference, the likelihood is combined with prior information to construct the posterior probability distribution for the unknown quantities. Markov Chain Monte Carlo (MCMC) methods can then be used to sample the posterior without relying on a local Gaussian approximation. This makes it possible to determine full probability distributions, credible intervals, and correlations among the CFFs, and to identify degeneracies or multiple allowed solutions that may not be visible in a Hessian analysis. Priors can also be used to incorporate theoretical information or to regularize directions that are only weakly constrained by the data, although their impact must be carefully assessed to distinguish information supplied by the measurements from assumptions introduced in the analysis.
A complementary strategy, first introduced in \cite{Adams:2024pxw,Pandey:2026rvn} is to first investigate the information content of the measurements locally, before imposing correlations among different kinematic points. In a likelihood-based local analysis, the multidimensional CFF probability distribution is determined independently at each kinematic point from the observables available there. This allows one to identify which CFF combinations are directly constrained by the measurements and to expose correlations, degeneracies, and multiple solutions that may otherwise be reduced or hidden when a common functional representation is imposed across kinematics. For a GFF likelihood-based extraction, see \cite{Guo:2025jiz}. 

%%%%%% FIGURE Corner & Surface Plot of Difference Method
\begin{figure}[h!]
   \centering
    \includegraphics[width=8cm]
    %{FiguresFinal/Corner_Diff.png}
    {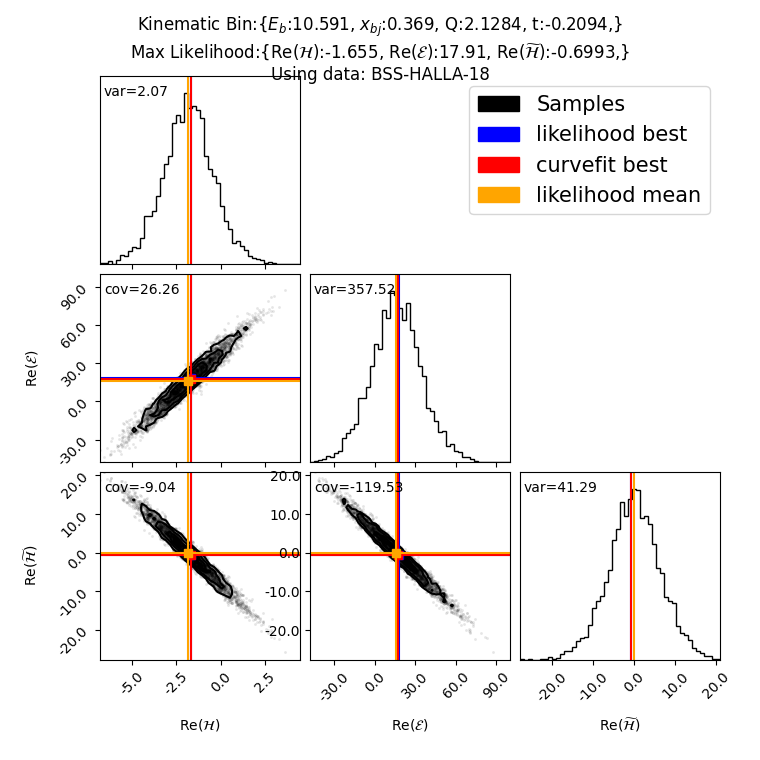}
    \hspace{0.1cm}
    \caption{ 
    Corner plot of 3 CFFs: $\Re e {\cal H}$, $\Re e {\cal E}$, $\Re e \widetilde{\cal{H}}$ using the Bayesian maximum likelihood analysis of \cite{Adams:2024pxw}.  
%    Right: Surface plot of the maximum likelihood parameters against the difference of two cross sections at all possible angles. 
%    The Blue points are all possible cross section differences. 
%    The Red points are those used in performing the fit to obtain the grey surface. 
%    The information contained in all blue points is fully contained in the red points only. 
    }
    \label{fig:3CFF}
\end{figure}

%A complementary approach to CFF extraction is provided by Bayesian inference . 
Rather than identifying a single best-fit solution and estimating uncertainties from the local curvature of the $\chi^2$ function, the Bayesian formulation treats the CFFs as parameters whose probability distribution is inferred from the measured observables. For a set of CFFs denoted collectively by $\boldsymbol{\mathcal F}$, Bayes' theorem gives,
%%%
\begin{equation}
p(\boldsymbol{\mathcal F}\mid D) \propto {\cal L}(D\mid\boldsymbol{\mathcal F}) \, \pi(\boldsymbol{\mathcal F}),
\end{equation}
%%%
where $D$ denotes the measured cross sections and asymmetries, ${\cal L}$ is their joint likelihood, and $\pi$ represents prior information or theoretical constraints.
This formulation is particularly useful for the first inverse problem because the mapping from CFFs to observables is nonlinear and generally not one-to-one. Several CFFs contribute simultaneously to the same observables, and the available measurements may constrain particular combinations much more strongly than the individual CFFs themselves. The resulting probability distribution in CFF space can therefore exhibit strong correlations, extended or nearly flat directions, asymmetric uncertainties, and, in some cases, distinct solutions with comparable probability.

MCMC methods provide a numerical means of exploring this multidimensional probability distribution without reducing it to the neighborhood of a single optimum. Samples generated by the Markov chains represent the joint posterior distribution of the CFFs and can be used to construct marginalized distributions, correlations, and Bayesian credible regions. In this way, the analysis provides not only estimates and uncertainties for individual CFFs, but also information about which combinations of CFFs are actually constrained by the measurements.
This distinction becomes especially important when the inverse problem is underconstrained. A conventional minimization may identify one parameter set that provides an excellent description of the data, while other, potentially distant regions of parameter space provide descriptions of comparable quality. A Hessian uncertainty analysis characterizes the curvature around the selected minimum and is most directly applicable when the probability distribution is approximately Gaussian in that region. MCMC instead explores the allowed parameter space more broadly and can reveal non-Gaussian or multimodal structures that cannot in general be represented by a covariance matrix alone.
In a local Bayesian analysis, this procedure can be carried out independently at each measured kinematic point. No functional relation between CFFs at neighboring values of \(x_B\), \(t\), or \(Q^2\) is required. The resulting posterior therefore provides a direct assessment of what the observables at that kinematic point determine, subject to the specified priors and theoretical description of the reaction. This makes it possible to separate, to the extent possible, constraints arising from the measurements themselves from those introduced through assumptions connecting different kinematic regions.

The role of the prior is consequently an important part of the analysis rather than merely a technical choice. For well-constrained CFF combinations, the likelihood can dominate the posterior and the influence of reasonable prior choices is small. For poorly constrained directions, however, the posterior may retain significant prior dependence. Examining this dependence provides useful information about identifiability: if the inferred distribution of a CFF changes substantially under reasonable variations of the prior, the available data do not determine that quantity independently. Conversely, posterior distributions that remain stable under such variations provide stronger evidence that the corresponding information is supplied by the measurements.
Bayesian/MCMC methods therefore do not remove the intrinsic underdetermination of the CFF extraction problem. Rather, their advantage is that they can make this underdetermination explicit. Instead of forcing the available information into a single solution with approximately Gaussian uncertainties, they provide a probabilistic representation of the family of CFF configurations compatible with the measurements and the stated assumptions.

Global analyses can subsequently exploit correlations among different kinematic regions to further constrain the CFFs, but the local analysis provides an important benchmark for distinguishing information supplied directly by the data from that introduced through the global parametrization or additional theoretical assumptions.
\\

\subsubsection{AI in GPD Phenomenology: From Interpretable Machine Learning to Agentic AI}
\label{subsec:AI}
The presence of strong correlations, degeneracies, and multiple solutions has motivated the development of AI-based approaches that go beyond conventional fitting strategies. Neural networks provide flexible representations capable of exploiting multidimensional correlations in the data without imposing restrictive functional forms, while generative architectures can address the ``one-to-many" character of the inverse problem by representing families of solutions compatible with the same measurements. Such methods therefore offer new possibilities not only for extracting CFFs and GPDs, but also for identifying degeneracies, quantifying the information actually constrained by the data, and ultimately improving uncertainty quantification. Several recent approaches along these lines are discussed below.

%The presence of strong correlations, degeneracies, and multiple solutions has also motivated the development of AI-based approaches, including neural-network and generative methods, designed to exploit the multidimensional correlations in the data while providing more flexible representations of the underlying inverse problem.

%While the underconstrained nature of the observable-to-CFF inverse problem has motivated a variety of extraction and uncertainty quantification strategies, from $\chi^2$ minimization and Hessian-based uncertainty estimates, to the statistically more sound Monte Carlo replica methods and neural-network parametrizations, introduced to reduce reliance on local Gaussian approximations and predetermined functional forms. More recently, likelihood-based Bayesian analyses combined with MCMC sampling have been used to explore the multidimensional probability distributions of the CFFs, retaining correlations, asymmetric uncertainties, and possible multiple solutions.
%%%
Generative inverse approaches, such as the VAIM, provide yet another strategy by learning the one-to-many mapping between observables and the CFF configurations compatible with them \cite{Almaeen:2024guo}. Although these methods differ substantially in their statistical interpretation, they address the common problem of determining and representing the range of CFF solutions allowed by incomplete and correlated experimental information. These methodological choices are distinct from the distinction between local and global extractions. Local analyses seek to determine what the measurements at an individual kinematic point allow without imposing correlations between different kinematic regions, whereas global approaches introduce a common representation across kinematics. Hessian, replica, Bayesian/MCMC, neural-network, and generative inverse methods concern instead how the allowed CFF information is inferred and represented. In practice, these two aspects are often intertwined, since the choice of parametrization and statistical framework determines how information is propagated between observables, CFFs, and different regions of kinematic space.
%%%%%%
%%%%%% VAIM FIGURE
\begin{figure}[ht]
\centering
    \includegraphics[width=0.80\textwidth ]{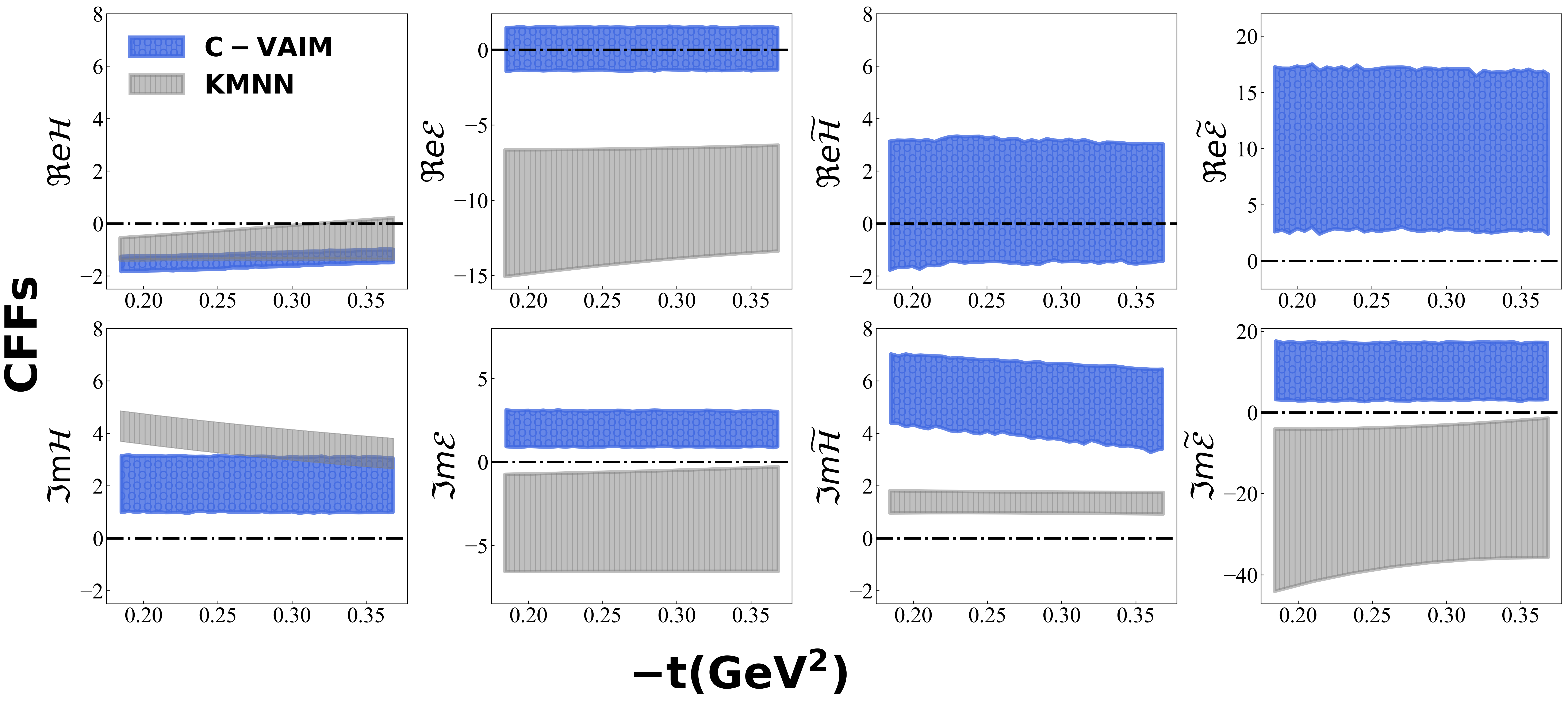}
    \caption{VAIM-based prediction of all eight CFFs as a function of $t$ for a fixed kinematics: $x_{Bj} = 0.35$ and $Q^{2}=1.9$ GeV$^{2}$, and initial electron energy of 6 GeV~\cite{JeffersonLabHallA:2015dwe} 
    Results are compared with the NN based extraction of Ref.~\cite{Cuic:2020iwt} (adapted from \cite{Almaeen:2024guo}).}
    \label{fig:cff_trends1}
%    \end{centering}
\end{figure}

%%%% NNGPD
A recent development in this direction is the Neural Network Generalized Parton Distribution (NNGPD) framework, which uses neural networks as flexible representations of multidimensional GPDs, \cite{Xu:2026lko}. In contrast to conventional parametrizations, where a specific analytic dependence on \(x\), \(\xi\), and \(t\) is assumed at the outset, NNGPD is designed to reduce functional-form bias while retaining the theoretical structure required of GPDs. Physical properties such as polynomiality and constraints on Mellin moments can be incorporated into the construction and training of the network, allowing the admissible function space to be restricted by QCD rather than primarily by a predetermined functional ansatz. Such a framework is particularly suited to the second inverse problem, where information from experimental observables can be combined with complementary constraints from PDFs, form factors, and lattice QCD.

This approach is particularly suited to the second inverse problem, where the limited information supplied by CFF convolution integrals must be used to reconstruct a multidimensional function. The flexibility of the neural-network representation allows correlations among the different GPD variables to emerge from the combined information rather than being fixed through an assumed factorized form. At the same time, the NNGPD framework provides a natural setting for incorporating complementary constraints from experimental observables, PDFs, form factors, and lattice-QCD calculations. The objective is therefore not simply to replace conventional parametrizations with a more flexible numerical representation, but to develop a physics-informed inference framework in which theoretical constraints and different sources of information can be consistently combined.

An important issue for such flexible representations is interpretability. Increasing the expressive power of the parametrization can enlarge the space of GPD solutions compatible with the available information, particularly in poorly constrained kinematic regions. NNGPD therefore also provides a useful setting for investigating which features of the reconstructed GPDs are determined by the input information and which arise from theoretical constraints or the architecture and training procedure. In this context, neural-network approaches can be complemented by interpretable AI methods, including symbolic regression, which can extract compact analytic structures from learned or lattice-constrained GPD representations and test whether physically motivated behaviors are supported by the inferred distributions.
\\

%%%%%
%%%%% NNGPD
\begin{figure}[ht]
\centering
\includegraphics[width=10cm]{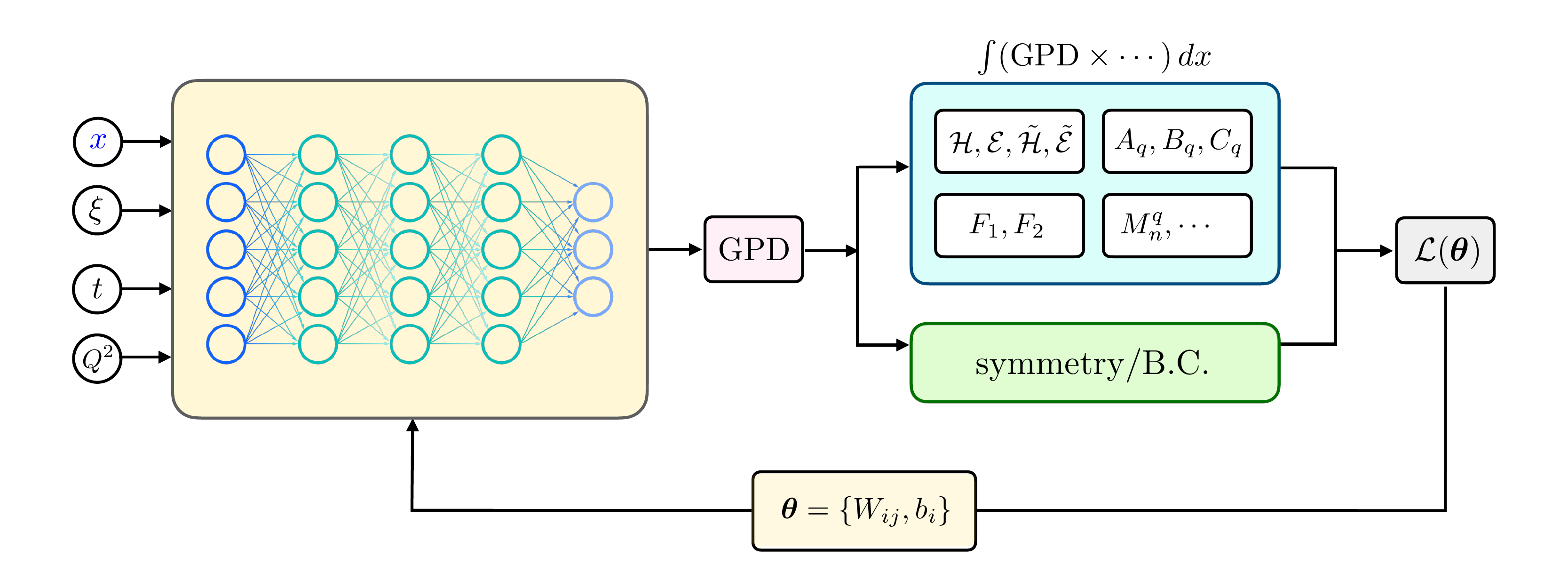}
\includegraphics[width=7cm]{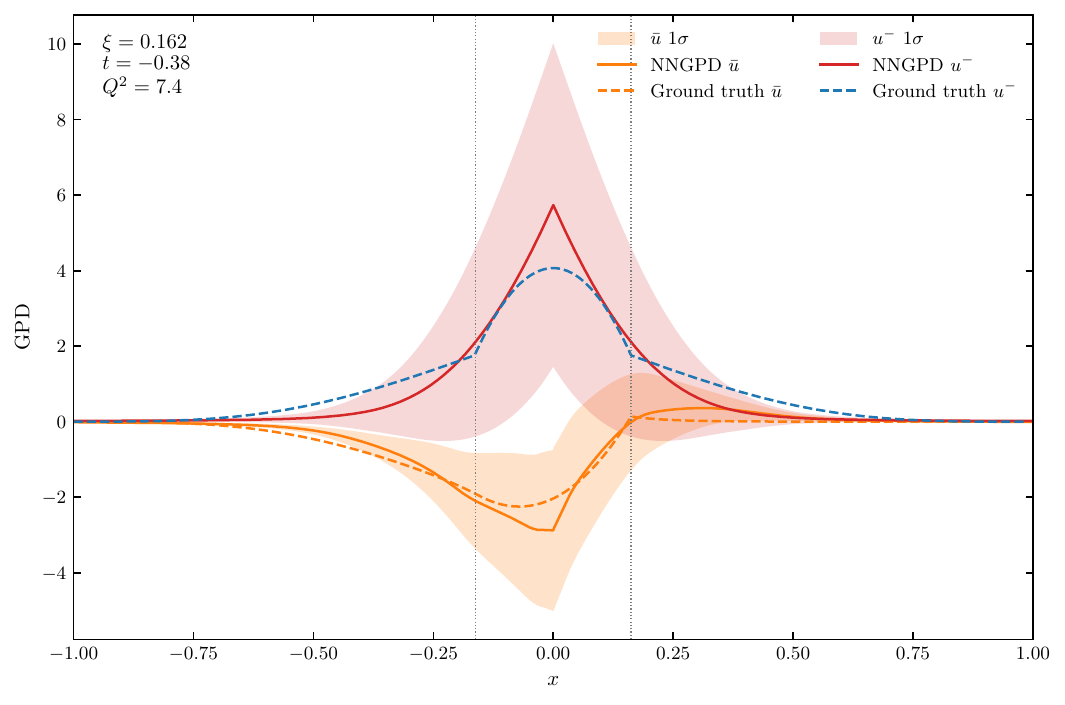}
\caption{(Adapted from \cite{Xu:2026lko}) {\it Upper panel}: Physics-informed machine-learning framework for GPDs.
A neural network maps the kinematic variables $(x,\xi,t,Q^2)$ to the leading-twist GPDs $H$, $E$, $\widetilde H$, and $\widetilde E$. The network is trained using a composite loss function encoding physical constraints and enforcing symmetry under $x \to -x$ and boundary conditions; 
%Constraints are provided by experimental measurements of hard exclusive processes and lattice QCD calculations, enabling a unified, data-driven extraction of GPDs; 
{\it Lower Panel}: the valence, $H_{u_v}(x,\xi,t)$, and sea quark, $H_{\bar{u}}(x,\xi,t)$, GPDs, for the kinematics: $\xi= 0.162$, $t=-0.38$ GeV$^2$, $Q^2=7.4$ GeV$^2$, chosen in a range compatible with JLab measurements.}
\label{fig:GPD-comparison}
\end{figure}

\begin{figure}[h!]
\includegraphics[width=15cm]{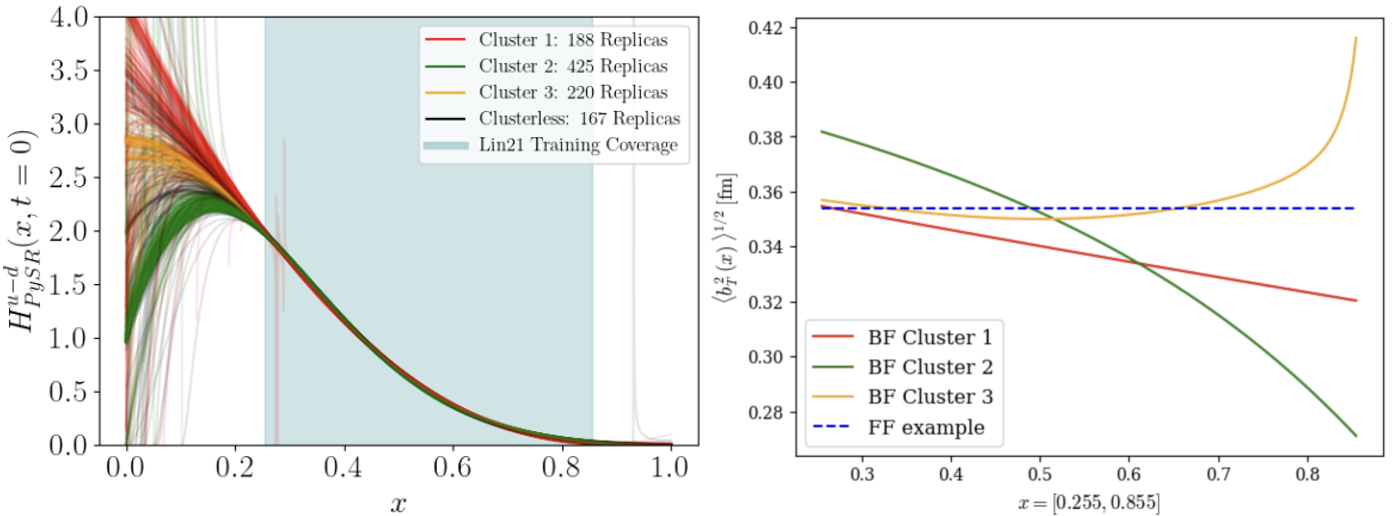}
\centering
\caption{(Adapted from \cite{Dotson:2025omi}). Symbolic regression fits on lattice QCD results clustered using an algorithm called HDBSCAN (left) and the average radii $\langle {\bf{b_{\perp}}}^2 \rangle^{1/2}$ as a function of $x$ of representative ``best fit" samples (right).}
\label{fig: SR}
\end{figure}

The increased flexibility of neural-network representations, however, also raises the question of interpretability: even when a network provides an accurate representation of the available information, it may be difficult to identify which functional structures are responsible for the resulting behavior. Symbolic regression (SR) provides a complementary approach by searching directly for compact analytic expressions that reproduce numerical data or learned distributions. In the context of GPDs, SR can be used not only as a parametrization tool, but also as a means of identifying and testing physically interpretable structures in their dependence on the partonic and kinematic variables. In particular, structures suggested by conventional phenomenology, such as factorized \(x\)- and \(t\)-dependence, Regge-inspired behavior, or correlations among these variables, can be confronted with expressions inferred directly from the input information rather than imposed a priori.
A recent application of SR to GPDs has demonstrated this strategy using phenomenological and lattice-QCD information, \cite{Dotson:2025omi}. By producing explicit analytic expressions, SR provides a bridge between flexible machine-learning representations and traditional QCD phenomenology: the resulting functions can be inspected, differentiated, integrated, evolved, and tested against theoretical constraints. This interpretability is particularly valuable for GPD studies, where the limited information available from experiment and lattice calculations makes it important to distinguish robust physical structures from features induced by a particular parametrization. See Fig. \ref{fig: SR} for a demonstration of solution clustering in SR and its reflection in the average partonic radii, a key quantity for proton imaging. 

Neural networks and symbolic regression can therefore play complementary roles in future GPD analyses. Neural networks provide the flexibility required to explore a broad space of admissible GPDs without imposing restrictive analytic forms, while symbolic regression can be used to identify compact structures emerging from such representations and to formulate new physically testable hypotheses. Their combination with Bayesian inference and uncertainty quantification offers a possible route toward GPD determinations that are simultaneously flexible and physically interpretable.

Large language models (LLMs) are also beginning to emerge as tools for scientific inference, knowledge integration, and the automated interpretation of complex phenomenological analyses; their potential role in GPD studies will be discussed further in Section \ref{sec:Limitations}.
\\

%%%%% LATTICE
\subsection{Physics Constraints}
\subsubsection{Incorporating Lattice QCD results}
\label{subsec:LQCD}
%%%%%
LQCD provides a complementary source of information for GPD phenomenology , that is conceptually distinct from the statistical methods used to extract CFFs from experimental observables but is particularly relevant to the second inverse problem. While exclusive measurements constrain GPDs indirectly through convolution integrals defining the CFFs, lattice calculations can provide first-principles information on Mellin moments and generalized form factors and, increasingly, on the $x$-dependence of GPDs through quasi- and pseudo-distribution approaches. The information provided by LQCD calculations is otherwise difficult to isolate experimentally, including for instance, the separation of connected and disconnected quark contributions. Incorporating these results consistently requires careful treatment of lattice systematic uncertainties, perturbative matching and evolution, and correlations among lattice observables. A unified statistical framework capable of combining these theoretical uncertainties with experimental uncertainties into phenomenological GPD analyses will therefore become increasingly important for future determinations.
%These quantities therefore constrain different projections of the same underlying GPDs.
%Since knowledge of the CFFs does not uniquely determine the underlying GPDs, lattice-QCD constraints can restrict directions in GPD function space that remain weakly determined by exclusive measurements. Conversely, experimental measurements provide information over kinematic regions and combinations that are difficult to access directly on the lattice. The two approaches are therefore complementary rather than redundant.
%%
%Moment-based frameworks provide a natural setting for combining generalized form factors calculated on the lattice with constraints from PDFs, elastic form factors, and deeply virtual exclusive measurements. Recent analyses have begun to implement this strategy within a common fit, illustrating the potential of combined experiment--lattice analyses for reducing the ambiguities inherent in GPD reconstruction.
%
%The increasing availability of lattice calculations of quasi- and pseudo-GPDs may eventually provide more direct constraints on the $x$-dependence itself. 
The main inputs from lattice in GPD analyses can be summarized as:
\\

\noindent {\it Mellin moments and generalized form factors}

\noindent Moments of GPDs are related to matrix elements of local QCD operators and can therefore be calculated directly in lattice QCD. These calculations provide generalized form factors and constraints on quantities such as momentum fractions, angular momentum, and the energy-momentum tensor. They are especially valuable because they constrain integrals of GPDs even in kinematic regions where experimental information is limited.
\\

\noindent {\it Quasi- and pseudo-GPD information}

\noindent Recent developments in large-momentum effective theory and related approaches make it possible to access the $x$ dependence of partonic distributions more directly through quasi- and pseudo-distributions. Their extension to off-forward kinematics opens a path toward lattice information on the shape of GPDs themselves, rather than only on a finite set of Mellin moments. These calculations remain technically demanding but provide an increasingly important bridge between lattice QCD and phenomenological reconstruction.
\\

\noindent {\it Connected and disconnected contributions}

\noindent Lattice calculations can separate connected and disconnected quark contractions, providing information on the different dynamical origins of valence-like and sea-quark contributions. Including disconnected contributions is essential for a complete flavor decomposition and becomes increasingly important when extending GPD analyses beyond the valence region. Their incorporation also offers a way to connect phenomenological descriptions of sea-quark structure with the underlying QCD dynamics.
\\

\noindent Lattice results can be included as additional constraints in fits or Bayesian analyses together with experimental observables. This is particularly useful for quantities that are poorly constrained by experiment alone, such as specific Mellin moments, flavor combinations, or regions of phase space with limited experimental coverage. A consistent treatment requires careful attention to lattice uncertainties, correlations, renormalization, matching, and the differences between lattice and experimental kinematics.
A long-term objective is the simultaneous determination of GPDs using experimental data and lattice-QCD information within a common statistical framework. Such analyses can exploit the complementary strengths of the two sources: experiments constrain physical scattering amplitudes over a range of kinematics, while lattice QCD provides first-principles information on moments and selected structural properties. The challenge is to combine these inputs without double counting information and with a consistent treatment of theoretical and statistical uncertainties.

\subsubsection{Phenomenological parametrizations}
\label{subsec:pheno_parametrizations}
There are various ways to parametrize GPDs, but here we will mainly discuss three different types of parametrization. Namely, the double distributions (DDs), conformal moment representations, and spectator models. 
\\

%\subsubsection
\noindent {\it Double distributions}
\\

\noindent In the DD representation, \cite{Radyushkin:1997ki, Teryaev:2001qm}, a quark GPD, such as $H^q(x,\xi,t)$, can be constructed by two functions, denoted by $F^q(\beta, \alpha, t)$ and $G^q(\beta, \alpha, t)$, where $\beta$ and $\alpha$ are two momentum fraction variables, as
\begin{equation}
    H^q(x,\xi,t) = \int_{\Omega}d\beta\,d\alpha\,\delta(x-\beta-\xi\alpha) \left[ F^q(\beta,\alpha,t) + \xi\,G^q(\beta,\alpha,t) \right] \, . 
\end{equation}
The DDs have support in the physical region
\begin{equation}
    \Omega=\left\{(\beta,\alpha):|\beta|+|\alpha|\leq1\right\}
\end{equation}
while the delta function sets $x=\beta + \xi\alpha$ and projects the DDs to the GPDs, which is mathematically a Radon transform, \cite{Teryaev:2001qm}. The functions $F^q$ and $G^q$, however, cannot be uniquely determined, as any regular function $\chi$ vanishing on the boundary of $\Omega$ can be used to transform the DDs 
\begin{equation}
    F^q\rightarrow F^q + \frac{\partial\chi^q}{\partial\alpha}, \qquad G^q\rightarrow G^q-\frac{\partial\chi^q}{\partial\beta},
\end{equation}
and still yield the same GPD $H^q$, \cite{Teryaev:2001qm, Tiburzi:2004qr}. One common choice of fixing this freedom is to choose the Polyakov-Weiss gauge, in which the highest allowed power of $\xi$ in the Mellin moments is isolated in the D-term, \cite{Polyakov:1999gs, Tiburzi:2004qr}. This can be achieved by localizing $\beta=0$, i.e., $G^q(\beta,\alpha,t) = \delta(\beta) D^q(\alpha, t)$, resulting in for $\xi>0$
\begin{equation}
    H^q(x,\xi,t) = \int_{\Omega}d\beta\,d\alpha\, \delta(x-\beta-\xi\alpha) F^q(\beta,\alpha,t) + \theta(\xi - |x|) D^q\left(\frac{x}{\xi},t\right) \, .
\end{equation}
Therefore, the D-term has support only in the ERBL region, i.e., $|x|\leq \xi$. The polynomiality property of the GPD $H^q$ parametrized by DDs can be seen from the direct evaluation of their Mellin moments
\begin{equation}
    \int_{-1}^{1}dx\,x^n H^q(x,\xi,t) = \int_{\Omega}d\beta\,d\alpha\,(\beta+\xi\alpha)^n F^q(\beta,\alpha,t) + \xi^{n+1} \int_{-1}^{1} dz \, z^n D^q(z,t)
\end{equation}
where the second DD $G^q$ generated the highest allowed power $\xi^{n+1}$ and we use the notation $z=x/\xi$. Although expanding $(\beta+\xi\alpha)^n$, together with the $\xi$ factor multiplying $G^q$, generates all powers up to $\xi^{n+1}$, time-reversal invariance of the GPD $H^q$ requires it to be invariant under the $\xi\to -\xi$ transformation, which can be implemented by taking $F$ to be even and $G$ to be odd under $\alpha \to -\alpha$, thereby eliminating the odd powers of $\xi$. 

Finally, we note that the unpolarized quark GPDs $H^q$ and $E^q$ have the same $D$-term entering in their DD representation with opposite signs and therefore the $D$-terms cancel in the sum $H^q+E^q$, consistent with Ji's sum rule. 
\\

\noindent {\it Conformal moment representation}

\noindent In contrast to the DD representation, the conformal moment representation describes GPDs in moment space. As shown in \cite{Mueller:2005ed}, GPDs admit a formal conformal partial wave expansion of the form 
\begin{equation} \label{conformal_expansion}
    H^q(x,\xi,t) = \sum_{j=0}^{\infty}(-1)^j p_j(x,\xi)\,H_j^q(\xi,t)
\end{equation}
where $p_j(x,\xi)$ denotes the conformal partial wave and $H_j^q(\xi,t)$ is the corresponding conformal moment. For integer $j$, and defining $z=x/\xi$, the quark conformal partial waves are constructed from the orthogonal basis of Gegenbauer polynomials $C_j^{3/2}(z)$ with respect to the weight $1-z^2$ on the interval $|z| \leq 1$. In the ERBL region, $|x| \leq \xi$, these partial waves are given by
\begin{equation}
    p_j(x,\xi) = \xi^{-j-1} \frac{2^j\Gamma(j+5/2)} {\Gamma(3/2)\Gamma(j+3)} \left(1-\frac{x^2}{\xi^2}\right) C_j^{3/2}\left(-\frac{x}{\xi}\right)
\end{equation}
The nonperturbative information of GPDs is contained in the conformal moments $H_j^q(\xi,t)$. This means that parametrizing a GPD amounts to parametrizing its conformal moments at a specific input scale. One can then constrain the parameters at an input scale by using exclusive scattering data, lattice data, and/or model input. 

However, the discrete conformal partial wave expansion is a formal expansion and not generally convergent over the full GPD support range. A convenient way to perform a resummation of this expansion is to analytically continue the conformal spin variable $j$, together with the partial waves and conformal moments, to complex values and perform the Mellin--Barnes integral \cite{Mueller:2005ed, Muller:2014wxa}
\begin{equation}
    H^q(x,\xi,t) = \frac{i}{2} \int_{c-i\infty}^{c+i\infty} \frac{dj}{\sin(\pi j)} p_j(x,\xi) H_j^q(\xi,t) \, .
\end{equation}
In this expression, once the contour is closed to the right, the residue contributions of the $1/\sin(\pi j)$ term come from the nonnegative integer values of $j$, hence, they reproduce the original expansion in Eq.~(\ref{conformal_expansion}). The integral is evaluated along a vertical contour crossing the real-valued point $c$ chosen such that the relevant poles of $1/\sin(\pi j)$ that constitute the series expansion lie to its right, while the singularities of the analytically continued conformal moments lie to the left of this contour. Hence, once partial waves and conformal moments are analytically continued to the complex plane consistently, the Mellin--Barnes integral reconstructs GPDs over the full $x$ range. Another advantage of the conformal representation of GPDs is that it simplifies the leading order QCD evolution \cite{Mueller:2005ed, Kumericki:2007sa}.

\begin{figure}[H]
    \centering
    \includegraphics[width=0.48\linewidth]{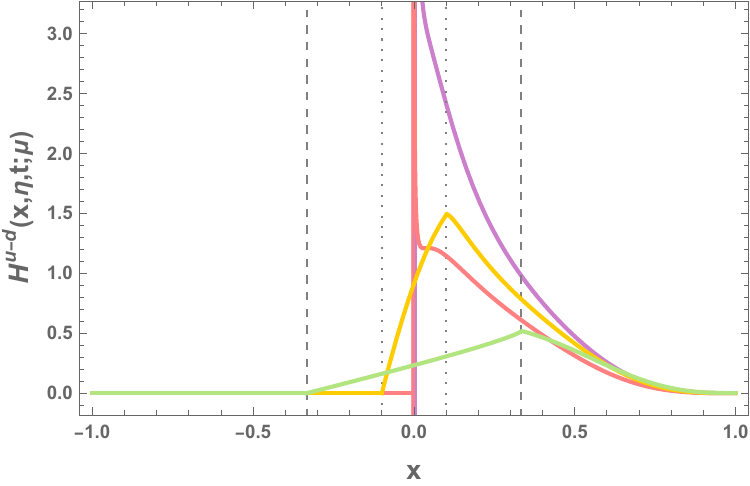}
    \hfill
    \includegraphics[width=0.48\linewidth]{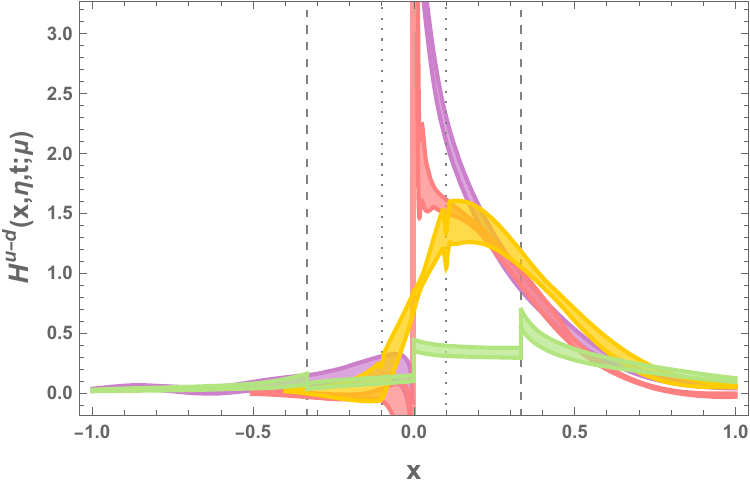}
    \caption{(Adapted from \cite{Mamo:2024vjh}). Left: Isovector ($u-d$) quark GPD $H^{u-d}(x, \eta, t; \mu)$ at various kinematic configurations based on a holographic string-based approach by \cite{Mamo:2024vjh}. Colors represent $\eta = 0$, $-t = 0$, $\mu = 2~\mathrm{GeV}$ (purple), $\eta = 1/3$, $-t = 0.69~\mathrm{GeV}^2$, $\mu = 2~\mathrm{GeV}$ (green), $\eta = 0.1$, $-t = 0.23~\mathrm{GeV}^2$, $\mu = 2~\mathrm{GeV}$ (yellow), and $\eta = 0$, $-t = 0.39~\mathrm{GeV}^2$, $\mu = 3~\mathrm{GeV}$ (pink). Right: Lattice results from \cite{Alexandrou:2020zbe} (purple and green), \cite{Holligan:2023jqh} (yellow), \cite{Lin:2020rxa} (pink) for $H^{u-d}(x, \eta, t; \mu)$ at the same color kinematics.} 
    \label{fig:stringparametrization}
\end{figure}

The application of the conformal moment representation and integral evaluation has been applied in a series of works, for instance, to GPDs in \cite{Mueller:2005ed, Kumericki:2007sa, Kumericki:2009uq, Mamo:2024jwp, Mamo:2024vjh, Hechenberger:2025wnz, Guo:2025muf}. We display a recent application, \cite{Mamo:2024vjh}, of Mellin--Barnes integrals to reconstruct the GPD $H^{u-d}(x, \eta, t; \mu)$ at various kinematic configurations in Fig.~\ref{fig:stringparametrization} (left), where the authors parametrize the conformal moments within a holographic string-based model. The resulting GPD $H^{u-d}(x, \eta, t; \mu)$ is compared with lattice results (right).

\noindent{\it Spectator models}

\begin{figure}[h!]
\includegraphics[width=15cm]{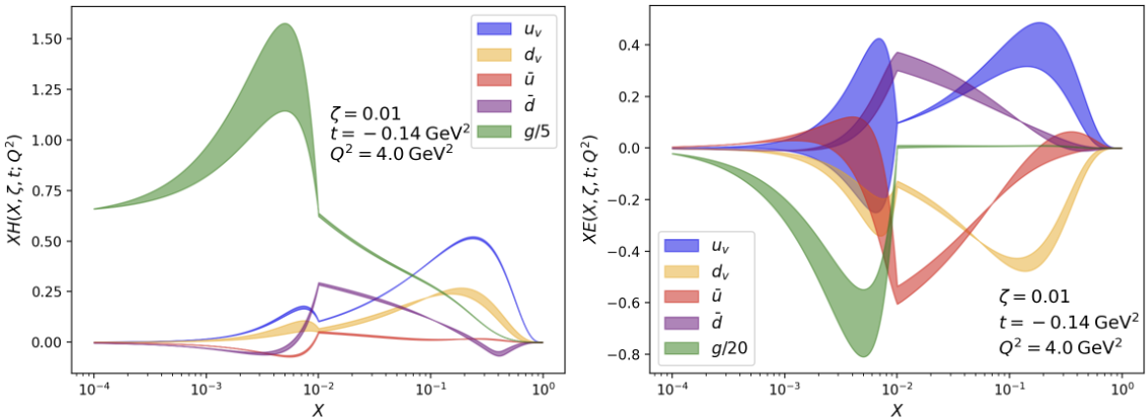}
\centering
\caption{Adapted from \cite{Panjsheeri:2025vpa}. The GPDs $H, E$ in the Reggeized spectator model-based parameterization for representative future EIC kinematics.}
\label{fig: UVA2}
\end{figure}

\noindent GPD spectator models provide a phenomenological, bottom-up realization of the covariant scattering-matrix approach first introduced in \cite{Landshoff:1970ff,Brodsky:1973hm}. In these models, the parton--proton amplitude is represented through quark and spectator propagators and proton--quark vertex functions. The pole structure of the amplitude determines the GPD in different kinematic regions: in the DGLAP region, \(|x|\geq\eta\), the spectator pole places the recoiling system on shell, whereas in the ERBL region, \(|x|\leq\eta\), the relevant contributions arise from the poles associated with the quark lines.
An early scalar-diquark realization was developed in \cite{Brodsky:2008qu}, with vertex functions chosen to reproduce the perturbative large-$|t|$ behavior of the Dirac form factor. More flexible spectator parametrizations were subsequently developed in Refs.~\cite{Ahmad:2006gn,Ahmad:2009fvg,Goldstein:2010gu,GonzalezHernandez:2012jv} by combining the diquark description with a Regge-inspired spectral distribution. Rather than assigning a fixed mass to the spectator, these ``reggeized diquark'' models integrate over the spectator invariant mass $M_X$ using a model spectral function $\rho(M_X^2)$. The low-mass region describes scalar and axial-vector diquark configurations, while the higher-mass continuum represents increasingly complex spectator states and generates the characteristic small-$x$ Regge behavior, $x^{-\alpha}$. This construction provides a flexible description of both the $x$ and \(t\) dependence of the GPDs, including the low-\(|t|\) behavior of the nucleon form factors. The extension to the ERBL region is constrained by crossing symmetry, continuity at \(x=\pm\eta\), and polynomiality.
The resulting parametrization can be summarized in the following expression for the quark sector, 
\begin{equation}
F_q(x,\eta,t)  = {\cal N}_q G_{M_X,m}^{M_\Lambda}(x,\eta,t) \,  
R^{\alpha,\alpha^\prime}_{p_q}(x,\eta,t) 
\label{fit_form}
\end{equation}
where $q=u,d$, $F_q \equiv H_q, E_q, \tilde{H}_q, \tilde{E}_q$; the functions $G_{M_X,m}^{M_\Lambda}$,   
$R^{\alpha,\alpha^\prime}_{p_q}$, is the Regge term given by,
\begin{equation}
R^{\alpha,\alpha^\prime}_{p_q}=  x^{-[\alpha_q + \alpha^\prime_q  (1-x)^{p_q} t  ]},
\label{regge}
\end{equation}
where  $\alpha_q$, $\alpha'_q$ and $p_q$ are tunable parameters; $G_{M_X,m}^{M_\Lambda}(x,\eta,t)$ is the quark-diquark/spectator term with mass parameters for the struck quark, $m_q$, the diquark/spectator, $M_X$, the diquark/spectator form factor cut-off parameter, $M_\Lambda$. 
The parametrization was extended to the valence quark, sea quark, and gluon contributions, \cite{Kriesten:2021sqc}. 
%For valence quarks the spectator is given by scalar and axial-vector diquarks, which, through the $SU(4)$ symmetry, allow one to perform a flavor analysis by distinguishing
%between isoscalar ($ud$) and isovector ($uu$) spectators. 
For scattering from sea quarks, the spectator is a tetra-quark state, namely a $uudq(\bar{q})$ state with $q=u,d,s,c$. For gluons it is a three quark system in a color octet state. The flavor dependence obtained in the model reflects the different spin--flavor configurations of the spectator system associated with (u) and (d) quarks.
%The possibility of distinguishing among different flavors in this model reflects the underlying color symmetry which can be seen as an indirect manifestation of chiral symmetry breaking. 
The latest update of this parameterization includes analytic, rather than integral, forms for the spectator model expressions and fits to new lattice QCD results that allow for a stronger constraint on the $t$ dependence of the antiquark distributions, \cite{Panjsheeri:2025vpa}. See Fig. \ref{fig: UVA2} for predictions at future EIC kinematics for the $H, E$ GPDs. 

A related class of GPD models is formulated directly in terms of light-front wave functions (LFWFs), with GPDs constructed from overlaps between the initial- and final-state proton wave functions. The general overlap representation was developed in \cite{Diehl:1998kh,Diehl:2000xz}, providing a direct connection between GPDs and the light-front Fock-state structure of the proton. This framework was subsequently used to construct phenomenological parametrizations constrained by nucleon form factors and parton distributions \cite{Diehl:2004cx}. Related light-front approaches have been extensively developed by Mukherjee and collaborators, 
%ranging from perturbative dressed-particle models to explicit light-front quark--diquark descriptions of the proton 
\cite{Chakrabarti:2005zm}. 
More recently, LFWFs motivated by soft-wall AdS/QCD have been used to describe GPDs, transverse densities, Wigner distributions, and their connections with TMDs \cite{Chakrabarti:2015ama,Chakrabarti:2024hwx}. GPDs have also been calculated within basis light-front quantization (BLFQ), where the proton LCWFs are obtained by diagonalizing an effective light-front Hamiltonian rather than introduced through a phenomenological ansatz. These calculations have been extended from quark GPDs at nonzero skewness to higher-twist distributions and, with the inclusion of a (qqqg) Fock sector, to gluon GPDs \cite{Liu:2024umn,Zhang:2025nll,Zhang:2026dzi}.
Such approaches provide a direct connection between GPDs and the light-front Fock-state structure of the proton, although Fock-space truncation generally limits the description of the ERBL region.
%These light-front constructions provide an alternative to covariant spectator models in which the partonic structure is encoded directly in the proton LFWFs rather than inferred from the pole structure of a covariant parton--proton amplitude.
\\
\vspace{0.5cm}

\begin{table}[H]
    \centering
    \begin{tabular}{|c|c|c|}
\hline
Approach     & Representation    & Main constraints      \\
\hline
 VGG  [1]   & Double distributions (DD)   & PDFs, FFs, polynomiality         \\
 UVA   [2]   & Spectator+ Regge  & PDFs, FFs, flavor      \\
 GK  [3]   & DD + Regge & PDFs, FFs, meson data   \\
 KM  [4]  & Conformal moments & DVCS, PDFs/FFs      \\
 Holographic QCD [5] & Conformal moments &PDFs, FFs, polynomiality   \\
 GUMP/QGT  [6]   & Conformal moments  & DVCS/DVMP + PDFs + FFs + lattice      \\
 NNGPD  [7]  & Neural networks   & Data + theoretical constraints             \\
 Symbolic regression [8]   & Learned analytic forms  & Phenomenology/lattice + constraints    
\\
\hline
\end{tabular}
    \caption{Summary of GPD parametrizations: [1] \cite{Vanderhaeghen:1999xj}; [2] \cite{Panjsheeri:2025vpa}; [3] \cite{Goloskokov:2007nt}; [4] \cite{Kumericki:2011zc}; [5] \cite{Mamo:2024vjh} [6] \cite{Guo:2025muf}; [7] \cite{Xu:2026lko}; [8] \cite{Dotson:2025omi}. }
    \label{tab:parametrizations}
\end{table}

%\begin{enumerate}
%\item Local versus global CFF extraction
%\item Least-squares/Hessian methods
%
%\item Likelihood-based analyses/Bayesian inference and MCMC: Correlations and degeneracies among CFFs; Priors and theoretical constraints
%\item Neural-network inverse methods
%\item AI and interpretable machine learning;
%\item Incorporating Lattice QCD
%\begin{itemize}
%\item Mellin moments and generalized form factors
%\item Quasi- and pseudo-GPD information
%\item Connected and disconnected contributions
%\item Combining lattice constraints with experimental data
%\item Global experiment–lattice analyses
%\end{itemize}
%\item Uncertainty quantification/Propagation of experimental uncertainties into GPD uncertainties
%\end{enumerate}

%%%%%%
%%%%%%
%%%%%%

\begin{figure}[H]
    \centering

    \begin{minipage}[c]{0.45\linewidth}
        \centering
        \includegraphics[width=\linewidth]{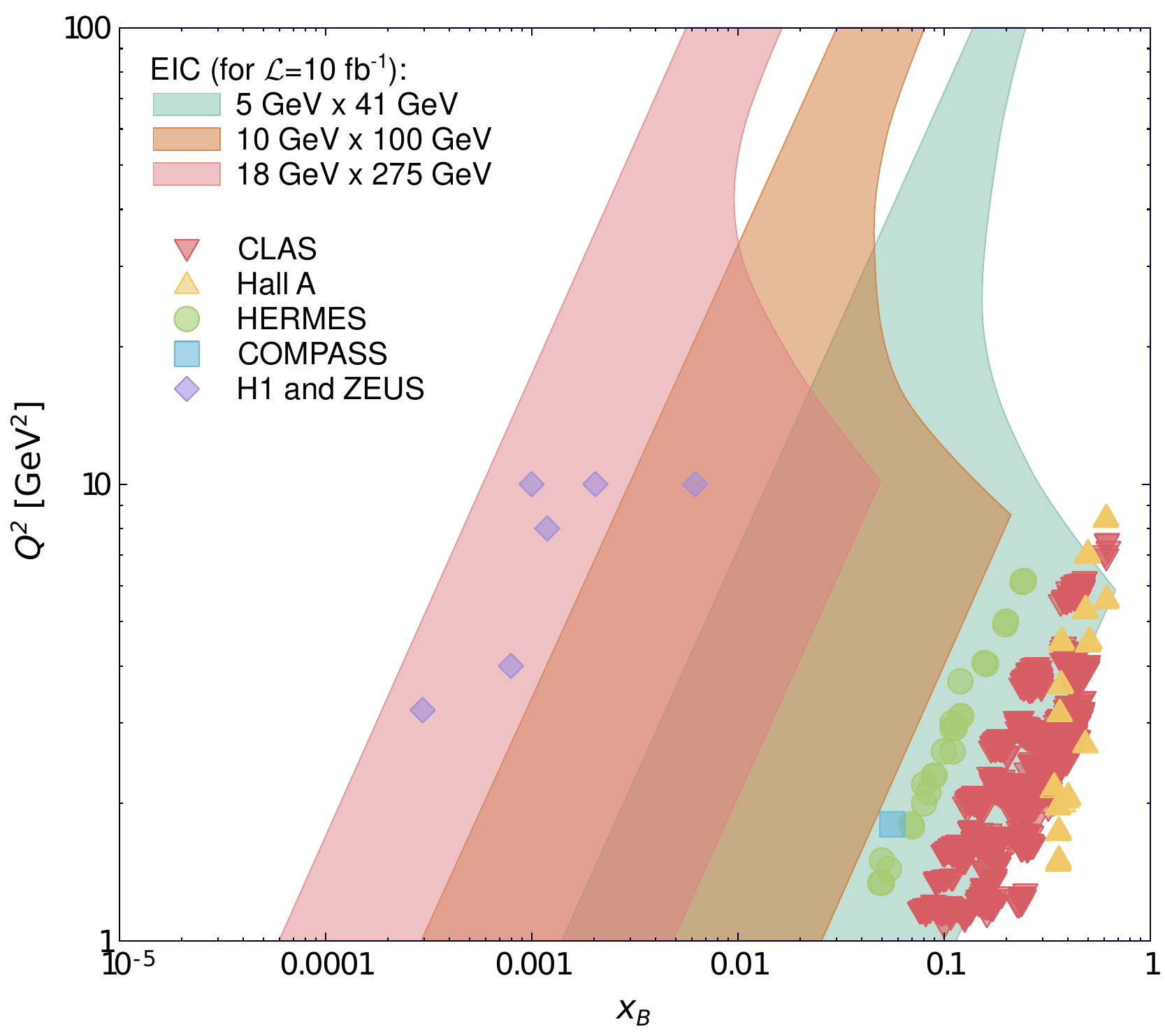}
    \end{minipage}
    \hfill
    \begin{minipage}[c]{0.52\linewidth}
        \centering
        \includegraphics[width=\linewidth]{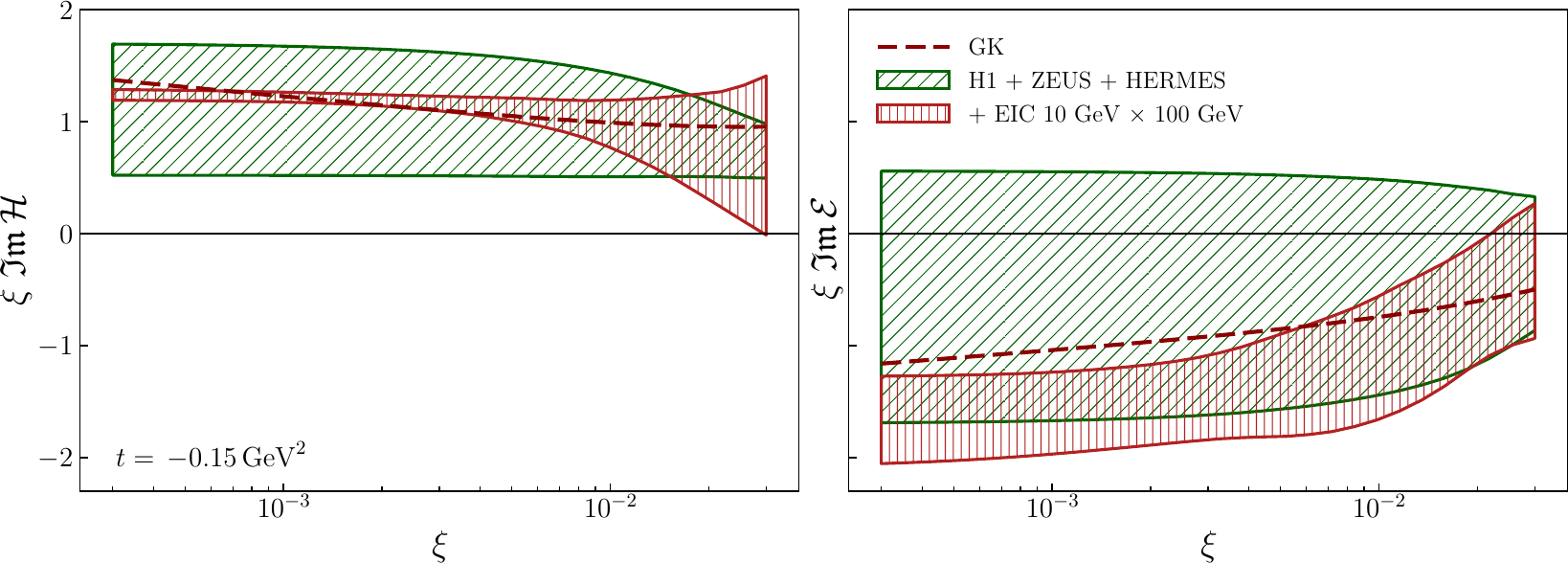}
        \vspace{0.5em}
        \includegraphics[width=\linewidth]{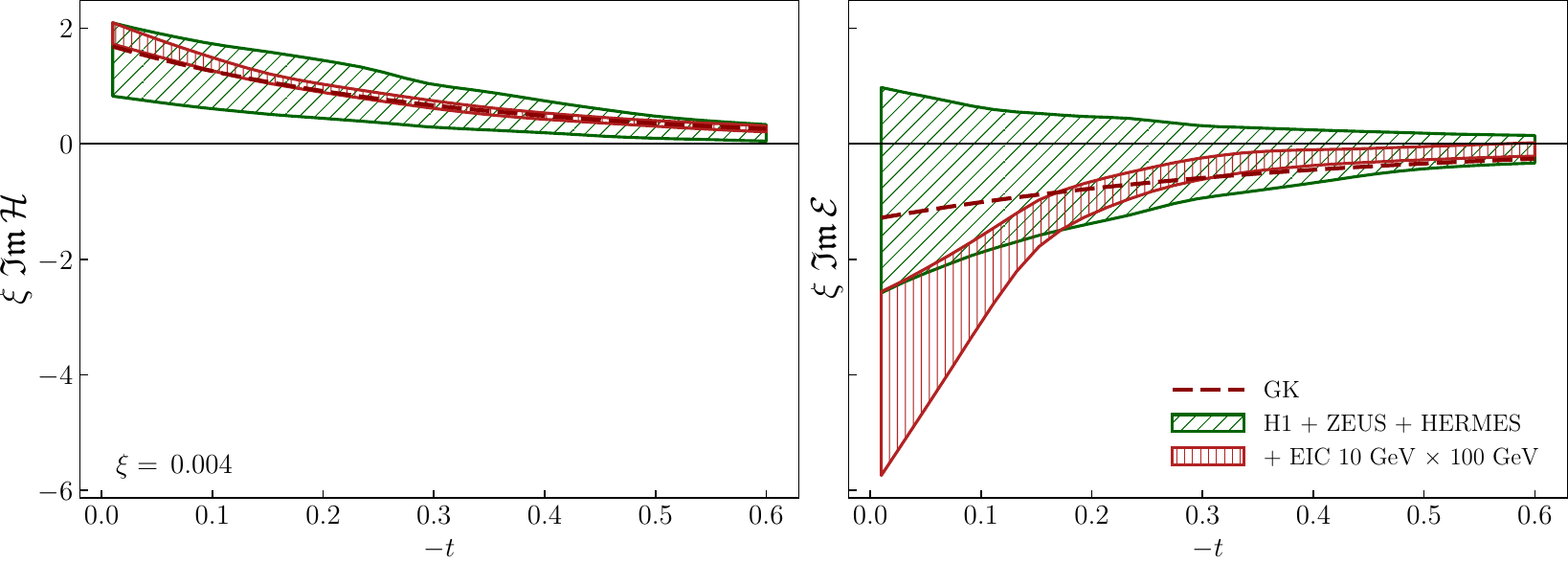}
    \end{minipage}
    \caption{Adapted from \cite{Aschenauer:2025cdq}. Existing DVCS data for proton and the ePIC detector ($x_{\mathrm{Bj}}, Q^2$) coverage in three energy configurations (left); projected impact of EIC data on the extraction of the imaginary part of the CFFs $\mathcal{H}$ and $\mathcal{E}$ as functions of $\xi$ (upper right) and $-t$ (lower right)}.
    \label{fig:EICimpact}
\end{figure}

%%%%%%%
%%%%%%%
\section{Current Limitations, Open Problems and Future Directions}
\label{sec:Limitations}

GPD phenomenology is entering a qualitatively new stage. The theoretical foundations of deeply virtual exclusive processes are well established, increasingly precise measurements are becoming available, and a variety of phenomenological approaches have demonstrated sensitivity to different aspects of quark and gluon structure. The central challenge is therefore shifting from establishing access to GPDs to determining how completely and reliably their multidimensional structure can be reconstructed from the available information.

Several limitations remain. Experimental coverage is still incomplete in kinematics, polarization observables, flavor, and particularly in the gluon sector, while finite-\(Q^2\), higher-twist, and other theoretical corrections must be controlled at the precision required by current and future measurements. At the phenomenological level, the successive inverse problems connecting observables to CFFs and CFFs to GPDs introduce correlations, degeneracies, and intrinsic ambiguities that cannot necessarily be resolved by increasing the flexibility of a parametrization alone. A central objective for the next generation of analyses will therefore be to distinguish information genuinely constrained by experiment from that supplied by theoretical assumptions, parametrizations, or correlations imposed across kinematics.
%%%
At the same time, the field is acquiring powerful new sources of information and new methodologies. Bayesian inference and uncertainty quantification, together with neural networks, generative models, symbolic regression, and other interpretable AI approaches, offer new possibilities for combining all constraints while representing the complex solution spaces characteristic of the inverse problem. 
\\

\noindent The principal limitations and opportunities that will shape the next stage of GPD phenomenology can be summarized as follows:
\\

\begin{enumerate}

\item \textbf{Limited kinematic coverage}: A major limitation of present GPD phenomenology is the restricted and uneven kinematic coverage of the available measurements. Existing fixed-target data predominantly constrain the valence-quark region over limited and correlated ranges of \(x_B\), \(t\), and \(Q^2\), while collider measurements at small \(x_B\) provide complementary but comparatively sparse information. As a consequence, large regions of the multidimensional GPD domain remain unconstrained, and correlations among the kinematic variables make it difficult to disentangle longitudinal momentum, transverse spatial, and scale dependence.
%
%This limitation has consequences beyond the statistical precision of present extractions. 
%A restricted $Q^2$ range limits tests of QCD evolution and the separation of leading- and higher-twist contributions; 
%incomplete $t$ coverage limits the transverse spatial information obtainable from GPDs; and the limited reach in \(x_B\) restricts the connection between the valence, sea-quark, and gluon regimes. 
Extrapolation into unmeasured regions can become strongly dependent on the assumed GPD parametrization or other theoretical constraints.
The EIC will substantially extend the kinematic reach and precision of deeply virtual exclusive measurements, see Fig.~\ref{fig:EICimpact} (left), particularly toward the sea-quark and gluon dominated regions, while providing a much broader range in $Q^2$. This enlarged phase space will be important not simply by increasing the number of data points, but because measurements of the same underlying distributions at different kinematics can provide new constraints on GPD extraction. For instance, the impact of future EIC data on the imaginary part of the $\mathcal{H}$ and $\mathcal{E}$ CFF extraction was discussed in \cite{Aschenauer:2025cdq} using DVCS beam spin asymmetry data. Including the pseudodata generated by the EpIC event generator, \cite{Aschenauer:2022aeb}, and then processing it in the ePIC detector simulation, improved the extraction of both CFFs significantly, see Fig.~\ref{fig:EICimpact} (right), compared to the extraction of CFFs with only existing measurements. This signifies the potential impact of future EIC experiments.
%%%%
%Nevertheless, broader kinematic coverage alone will not remove the intrinsic ambiguities of GPD reconstruction. A comprehensive determination will require complementary observables, targets, and exclusive processes, together with information from lattice QCD and theoretical constraints. Future analyses will therefore need to exploit both the expanded experimental coverage and increasingly flexible statistical frameworks capable of propagating correlations and uncertainties across the multidimensional kinematic domain.

\item \textbf{Deconvolution ambiguity}: 
Recent global GPD analyses have begun to combine deeply virtual exclusive measurements with constraints from PDFs, elastic form factors, and lattice QCD within a common fit. The GUMP1.0 analysis, for example, employs a conventional \(\chi^2\) minimization, with uncertainties estimated from the Hessian matrix and subsequently propagated through a Monte Carlo procedure. The authors note that the present uncertainty analysis neglects correlations and several sources of systematic uncertainty, and identify Bayesian inference and replica-based approaches as directions for a more comprehensive statistical treatment \cite{Guo:2025muf}.
Complementary likelihood-based analyses emphasize the importance of exploring the full multidimensional parameter space rather than relying exclusively on the local behavior around a best-fit solution \cite{Adams:2024pxw,Pandey:2026rvn}. Such analyses can reveal extended correlated directions, non-Gaussian probability distributions, and multiple solutions that may not be adequately represented by Hessian uncertainties. 
%They also provide a means of assessing which parameter combinations are actually constrained by the available information before additional assumptions or correlations across kinematics are imposed.
%%
These considerations illustrate a broader challenge for future global GPD analyses: as increasingly diverse experimental and theoretical inputs are combined, a consistent treatment of correlations, non-Gaussian uncertainties, systematic effects, and possible degeneracies will become essential. 
%Incorporating likelihood-based and Bayesian inference within global analyses offers a natural path toward determining not only the preferred GPD solution, but also the full range of solutions compatible with the combined information.

\item \textbf{Parametrization and theoretical bias}:
Because present data do not fully constrain the multidimensional dependence of GPDs, phenomenological extractions necessarily rely on assumptions about their functional form and correlations among \(x\), \(\xi\), and \(t\), as well as on theoretical constraints and input from PDFs and form factors. Such assumptions are essential for making the inverse problem tractable, but can also bias the information extracted from the data.
%restrict the allowed solution space and lead to uncertainties that reflect the chosen parametrization as much as the information contained in the data.
%
Future analyses should therefore assess the stability of extracted GPD features against changes in parametrization and theoretical assumptions. Flexible representations, including neural networks, can reduce functional-form bias, while Bayesian and generative approaches can provide a broader exploration of admissible solutions. A future goal is to distinguish clearly between features required by the data and those resulting from the assumptions used in the extraction.

\item \textbf{CFF correlations and multiple solutions}:
As discussed above, the simultaneous contribution of several CFFs to the measured observables can generate strong correlations, weakly constrained directions, and multiple solutions. 
%The challenge for future analyses is therefore not simply to reduce the uncertainties assigned to individual CFFs, but to determine which combinations are genuinely identifiable from the available measurements and which remain constrained primarily through assumptions or correlations across kinematics.
%%
Progress will require both new experimental information and more complete statistical treatments. Measurements of complementary polarization observables, beam energies, targets, and exclusive channels can probe different combinations of CFFs and thereby help break existing degeneracies. At the same time, likelihood-based and Bayesian approaches capable of representing non-Gaussian and multimodal probability distributions will be important for preserving ambiguities that remain in the data. Such analyses can also help identify the observables and kinematic regions that provide the greatest additional information, establishing a direct connection between CFF extraction and the optimization of future measurements.

\item \textbf{Higher-twist effects}: At finite \(Q^2\), the leading-twist description of deeply virtual exclusive processes receives corrections from both higher-twist partonic contributions and kinematic power corrections. Schematically, the amplitude can be organized as
\[
{\cal A}
=
{\cal A}^{(2)}
+\frac{1}{Q}{\cal A}^{(3)}
+\frac{1}{Q^2}{\cal A}^{(4)}
+\cdots ,
\]
where the superscript denotes the corresponding twist. While the leading twist-two amplitudes provide the basis for most phenomenological GPD analyses, the moderate values of \(Q^2\) covered by much of the existing fixed-target data make power-suppressed contributions potentially relevant.
At twist three, additional helicity amplitudes and CFFs enter the reaction. 
While total quark angular momentum is accessible through twist-two GPDs via the Ji sum rule (Section \ref{sec:definitions}), isolating its orbital component brings in twist-three GPDs and the underlying quark--gluon correlations.
%These contain both contributions related to twist-two GPDs through Wandzura--Wilczek-type relations and genuine quark--gluon--quark correlations. Their presence modifies the azimuthal and polarization dependence of the observables and can therefore affect the extraction of the leading-twist CFFs if they are neglected. 
Measurements of different azimuthal harmonics and polarization observables provide, in principle, sensitivity to these subleading amplitudes.
Power corrections can also arise from purely kinematic effects associated with the finite nucleon mass and momentum transfer, generating terms proportional to \(M^2/Q^2\) and \(t/Q^2\).  
%Such effects can be numerically important in the kinematic regime of Jefferson Lab and other fixed-target experiments and should be treated systematically.
%
%Higher-twist effects therefore introduce an additional source of uncertainty into the first inverse problem. If the theoretical description used in the extraction contains only leading-twist CFFs, contributions from omitted subleading amplitudes may be absorbed into the fitted leading-twist quantities. Conversely, 
Allowing for additional higher-twist CFFs increases the dimensionality of an already underconstrained inverse problem and therefore requires a dedicated treatment of the resulting correlations, degeneracies, and uncertainties.

\item \textbf{Gluon GPDs and small-$x$ behavior}:
One of the missing pieces to the proton spin puzzle is the low $x$ and gluon dynamics. Experimental probes have thus far not probed particularly deeply into the low $x$ region, especially for exclusive processes, hence the need for the EIC. Studies of GPD low $x$ evolution include, \cite{Hatta:2016aoc, Hatta:2022bxn, Kovchegov:2025yyl}. The gluon gravitational form factors calculated on the lattice provide a constraint on phenomenology for studying these dynamics, \cite{Shanahan:2018pib, Hackett:2023rif}. 
%Without appealing directly to low $x$ evolution equations and gluon distributions, however, one can leverage the interplay of $x$ and $t$ correlations to study low $x$ with GPDs. Integrating the GPDs over varying regions of $x$ allows for a study of how much of the electromagnetic form factors, at different values of $t$, are generated by different regions of $x$, Refs. \cite{Gonzalez-Hernandez:2012xap, Diehl:2013xca}.
%The determination of gluon GPDs remains considerably less developed than that of their quark counterparts. 
In DVCS, gluons enter beyond leading order and through QCD evolution, so that their extraction requires a broad $Q^2$ range to disentangle gluon contributions from the quark singlet sector. Present fixed-target measurements provide only limited leverage for such a separation, whereas
a major opportunity is provided by processes with enhanced direct sensitivity to gluons. Exclusive heavy-vector-meson production, particularly $J/\psi$ and $\Upsilon$ production, provides access to gluon GPDs and their momentum-transfer dependence, while measurements in ultra-peripheral collisions at RHIC and the LHC extend this sensitivity to very small $x$. Near-threshold quarkonium production probes a complementary large-skewness regime and may provide information on moments of gluon GPDs and the gluonic energy--momentum tensor. At the EIC, the large available $Q^2$ and energy ranges should substantially improve the separation and multidimensional imaging of the gluon sector.
An important theoretical challenge will be to establish a consistent description across these different kinematic regimes. At increasingly small $x$, the conventional collinear GPD framework interfaces with high-energy descriptions in terms of dipoles, gluon-density fluctuations, and eventually nonlinear or saturation dynamics. Understanding this connection and the associated uncertainty quantification,  will be essential for extending three-dimensional hadron imaging from the predominantly quark-dominated region of present measurements to the gluon-dominated regime.

\item \textbf{Chiral-odd GPDs}:
The chiral-odd GPDs \(H_T\), \(E_T\), \(\widetilde H_T\), and \(\widetilde E_T\) remain among the least constrained components of nucleon structure. Unlike the chiral-even GPDs, they do not contribute to DVCS at leading twist because the electromagnetic hard scattering conserves quark chirality. Their experimental determination has therefore focused primarily on exclusive pseudoscalar-meson production, particularly \(\pi^0\) and \(\eta\) electroproduction, where transverse-photon amplitudes can provide sensitivity to chiral-odd GPD combinations.
A substantial body of JLab measurements shows the phenomenological importance of transverse amplitudes in these channels, motivating descriptions in terms of chiral-odd GPDs and providing first constraints on the corresponding amplitudes and CFFs. Their quantitative extraction, however, remains considerably more challenging than in DVCS. The interpretation depends on the factorization mechanism and on the treatment of the meson-production subprocess, while several chiral-odd GPDs contribute simultaneously to the measured observables. The limited number of polarization observables further leads to strong correlations and ambiguities among the corresponding amplitudes.
Of particular interest is $H_T$, whose forward limit is the quark transversity distribution, $H_T^q(x,0,0)=h_1^q(x))$. Its first moment therefore gives the tensor charge,
\[
\delta q=\int_{-1}^{1}dx,H_T^q(x,0,0),
\]
providing a potential connection between exclusive measurements of chiral-odd GPDs and an independent determination of the nucleon tensor charge.
Future progress will require a broader set of polarization measurements, improved control of the reaction mechanism and its \(Q^2\) dependence, and statistically complete analyses identifying which chiral-odd CFF combinations are actually constrained by the data. Combining such measurements with lattice-QCD information and constraints from transversity and the tensor charge may provide important complementary information. Establishing a quantitatively  extraction of the chiral-odd sector provides an alternative source of the tensor hadronic matrix elements needed in Beyond-Standard-Model (BSM) analyses, \cite{Courtoy:2015haa}. 

\item \textbf{Nuclear GPDs}:
%Extending GPD phenomenology from the nucleon to nuclei introduces both new opportunities and additional theoretical challenges. Coherent deeply virtual exclusive scattering provides access to GPDs of the nucleus as a whole and thus to the multidimensional spatial and momentum structure of nuclear partons, whereas incoherent processes can probe the partonic structure of bound nucleons. Separating genuine modifications of nucleon structure from conventional nuclear effects, such as binding, Fermi motion, and final-state interactions, remains a central challenge.
At the EIC, measurements over a broad range of nuclear targets will extend to higher energies and into the sea-quark and gluon dominated regimes. In particular, coherent and incoherent exclusive production will provide complementary sensitivity to the average spatial distribution and fluctuations of gluons in nuclei. A consistent framework connecting nucleon and nuclear GPDs, nuclear dynamics, and the small-$x$ description of gluon structure will therefore be an important component of future multidimensional imaging.

\item \textbf{Systematic treatment of lattice and experimental uncertainties}:
The uncertainties entering a global GPD analysis can be broadly separated into aleatoric and epistemic components. Aleatoric uncertainties arise from the intrinsic statistical variability of the measurements, whereas epistemic uncertainties reflect incomplete knowledge of the theoretical description, parametrization, or inference procedure and can, in principle, be reduced as additional information becomes available. 
Experimental measurements contain statistical and correlated systematic uncertainties, normalization effects, and correlations among observables and kinematic bins. Lattice-QCD calculations introduce a different set of uncertainties, including statistical correlations, finite-volume and discretization effects, renormalization and matching uncertainties, excited-state contamination, and extrapolations to physical parameters and kinematics. While statistical fluctuations are predominantly aleatoric, theoretical, model, and parametrization uncertainties are largely epistemic, with some sources of systematic uncertainty not falling uniquely into either category.
Future global analyses will therefore require statistical frameworks capable of incorporating both classes of uncertainty and the full covariance information from experiment and lattice QCD, together with theoretical and parametrization uncertainties. A natural setting for such multimodal inference, allowing the information supplied by different sources to be quantified was provided {\it e.g.} in \cite{Almaeen:2024guo}. A future goal should be a joint determination in which experimental and lattice information constrain complementary directions in GPD space, while their respective aleatoric and epistemic uncertainties and correlations remain explicitly represented.

\item \textbf{Advanced inference and AI.}
The increasing dimensionality and diversity of information entering GPD analyses will require inference methods capable of exploiting correlations while retaining a rigorous treatment of uncertainties and degeneracies. Neural-network, generative, and interpretable AI approaches provide promising complementary tools for flexible GPD representations, exploration of non-unique inverse solutions, and reduction of parametrization bias. 
%Their role, together with symbolic regression and emerging AI-assisted scientific workflows, is discussed further in the Conclusions.
AI methods are becoming increasingly relevant to the inverse problems encountered in GPD phenomenology. Neural networks allow flexible GPD representations with reduced dependence on predetermined functional forms, while generative approaches can describe multiple solutions compatible with the same experimental information. Symbolic regression provides a complementary direction, seeking analytic expressions that can be directly compared with known QCD constraints and interpreted physically. An important open question is how these approaches can be combined with well-defined uncertainty quantification.
Further developments may involve LLMs and agentic AI, particularly as tools for connecting different components of an analysis. Coupled to validated numerical codes and databases, such systems could coordinate experimental data analysis, lattice-QCD input, perturbative calculations, statistical inference, and symbolic methods, and could be used to formulate and test hypotheses across these different sources of information. 
%For scientific applications, however, such developments will require quantitative uncertainty estimates, reproducibility, enforcement of physical constraints, and a clear distinction between established results and generated hypotheses. 
Recent developments in experimental contexts, \textit{e.g.},  \cite{Moreno:2026mqk}, high energy theory, \cite{Schwartz:2026ekw}, and for interdisciplinary physics studies, \cite{Bakshi:2025fgx}, have already been offering potential effective uses of agentic AI techniques applicable to QCD phenomenology.
\end{enumerate}
\vspace{0.5cm}

The objective of future GPD analyses is not simply to obtain smaller uncertainty bands, but to determine which features of hadron structure are actually constrained by the data and which depend on theoretical or phenomenological assumptions. The combination of the JLab and EIC programs with lattice QCD, precision theory, improved statistical inference, and new AI methodologies should make this distinction increasingly quantitative.

%%%%%%%%
%%%%%%%%
%\section{Conclusions and Outlook}
%\label{sec:conclusions}
%\section{Conclusions and Outlook}
%
%The next stage of GPD phenomenology will be driven by the transition from demonstrating multidimensional hadron imaging to achieving its quantitative and statistically sound realization. 
%%
%The high-luminosity program at Jlab and the future EIC will expand the kinematic reach, polarization information, and access to sea-quark and gluon structure. 
%Fully exploiting these measurements will require global multichannel analyses combined with with increasingly precise theoretical calculations.
%The integration of experimental measurements with lattice-QCD calculations and theoretical constraints will be particularly important for constraining directions in GPD space that cannot be determined by any single source of information. Progress will depend equally on advances in inference. Bayesian and likelihood-based methods provide a framework for treating correlations, non-Gaussian uncertainties, and multiple solutions in the inverse problems connecting observables, CFFs, and GPDs. 

\begin{ack}[Acknowledgments] 
 TThe authors thank the members and affiliates of the ExclAIm collaboration
 %Douglas Adams, Prasanna Balachandran, 
 %Marie Bo\"{e}r, Gia-Wei Chern, Michael Engelhardt, Geoffrey Fox, Gary Goldstein,  Liam Hockley, Chunyu Hu,   
 %Ho (Jason) Jang, Adil Khawaja, Megan Kralj, Yaohang Li,  Huey-Wen Lin, ShunShun Liu, Emmanuel Ortiz Pacheco, Saraswati Pandey, Brannon Semp, Dennis Sivers, Matthew Sievert, Jitao Xu. 
 We are also grateful for lively discussions with Geoffrey Fox, Ke-Fei Liu, with the organizers and participants in the EIC Theory Institute at Brookhaven National Laboratory, Summer 2026, and with the participants in the ``Towards Improved Hadron Femtography with Hard Exclusive Reactions Workshop" Edition V, University of Virginia, 2026. 
This work was completed by the ExclAIm collaboration under the Department of Energy grant DE-SC0024644. We also acknowledge DOE grant DE-SC0016286. 
\end{ack}

\bibliographystyle{Harvard}
\bibliography{DVCS}

@article{Polyakov:2018zvc,
    author = "Polyakov, Maxim V. and Schweitzer, Peter",
    title = "{Forces inside hadrons: pressure, surface tension, mechanical radius, and all that}",
    eprint = "1805.06596",
    archivePrefix = "arXiv",
    primaryClass = "hep-ph",
    doi = "10.1142/S0217751X18300259",
    journal = "Int. J. Mod. Phys. A",
    volume = "33",
    number = "26",
    pages = "1830025",
    year = "2018"
}

@article{Goeke:2007fp,
    author = "Goeke, K. and Grabis, J. and Ossmann, J. and Polyakov, M. V. and Schweitzer, P. and Silva, A. and Urbano, D.",
    title = "{Nucleon form-factors of the energy momentum tensor in the chiral quark-soliton model}",
    eprint = "hep-ph/0702030",
    archivePrefix = "arXiv",
    reportNumber = "RUB-TP2-05-2006",
    doi = "10.1103/PhysRevD.75.094021",
    journal = "Phys. Rev. D",
    volume = "75",
    pages = "094021",
    year = "2007"
}

@Article{Belitsky:2010jw,
     author    = "Belitsky, A. V. and Mueller, D.",
     title     = "{Exclusive electroproduction revisited: treating
                  kinematical effects}",
     journal   = "Phys. Rev.",
     volume    = "D82",
     year      = "2010",
     pages     = "074010",
     eprint    = "1005.5209",
     archivePrefix = "arXiv",
     primaryClass  =  "hep-ph",
     doi       = "10.1103/PhysRevD.82.074010",
     SLACcitation  = "%%CITATION = 1005.5209;%%"
}

@Article{Chen:2006na,
     author    = "Chen, S. and Avakian, H. and Burkert, V. and others",
 collaboration = "CLAS",
     title     = "Measurement of deeply virtual Compton scattering with a
                  polarized  proton target",
     journal   = "Phys. Rev. Lett.",
     volume    = "97",
     year      = "2006",
     pages     = "072002",
     eprint    = "hep-ex/0605012",
     SLACcitation  = "%%CITATION = HEP-EX/0605012;%%"
}

@Article{Vanderhaeghen:1999xj,
     author    = "Vanderhaeghen, M. and Guichon, P. A. M. and Guidal, M.",
     title     = "Deeply virtual electroproduction of photons and mesons on
                  the nucleon:  Leading order amplitudes and power
                  corrections",
     journal   = "Phys. Rev.",
     volume    = "D60",
     year      = "1999",
     pages     = "094017",
     eprint    = "hep-ph/9905372",
     SLACcitation  = "%%CITATION = HEP-PH 9905372;%%"
}

@article{Shiells:2021xqo,
    author = "Shiells, Kyle and Guo, Yuxun and Ji, Xiangdong",
    title = "{On extraction of twist-two Compton form factors from DVCS observables through harmonic analysis}",
    eprint = "2112.15144",
    archivePrefix = "arXiv",
    primaryClass = "hep-ph",
    doi = "10.1007/JHEP08(2022)048",
    journal = "JHEP",
    volume = "08",
    pages = "048",
    year = "2022"
}

@article{Kumericki:2013br,
    author = {Kumeri{\v{c}}ki, Kresimir and M{\"u}ller, Dieter and Murray, Morgan},
    title = "{HERMES impact for the access of Compton form factors}",
    eprint = "1301.1230",
    archivePrefix = "arXiv",
    primaryClass = "hep-ph",
    doi = "10.1134/S1063779614040108",
    journal = "Phys. Part. Nucl.",
    volume = "45",
    number = "4",
    pages = "723--755",
    year = "2014"
}

@Article{Diehl:2005jf,
     author    = "Diehl, M. and Hagler, Ph.",
     title     = "Spin densities in the transverse plane and generalized
                  transversity  distributions",
     journal   = "Eur. Phys. J.",
     volume    = "C44",
     year      = "2005",
     pages     = "87-101",
     eprint    = "hep-ph/0504175",
     SLACcitation  = "%%CITATION = HEP-PH/0504175;%%"
}

@Article{Ahmad:2008hp,
     author    = "Ahmad, Saeed and Goldstein, Gary R. and Liuti, Simonetta",
     title     = "{Nucleon Tensor Charge from Exclusive $\pi^o$
                  Electroproduction}",
     journal   = "Phys. Rev.",
     volume    = "D79",
     year      = "2009",
     pages     = "054014",
     eprint    = "0805.3568",
     archivePrefix = "arXiv",
     primaryClass  =  "hep-ph",
     doi       = "10.1103/PhysRevD.79.054014",
     SLACcitation  = "%%CITATION = 0805.3568;%%"
}

@Article{Polyakov:2002yz,
     author    = "Polyakov, M. V.",
     title     = "Generalized parton distributions and strong forces inside
                  nucleons and nuclei",
     journal   = "Phys. Lett.",
     volume    = "B555",
     year      = "2003",
     pages     = "57-62",
     eprint    = "hep-ph/0210165",
     SLACcitation  = "%%CITATION = HEP-PH/0210165;%%"
}

@Article{Airapetian:2008aa,
     author    = "Airapetian, A. and others",
 collaboration = "HERMES",
     title     = "{Measurement of Azimuthal Asymmetries With Respect To Both
                  Beam Charge and Transverse Target Polarization in Exclusive
                  Electroproduction of Real Photons}",
     journal   = "JHEP",
     volume    = "06",
     year      = "2008",
     pages     = "066",
     eprint    = "0802.2499",
     archivePrefix = "arXiv",
     primaryClass  =  "hep-ex",
     doi       = "10.1088/1126-6708/2008/06/066",
     SLACcitation  = "%%CITATION = 0802.2499;%%"
}

@Article{Ye:2006gza,
     author    = "Ye, Zhenyu",
 collaboration = "HERMES",
     title     = "Transverse target-spin asymmetry associated with DVCS on
                  the proton and a resulting model-dependent constraint on
                  the total angular momentum of quarks in the nucleon",
     year      = "2006",
     eprint    = "hep-ex/0606061",
     SLACcitation  = "%%CITATION = HEP-EX/0606061;%%"
}

@Article{Kopytin:2005vv,
     author    = "Kopytin, M.",
 collaboration = "HERMES",
     title     = "Measurement of deeply virtual Compton scattering at
                  HERMES",
     journal   = "AIP Conf. Proc.",
     volume    = "792",
     year      = "2005",
     pages     = "424-427",
     SLACcitation  = "%%CITATION = APCPC,792,424;%%"
}

@Article{MunozCamacho:2006hx,
     author    = "Munoz Camacho, C. and others",
 collaboration = "Jefferson Lab Hall A",
     title     = "Scaling tests of the cross section for deeply virtual
                  Compton  scattering",
     journal   = "Phys. Rev. Lett.",
     volume    = "97",
     year      = "2006",
     pages     = "262002",
     eprint    = "nucl-ex/0607029",
     SLACcitation  = "%%CITATION = NUCL-EX/0607029;%%"
}

@Article{Belitsky:2001ns,
     author    = "Belitsky, Andrei V. and Mueller, Dieter and Kirchner, A.",
     title     = "Theory of deeply virtual Compton scattering on the
                  nucleon",
     journal   = "Nucl. Phys.",
     volume    = "B629",
     year      = "2002",
     pages     = "323-392",
     eprint    = "hep-ph/0112108",
     SLACcitation  = "%%CITATION = HEP-PH 0112108;%%"
}

@Article{Radyushkin:1997ki,
     author    = "Radyushkin, A. V.",
     title     = "Nonforward parton distributions",
     journal   = "Phys. Rev.",
     volume    = "D56",
     year      = "1997",
     pages     = "5524-5557",
     eprint    = "hep-ph/9704207",
     SLACcitation  = "%%CITATION = HEP-PH 9704207;%%"
}

@article{Polyakov:1999gs,
    author = "Polyakov, Maxim V. and Weiss, C.",
    title = "{Skewed and double distributions in pion and nucleon}",
    eprint = "hep-ph/9902451",
    archivePrefix = "arXiv",
    reportNumber = "RUB-TPII-1-99",
    doi = "10.1103/PhysRevD.60.114017",
    journal = "Phys. Rev. D",
    volume = "60",
    pages = "114017",
    year = "1999"
}

@article{Teryaev:2001qm,
    author = "Teryaev, O. V.",
    title = "{Crossing and radon tomography for generalized parton distributions}",
    eprint = "hep-ph/0102303",
    archivePrefix = "arXiv",
    reportNumber = "TPR-01-02",
    doi = "10.1016/S0370-2693(01)00564-0",
    journal = "Phys. Lett. B",
    volume = "510",
    pages = "125--132",
    year = "2001"
}

@article{Tiburzi:2004qr,
    author = "Tiburzi, B. C.",
    title = "{Double distributions: Loose ends}",
    eprint = "hep-ph/0405211",
    archivePrefix = "arXiv",
    reportNumber = "NT-UW-04-09",
    doi = "10.1103/PhysRevD.70.057504",
    journal = "Phys. Rev. D",
    volume = "70",
    pages = "057504",
    year = "2004"
}

@article{Muller:2014wxa,
    author = {M{\"u}ller, Dieter and Polyakov, Maxim V. and Semenov-Tian-Shansky, Kirill M.},
    title = "{Dual parametrization of generalized parton distributions in two equivalent representations}",
    eprint = "1412.4165",
    archivePrefix = "arXiv",
    primaryClass = "hep-ph",
    doi = "10.1007/JHEP03(2015)052",
    journal = "JHEP",
    volume = "03",
    pages = "052",
    year = "2015"
}

@Article{Ji:1996ek,
     author    = "Ji, Xiang-Dong",
     title     = "{Gauge invariant decomposition of nucleon spin}",
     journal   = "Phys. Rev. Lett.",
     volume    = "78",
     year      = "1997",
     pages     = "610-613",
     eprint    = "hep-ph/9603249",
     archivePrefix = "arXiv",
     doi       = "10.1103/PhysRevLett.78.610",
     SLACcitation  = "%%CITATION = HEP-PH/9603249;%%"
}

@Article{Diehl:2003ny,
     author    = "Diehl, M.",
     title     = "{Generalized parton distributions}",
     journal   = "Phys. Rept.",
     volume    = "388",
     year      = "2003",
     pages     = "41-277",
     eprint    = "hep-ph/0307382",
     archivePrefix = "arXiv",
     doi       = "10.1016/j.physrep.2003.08.002",
     SLACcitation  = "%%CITATION = HEP-PH/0307382;%%"
}

@Article{Burkardt:2002hr,
     author    = "Burkardt, Matthias",
     title     = "{Impact parameter space interpretation for generalized
                  parton  distributions}",
     journal   = "Int. J. Mod. Phys.",
     volume    = "A18",
     year      = "2003",
     pages     = "173-208",
     eprint    = "hep-ph/0207047",
     archivePrefix = "arXiv",
     doi       = "10.1142/S0217751X03012370",
     SLACcitation  = "%%CITATION = HEP-PH/0207047;%%"
}

@Article{Burkardt:2000za,
     author    = "Burkardt, Matthias",
     title     = "{Impact parameter dependent parton distributions and off-
                  forward parton  distributions for zeta --> 0}",
     journal   = "Phys. Rev.",
     volume    = "D62",
     year      = "2000",
     pages     = "071503",
     eprint    = "hep-ph/0005108",
     archivePrefix = "arXiv",
     doi       = "10.1103/PhysRevD.62.071503",
     SLACcitation  = "%%CITATION = HEP-PH/0005108;%%"
}

@Article{Belitsky:2003nz,
     author    = "Belitsky, Andrei V. and Ji, Xiang-dong and Yuan, Feng",
     title     = "{Quark imaging in the proton via quantum phase-space
                  distributions}",
     journal   = "Phys. Rev.",
     volume    = "D69",
     year      = "2004",
     pages     = "074014",
     eprint    = "hep-ph/0307383",
     archivePrefix = "arXiv",
     doi       = "10.1103/PhysRevD.69.074014",
     SLACcitation  = "%%CITATION = HEP-PH/0307383;%%"
}

@Article{Ji:1996nm,
     author    = "Ji, Xiang-Dong",
     title     = "Deeply-virtual Compton scattering",
     journal   = "Phys. Rev.",
     volume    = "D55",
     year      = "1997",
     pages     = "7114-7125",
     eprint    = "hep-ph/9609381",
     SLACcitation  = "%%CITATION = HEP-PH 9609381;%%"
}

@Article{Goloskokov:2007nt,
     author    = "Goloskokov, S. V. and Kroll, P.",
     title     = "{The role of the quark and gluon GPDs in hard vector-meson
                  electroproduction}",
     journal   = "Eur. Phys. J.",
     volume    = "C53",
     year      = "2008",
     pages     = "367-384",
     eprint    = "hep-ph/0708.3569",
     archivePrefix = "arXiv",
     primaryClass  =  "hep-ph",
     SLACcitation  = "%%CITATION = 0708.3569;%%"
}

@Article{Goloskokov:2011rd,
     author    = "Goloskokov, S. V. and Kroll, P.",
     title     = "{Transversity in hard exclusive electroproduction of
                  pseudoscalar mesons}",
     journal   = "Eur. Phys. J.",
     volume    = "A47",
     year      = "2011",
     pages     = "112",
     eprint    = "1106.4897",
     archivePrefix = "arXiv",
     primaryClass  =  "hep-ph",
     doi       = "10.1140/epja/i2011-11112-6",
     SLACcitation  = "%%CITATION = 1106.4897;%%"
}

@article{Lorce:2017wkb,
    author = "Lorc{\'e}, C{\'e}dric and Mantovani, Luca and Pasquini, Barbara",
    title = "{Spatial distribution of angular momentum inside the nucleon}",
    eprint = "1704.08557",
    archivePrefix = "arXiv",
    primaryClass = "hep-ph",
    doi = "10.1016/j.physletb.2017.11.018",
    journal = "Phys. Lett. B",
    volume = "776",
    pages = "38--47",
    year = "2018"
}

@article{Schweitzer:2019kkd,
    author = "Schweitzer, Peter and Tezgin, Kemal",
    title = "{Monopole and quadrupole contributions to the angular momentum density}",
    eprint = "1905.12336",
    archivePrefix = "arXiv",
    primaryClass = "hep-ph",
    doi = "10.1016/j.physletb.2019.07.033",
    journal = "Phys. Lett. B",
    volume = "796",
    pages = "47--51",
    year = "2019"
}

@article{JeffersonLabHallA:2016wye,
    author = "Defurne, M. and others",
    collaboration = "Jefferson Lab Hall A",
    title = "{Rosenbluth separation of the $\pi^0$ electroproduction cross section}",
    eprint = "1608.01003",
    archivePrefix = "arXiv",
    primaryClass = "hep-ex",
    reportNumber = "JLAB-PHY-16-2309",
    doi = "10.1103/PhysRevLett.117.262001",
    journal = "Phys. Rev. Lett.",
    volume = "117",
    number = "26",
    pages = "262001",
    year = "2016"
}

@article{AbdulKhalek:2021gbh,
    author = "Abdul Khalek, R. and others",
    title = "{Science Requirements and Detector Concepts for the Electron-Ion Collider}: {EIC Yellow Report}",
    eprint = "2103.05419",
    archivePrefix = "arXiv",
    primaryClass = "physics.ins-det",
    reportNumber = "BNL-220990-2021-FORE, JLAB-PHY-21-3198, LA-UR-21-20953",
    doi = "10.1016/j.nuclphysa.2022.122447",
    journal = "Nucl. Phys. A",
    volume = "1026",
    pages = "122447",
    year = "2022"
}

@article{CLAS:2021lky,
    author = "Chatagnon, P. and others",
    collaboration = "CLAS",
    title = "{First Measurement of Timelike Compton Scattering}",
    eprint = "2108.11746",
    archivePrefix = "arXiv",
    primaryClass = "hep-ex",
    reportNumber = "JLAB-PHY-21-3488",
    doi = "10.1103/PhysRevLett.127.262501",
    journal = "Phys. Rev. Lett.",
    volume = "127",
    number = "26",
    pages = "262501",
    year = "2021"
}

@article{Collins:1996fb,
    author = "Collins, John C. and Frankfurt, Leonid and Strikman, Mark",
    title = "{Factorization for hard exclusive electroproduction of mesons in QCD}",
    eprint = "hep-ph/9611433",
    archivePrefix = "arXiv",
    reportNumber = "CERN-TH-96-314, PSU-TH-168",
    doi = "10.1103/PhysRevD.56.2982",
    journal = "Phys. Rev. D",
    volume = "56",
    pages = "2982--3006",
    year = "1997"
}

@article{NNPDF:2021uiq,
    author = "Ball, Richard D. and others",
    collaboration = "NNPDF",
    title = "{An open-source machine learning framework for global analyses of parton distributions}",
    eprint = "2109.02671",
    archivePrefix = "arXiv",
    primaryClass = "hep-ph",
    reportNumber = "Edinburgh 2021/13, Nikhef-2021-020, TIF-UNIMI-2021-12",
    doi = "10.1140/epjc/s10052-021-09747-9",
    journal = "Eur. Phys. J. C",
    volume = "81",
    number = "10",
    pages = "958",
    year = "2021"
}

@article{Ball:2008by,
    author = "Ball, Richard D. and Del Debbio, Luigi and Forte, Stefano and Guffanti, Alberto and Latorre, Jose I. and Piccione, Andrea and Rojo, Juan and Ubiali, Maria",
    collaboration = "NNPDF",
    title = "{A Determination of parton distributions with faithful uncertainty estimation}",
    eprint = "0808.1231",
    archivePrefix = "arXiv",
    primaryClass = "hep-ph",
    reportNumber = "EDINBURGH-2008-25, IFUM-923-FT, FREIBURG-2008-08",
    doi = "10.1016/j.nuclphysb.2008.09.037",
    journal = "Nucl. Phys. B",
    volume = "809",
    pages = "1--63",
    year = "2009",
    note = "[Erratum: Nucl.Phys.B 816, 293 (2009)]"
}

@Article{Guidal:2010de,
     author    = "Guidal, M.",
     title     = "{Constraints on the $\tilde{H}$ Generalized Parton
                  Distribution from Deep Virtual Compton Scattering Measured
                  at HERMES}",
     journal   = "Phys. Lett.",
     volume    = "B693",
     year      = "2010",
     pages     = "17-23",
     eprint    = "1005.4922",
     archivePrefix = "arXiv",
     primaryClass  =  "hep-ph",
     doi       = "10.1016/j.physletb.2010.07.059",
     SLACcitation  = "%%CITATION = 1005.4922;%%"
}

@Article{Guidal:2010ig,
     author    = "Guidal, M.",
     title     = "{Generalized Parton Distributions from Deep Virtual Compton
                  Scattering at CLAS}",
     journal   = "Phys. Lett.",
     volume    = "B689",
     year      = "2010",
     pages     = "156-162",
     eprint    = "1003.0307",
     archivePrefix = "arXiv",
     primaryClass  =  "hep-ph",
     doi       = "10.1016/j.physletb.2010.04.053",
     SLACcitation  = "%%CITATION = 1003.0307;%%"
}

@Article{Guidal:2004nd,
     author    = "Guidal, M. and Polyakov, M. V. and Radyushkin, A. V. and
                  Vanderhaeghen, M.",
     title     = "{Nucleon form factors from generalized parton
                  distributions}",
     journal   = "Phys. Rev.",
     volume    = "D72",
     year      = "2005",
     pages     = "054013",
     eprint    = "hep-ph/0410251",
     archivePrefix = "arXiv",
     doi       = "10.1103/PhysRevD.72.054013",
     SLACcitation  = "%%CITATION = HEP-PH/0410251;%%"
}

@Article{Guidal:2009aa,
     author    = "Guidal, M. and Moutarde, H.",
     title     = "{Generalized Parton Distributions from Deeply Virtual
                  Compton Scattering at HERMES}",
     journal   = "Eur. Phys. J.",
     volume    = "A42",
     year      = "2009",
     pages     = "71-78",
     eprint    = "0905.1220",
     archivePrefix = "arXiv",
     primaryClass  =  "hep-ph",
     doi       = "10.1140/epja/i2009-10840-4",
     SLACcitation  = "%%CITATION = 0905.1220;%%"
}

@article{Kumericki:2007sa,
    author = "Kumericki, K. and Mueller, Dieter and Passek-Kumericki, K.",
    title = "{Towards a fitting procedure for deeply virtual Compton scattering at next-to-leading order and beyond}",
    eprint = "hep-ph/0703179",
    archivePrefix = "arXiv",
    doi = "10.1016/j.nuclphysb.2007.10.029",
    journal = "Nucl. Phys. B",
    volume = "794",
    pages = "244--323",
    year = "2008"
}

@article{Kumericki:2011rz,
    author = "Kumericki, Kresimir and Mueller, Dieter and Schafer, Andreas",
    title = "{Neural network generated parametrizations of deeply virtual Compton form factors}",
    eprint = "1106.2808",
    archivePrefix = "arXiv",
    primaryClass = "hep-ph",
    doi = "10.1007/JHEP07(2011)073",
    journal = "JHEP",
    volume = "07",
    pages = "073",
    year = "2011"
}

@Article{Guidal:2008ie,
     author    = "Guidal, M.",
     title     = "{A fitter code for Deep Virtual Compton Scattering and
                  Generalized Parton Distributions}",
     journal   = "Eur. Phys. J.",
     volume    = "A37",
     year      = "2008",
     pages     = "319-332",
     eprint    = "0807.2355",
     archivePrefix = "arXiv",
     primaryClass  =  "hep-ph",
     doi       = "10.1140/epja/i2008-10630-6",
     SLACcitation  = "%%CITATION = 0807.2355;%%"
}

@Article{Kumericki:2009uq,
     author    = "Kumericki, Kresimir and Mueller, Dieter",
     title     = "{Deeply virtual Compton scattering at small $x_B$ and the
                  access to the GPD H}",
     journal   = "Nucl. Phys.",
     volume    = "B841",
     year      = "2010",
     pages     = "1-58",
     eprint    = "0904.0458",
     archivePrefix = "arXiv",
     primaryClass  =  "hep-ph",
     doi       = "10.1016/j.nuclphysb.2010.07.015",
     SLACcitation  = "%%CITATION = 0904.0458;%%"
}

@article{Moutarde:2019tqa,
    author = "Moutarde, H. and Sznajder, P. and Wagner, J.",
    title = "{Unbiased determination of DVCS Compton Form Factors}",
    eprint = "1905.02089",
    archivePrefix = "arXiv",
    primaryClass = "hep-ph",
    doi = "10.1140/epjc/s10052-019-7117-5",
    journal = "Eur. Phys. J. C",
    volume = "79",
    number = "7",
    pages = "614",
    year = "2019"
}

@article{Adams:2024pxw,
    author = "Adams, Douglas Q. and others",
    title = "{Likelihood and Correlation Analysis of Compton Form Factors for Deeply Virtual Exclusive Scattering on the Nucleon}",
    eprint = "2410.23469",
    archivePrefix = "arXiv",
    primaryClass = "hep-ph",
    month = "10",
    year = "2024"
}

@article{Bertone:2021yyz,
    author = "Bertone, V. and Dutrieux, H. and Mezrag, C. and Moutarde, H. and Sznajder, P.",
    title = "{Deconvolution problem of deeply virtual Compton scattering}",
    eprint = "2104.03836",
    archivePrefix = "arXiv",
    primaryClass = "hep-ph",
    doi = "10.1103/PhysRevD.103.114019",
    journal = "Phys. Rev. D",
    volume = "103",
    number = "11",
    pages = "114019",
    year = "2021"
}

@article{Qiu:2022bpq,
    author = "Qiu, Jian-Wei and Yu, Zhite",
    title = "{Exclusive production of a pair of high transverse momentum photons in pion-nucleon collisions for extracting generalized parton distributions}",
    eprint = "2205.07846",
    archivePrefix = "arXiv",
    primaryClass = "hep-ph",
    reportNumber = "JLAB-THY-22-3617, MSUHEP-22-018",
    doi = "10.1007/JHEP08(2022)103",
    journal = "JHEP",
    volume = "08",
    pages = "103",
    year = "2022"
}

@article{Siddikov:2025orq,
    author = "Siddikov, M. and Zemlyakov, I. and Roa, M.",
    title = "{Exclusive photoproduction of {\ensuremath{\chi}}c{\ensuremath{\gamma}} pairs in the small-x kinematics}",
    eprint = "2510.14767",
    archivePrefix = "arXiv",
    primaryClass = "hep-ph",
    doi = "10.1103/y73d-hzdd",
    journal = "Phys. Rev. D",
    volume = "113",
    number = "3",
    pages = "034024",
    year = "2026"
}

@article{Siddikov:2025kah,
    author = "Siddikov, Marat and Zemlyakov, Ivan",
    title = "{Exclusive photoproduction of {\ensuremath{\chi}}c{\ensuremath{\gamma}} pairs}",
    eprint = "2503.10848",
    archivePrefix = "arXiv",
    primaryClass = "hep-ph",
    doi = "10.1103/7wnc-p9h2",
    journal = "Phys. Rev. D",
    volume = "112",
    number = "1",
    pages = "014021",
    year = "2025"
}

@article{Crnkovic:2025man,
    author = "Crnkovi{\'c}, Nikola and Duplan{\v{c}}i{\'c}, Goran and Nabeebaccus, Saad and Passek-K., Kornelija and Pire, Bernard and Szymanowski, Lech and Wallon, Samuel",
    title = "{Hard exclusive photoproduction of photon-meson pairs: Pseudoscalar channels {\ensuremath{\pi}}, {\ensuremath{\eta}}, and {\ensuremath{\eta}}'}",
    eprint = "2511.19720",
    archivePrefix = "arXiv",
    primaryClass = "hep-ph",
    reportNumber = "RBI-ThPhys-2025-49",
    doi = "10.1103/g9m4-9r9c",
    journal = "Phys. Rev. D",
    volume = "113",
    number = "3",
    pages = "034001",
    year = "2026"
}

@article{Nabeebaccus:2026rco,
    author = "Nabeebaccus, Saad and Perez, David and Szymanowski, Lech and Wallon, Samuel",
    title = "{Exclusive photoproduction of a di-meson pair with large invariant mass}",
    eprint = "2605.03880",
    archivePrefix = "arXiv",
    primaryClass = "hep-ph",
    month = "5",
    year = "2026"
}

@article{Deja:2023ahc,
    author = "Deja, K. and Martinez-Fernandez, V. and Pire, B. and Sznajder, P. and Wagner, J.",
    title = "{Phenomenology of double deeply virtual Compton scattering in the era of new experiments}",
    eprint = "2303.13668",
    archivePrefix = "arXiv",
    primaryClass = "hep-ph",
    reportNumber = "CPHT-RR012.032022",
    doi = "10.1103/PhysRevD.107.094035",
    journal = "Phys. Rev. D",
    volume = "107",
    number = "9",
    pages = "094035",
    year = "2023"
}

@article{Huang:2026eai,
    author = "Huang, Yuan-Yuan and Cao, Xu",
    title = "{Constraining DVCS Compton Form Factors Using Lattice QCD informed Neural Network}",
    eprint = "2606.09152",
    archivePrefix = "arXiv",
    primaryClass = "hep-ph",
    month = "6",
    year = "2026"
}

@article{JeffersonLabHallA:2022pnx,
    author = "Georges, F. and others",
    collaboration = "Jefferson Lab Hall A",
    title = "{Deeply Virtual Compton Scattering Cross Section at High Bjorken xB}",
    eprint = "2201.03714",
    archivePrefix = "arXiv",
    primaryClass = "hep-ph",
    doi = "10.1103/PhysRevLett.128.252002",
    journal = "Phys. Rev. Lett.",
    volume = "128",
    number = "25",
    pages = "252002",
    year = "2022"
}

@article{Mezrag:2026wcf,
    author = "Mezrag, C. and Sznajder, P. and Wagner, J.",
    title = "{Extraction of DVCS amplitudes off the nucleon}",
    eprint = "2608.20244",
    archivePrefix = "arXiv",
    primaryClass = "hep-ph",
    month = "8",
    year = "2026"
}

@article{Dotson:2025omi,
    author = "Dotson, Andrew and others",
    title = "{Generalized Parton Distributions from Symbolic Regression}",
    eprint = "2504.13289",
    archivePrefix = "arXiv",
    primaryClass = "hep-ph",
    month = "4",
    year = "2025"
}

@article{Xu:2026lko,
    author = "Xu, Jitao and others",
    title = "{Neural Network Representation of Generalized Parton Distributions (NNGPD)}",
    eprint = "2605.06994",
    archivePrefix = "arXiv",
    primaryClass = "hep-ph",
    month = "5",
    year = "2026"
}

@article{Cuic:2020iwt,
    author = {{\v{C}}ui{\'c}, Marija and Kumeri{\v{c}}ki, Kre{\v{s}}imir and Sch{\"a}fer, Andreas},
    title = "{Separation of Quark Flavors Using Deeply Virtual Compton Scattering Data}",
    eprint = "2007.00029",
    archivePrefix = "arXiv",
    primaryClass = "hep-ph",
    reportNumber = "ZTF-EP-20-04",
    doi = "10.1103/PhysRevLett.125.232005",
    journal = "Phys. Rev. Lett.",
    volume = "125",
    number = "23",
    pages = "232005",
    year = "2020"
}

@article{Braun:2021grd,
    author = "Braun, V. M. and Manashov, A. N. and Moch, S. and Schoenleber, J.",
    title = "{Axial-vector contributions in two-photon reactions: Pion transition form factor and deeply-virtual Compton scattering at NNLO in QCD}",
    eprint = "2106.01437",
    archivePrefix = "arXiv",
    primaryClass = "hep-ph",
    reportNumber = "DESY 21--068, DESY-21-068",
    doi = "10.1103/PhysRevD.104.094007",
    journal = "Phys. Rev. D",
    volume = "104",
    number = "9",
    pages = "094007",
    year = "2021"
}

@article{Braun:2020yib,
    author = "Braun, V. M. and Manashov, A. N. and Moch, S. and Schoenleber, J.",
    title = "{Two-loop coefficient function for DVCS: vector contributions}",
    eprint = "2007.06348",
    archivePrefix = "arXiv",
    primaryClass = "hep-ph",
    reportNumber = "DESY-20-116, DESY 20-116",
    doi = "10.1007/JHEP09(2020)117",
    journal = "JHEP",
    volume = "09",
    pages = "117",
    year = "2020",
    note = "[Erratum: JHEP 02, 115 (2022)]"
}

@article{Bertone:2022frx,
    author = "Bertone, Valerio and Dutrieux, Herv{\'e} and Mezrag, C{\'e}dric and Morgado, Jos{\'e} M. and Moutarde, Herv{\'e}",
    title = "{Revisiting evolution equations for generalised parton distributions}",
    eprint = "2206.01412",
    archivePrefix = "arXiv",
    primaryClass = "hep-ph",
    doi = "10.1140/epjc/s10052-022-10793-0",
    journal = "Eur. Phys. J. C",
    volume = "82",
    number = "10",
    pages = "888",
    year = "2022"
}

@article{Braun:2022bpn,
    author = "Braun, V. M. and Ji, Yao and Schoenleber, Jakob",
    title = "{Deeply Virtual Compton Scattering at Next-to-Next-to-Leading Order}",
    eprint = "2207.06818",
    archivePrefix = "arXiv",
    primaryClass = "hep-ph",
    reportNumber = "TUM-HEP-1407/22",
    doi = "10.1103/PhysRevLett.129.172001",
    journal = "Phys. Rev. Lett.",
    volume = "129",
    number = "17",
    pages = "172001",
    year = "2022"
}

@article{Braun:2025noa,
    author = "Braun, Vladimir M. and Gotzler, Patrick and Manashov, Alexander N.",
    title = "{Conformal moments of the two-loop coefficient functions in DVCS}",
    eprint = "2512.14295",
    archivePrefix = "arXiv",
    primaryClass = "hep-ph",
    reportNumber = "DESY-25-187",
    doi = "10.1103/p5rk-497z",
    journal = "Phys. Rev. D",
    volume = "113",
    number = "7",
    pages = "074005",
    year = "2026"
}

@article{Ji:2023xzk,
    author = "Ji, Yao and Schoenleber, Jakob",
    title = "{Two-loop coefficient functions in deeply virtual Compton scattering: flavor-singlet axial-vector and transversity case}",
    eprint = "2310.05724",
    archivePrefix = "arXiv",
    primaryClass = "hep-ph",
    reportNumber = "TUM-HEP-1474/23",
    doi = "10.1007/JHEP01(2024)053",
    journal = "JHEP",
    volume = "01",
    pages = "053",
    year = "2024"
}

@article{Almaeen:2024guo,
    author = "Almaeen, Manal and Alghamdi, Tareq and Kriesten, Brandon and Adams, Douglas and Li, Yaohang and Lin, Huey-Wen and Liuti, Simonetta",
    title = "{VAIM-CFF: a variational autoencoder inverse mapper solution to Compton form factor extraction from deeply virtual exclusive reactions}",
    eprint = "2405.05826",
    archivePrefix = "arXiv",
    primaryClass = "hep-ph",
    doi = "10.1140/epjc/s10052-025-14091-3",
    journal = "Eur. Phys. J. C",
    volume = "85",
    number = "5",
    pages = "499",
    year = "2025"
}

@article{Hossen:2024qwo,
    author = "Hossen, Fayaz and others",
    title = "{Variational autoencoder inverse mapper for extraction of Compton form factors: Benchmarks and conditional learning}",
    eprint = "2408.11681",
    archivePrefix = "arXiv",
    primaryClass = "hep-ph",
    month = "8",
    year = "2024"
}

@article{Kriesten:2021sqc,
    author = "Kriesten, Brandon and Velie, Philip and Yeats, Emma and Lopez, Fernanda Yepez and Liuti, Simonetta",
    title = "{Parametrization of quark and gluon generalized parton distributions in a dynamical framework}",
    eprint = "2101.01826",
    archivePrefix = "arXiv",
    primaryClass = "hep-ph",
    doi = "10.1103/PhysRevD.105.056022",
    journal = "Phys. Rev. D",
    volume = "105",
    number = "5",
    pages = "056022",
    year = "2022"
}

@article{Kriesten:2020apm,
    author = "Kriesten, Brandon and Liuti, Simonetta and Meyer, Andrew",
    title = "{Novel Rosenbluth extraction framework for Compton form factors from deeply virtual exclusive experiments}",
    eprint = "2011.04484",
    archivePrefix = "arXiv",
    primaryClass = "hep-ph",
    doi = "10.1016/j.physletb.2022.137051",
    journal = "Phys. Lett. B",
    volume = "829",
    pages = "137051",
    year = "2022"
}

@article{Kriesten:2019jep,
    author = "Kriesten, Brandon and Liuti, Simonetta and Calero-Diaz, Liliet and Keller, Dustin and Meyer, Andrew and Goldstein, Gary R. and Osvaldo Gonzalez-Hernandez, J.",
    title = "{Extraction of generalized parton distribution observables from deeply virtual electron proton scattering experiments}",
    eprint = "1903.05742",
    archivePrefix = "arXiv",
    primaryClass = "hep-ph",
    doi = "10.1103/PhysRevD.101.054021",
    journal = "Phys. Rev. D",
    volume = "101",
    number = "5",
    pages = "054021",
    year = "2020"
}

@article{Freese:2024ypk,
    author = {Freese, A. and Adamiak, D. and Clo{\"e}t, I. and Melnitchouk, W. and Qiu, J. -W. and Sato, N. and Zaccheddu, M.},
    title = "{Kernel methods for evolution of generalized parton distributions}",
    eprint = "2412.13450",
    archivePrefix = "arXiv",
    primaryClass = "hep-ph",
    reportNumber = "JLAB-THY-24-4249",
    doi = "10.1016/j.cpc.2025.109552",
    journal = "Comput. Phys. Commun.",
    volume = "311",
    pages = "109552",
    year = "2025"
}

@article{Guo:2025muf,
    author = "Guo, Yuxun and Aslan, Fatma P. and Ji, Xiangdong and Santiago, M. Gabriel",
    title = "{First Global Extraction of Generalized Parton Distributions from Experiment and Lattice Data with Next-to-Leading-Order Accuracy}",
    eprint = "2509.08037",
    archivePrefix = "arXiv",
    primaryClass = "hep-ph",
    doi = "10.1103/qct5-y7rp",
    journal = "Phys. Rev. Lett.",
    volume = "135",
    number = "26",
    pages = "261903",
    year = "2025"
}

@article{Moffat:2023svr,
    author = {Moffat, Eric and Freese, Adam and Clo{\"e}t, Ian and Donohoe, Thomas and Gamberg, Leonard and Melnitchouk, Wally and Metz, Andreas and Prokudin, Alexei and Sato, Nobuo},
    title = "{Shedding light on shadow generalized parton distributions}",
    eprint = "2303.12006",
    archivePrefix = "arXiv",
    primaryClass = "hep-ph",
    reportNumber = "JLAB-THY-23-3786",
    doi = "10.1103/PhysRevD.108.036027",
    journal = "Phys. Rev. D",
    volume = "108",
    number = "3",
    pages = "036027",
    year = "2023"
}

@article{Qiu:2025ksq,
    author = "Qiu, Jian-Wei and Sato, Nobuo and Yu, Zhite",
    title = "{New framework for extracting GPDs from exclusive photon electroproduction}",
    eprint = "2511.20402",
    archivePrefix = "arXiv",
    primaryClass = "hep-ph",
    reportNumber = "JLAB-THY-25-4591",
    doi = "10.1103/36xc-mv5r",
    journal = "Phys. Rev. D",
    volume = "113",
    number = "5",
    pages = "054037",
    year = "2026"
}

@article{Qiu:2022pla,
    author = "Qiu, Jian-Wei and Yu, Zhite",
    title = "{Single diffractive hard exclusive processes for the study of generalized parton distributions}",
    eprint = "2210.07995",
    archivePrefix = "arXiv",
    primaryClass = "hep-ph",
    reportNumber = "MSUHEP-22-032, JLAB-THY-22-3742, JLAB-THY-22-3742, MSUHEP-22-032",
    doi = "10.1103/PhysRevD.107.014007",
    journal = "Phys. Rev. D",
    volume = "107",
    number = "1",
    pages = "014007",
    year = "2023"
}

@article{Pandey:2026rvn,
    author = "Pandey, Saraswati and Adams, Douglas Q. and Liuti, Simonetta",
    title = "{Markov Chain Monte Carlo (MCMC) based likelihood extraction of Chiral-Odd Compton Form Factors from Deeply Virtual Exclusive experiments}",
    eprint = "2605.18589",
    archivePrefix = "arXiv",
    primaryClass = "hep-ph",
    doi = "10.1088/1748-0221/21/08/C08008",
    journal = "JINST",
    volume = "21",
    number = "08",
    pages = "C08008",
    year = "2026"
}

@article{JeffersonLabHallA:2020dhq,
    author = "Dlamini, M. and others",
    collaboration = "Jefferson Lab Hall A",
    title = "{Deep Exclusive Electroproduction of {\ensuremath{\pi}}0 at High Q2 in the Quark Valence Regime}",
    eprint = "2011.11125",
    archivePrefix = "arXiv",
    primaryClass = "hep-ex",
    reportNumber = "
	JLAB-PHY-21-3308",
    doi = "10.1103/PhysRevLett.127.152301",
    journal = "Phys. Rev. Lett.",
    volume = "127",
    number = "15",
    pages = "152301",
    year = "2021"
}

@article{Courtoy:2015haa,
    author = "Courtoy, Aurore and Bae{\ss}ler, Stefan and Gonz{\'a}lez-Alonso, Mart{\'\i}n and Liuti, Simonetta",
    title = "{Beyond-Standard-Model Tensor Interaction and Hadron Phenomenology}",
    eprint = "1503.06814",
    archivePrefix = "arXiv",
    primaryClass = "hep-ph",
    doi = "10.1103/PhysRevLett.115.162001",
    journal = "Phys. Rev. Lett.",
    volume = "115",
    pages = "162001",
    year = "2015"
}

@article{Mankiewicz:1998kg,
    author = "Mankiewicz, L. and Piller, G. and Radyushkin, A.",
    title = "{Hard exclusive electroproduction of pions}",
    eprint = "hep-ph/9812467",
    archivePrefix = "arXiv",
    reportNumber = "TUM-T39-98-33, JLAB-THY-99-02",
    doi = "10.1007/s100529900045",
    journal = "Eur. Phys. J. C",
    volume = "10",
    pages = "307--312",
    year = "1999"
}

@article{CLAS:2014jpc,
    author = "Bedlinskiy, I. and others",
    collaboration = "CLAS",
    title = "{Exclusive ${\pi}^0$ electroproduction at $W>2$ GeV with CLAS}",
    eprint = "1405.0988",
    archivePrefix = "arXiv",
    primaryClass = "nucl-ex",
    reportNumber = "JLAB-PHY-14-1871",
    doi = "10.1103/PhysRevC.90.039901",
    journal = "Phys. Rev. C",
    volume = "90",
    number = "2",
    pages = "025205",
    year = "2014",
    note = "[Addendum: Phys.Rev.C 90, 039901 (2014)]"
}

@article{CLAS:2016tqs,
    author = "Bosted, P. E. and others",
    collaboration = "CLAS",
    title = "{Target and beam-target spin asymmetries in exclusive pion electroproduction for $Q^2>1$ GeV$^2$. II. $e p \rightarrow e \pi^0 p$}",
    eprint = "1611.04987",
    archivePrefix = "arXiv",
    primaryClass = "nucl-ex",
    reportNumber = "JLAB-PHY-16-2388",
    doi = "10.1103/PhysRevC.95.035207",
    journal = "Phys. Rev. C",
    volume = "95",
    number = "3",
    pages = "035207",
    year = "2017"
}

@article{Bedlinskiy:2012be,
      author         = "Bedlinskiy, I. and others",
      title          = "{Measurement of Exclusive $\pi^0$ Electroproduction
                        Structure Functions and their Relationship to Transversity
                        GPDs}",
      collaboration  = "CLAS Collaboration",
      journal        = "Phys.Rev.Lett.",
      volume         = "109",
      pages          = "112001",
      doi            = "10.1103/PhysRevLett.109.112001",
      year           = "2012",
      eprint         = "1206.6355",
      archivePrefix  = "arXiv",
      primaryClass   = "hep-ex",
      reportNumber   = "JLAB-PHY-12-1595",
      SLACcitation   = "%%CITATION = ARXIV:1206.6355;%%",
}

@article{Lorce:2011kd,
      author         = "Lorce, C. and Pasquini, B.",
      title          = "{Quark Wigner Distributions and Orbital Angular
                        Momentum}",
      journal        = "Phys.Rev.",
      volume         = "D84",
      pages          = "014015",
      doi            = "10.1103/PhysRevD.84.014015",
      year           = "2011",
      eprint         = "1106.0139",
      archivePrefix  = "arXiv",
      primaryClass   = "hep-ph",
      SLACcitation   = "%%CITATION = ARXIV:1106.0139;%%",
}

@article{Guidal:2013rya,
      author         = "Guidal, Michel and Moutarde, Hervé and Vanderhaeghen,
                        Marc",
      title          = "{Generalized Parton Distributions in the valence region
                        from Deeply Virtual Compton Scattering}",
      journal        = "Rept.Prog.Phys.",
      volume         = "76",
      pages          = "066202",
      doi            = "10.1088/0034-4885/76/6/066202",
      year           = "2013",
      eprint         = "1303.6600",
      archivePrefix  = "arXiv",
      primaryClass   = "hep-ph",
      SLACcitation   = "%%CITATION = ARXIV:1303.6600;%%",
}

@article{Airapetian:2009ac,
      author         = "Airapetian, A. and others",
      title          = "{Single-spin azimuthal asymmetry in exclusive
                        electroproduction of pi+ mesons on transversely polarized
                        protons}",
      collaboration  = "HERMES Collaboration",
      journal        = "Phys.Lett.",
      volume         = "B682",
      pages          = "345-350",
      doi            = "10.1016/j.physletb.2009.11.039",
      year           = "2010",
      eprint         = "0907.2596",
      archivePrefix  = "arXiv",
      primaryClass   = "hep-ex",
      reportNumber   = "DESY-09-106",
      SLACcitation   = "%%CITATION = ARXIV:0907.2596;%%",
}

@article{Goldstein:2013gra,
      author         = "Goldstein, G.R. and Hernandez, J. O. Gonzalez and Liuti,
                        S.",
      title          = "Flexible Parametrization of Generalized Parton
                        Distributions: The Chiral-Odd Sector",
      year           = "2013",
      eprint         = "1311.0483",
      archivePrefix  = "arXiv",
      primaryClass   = "hep-ph",
      SLACcitation   = "%%CITATION = ARXIV:1311.0483;%%",
		journal			= "to be published",
}

@article{Goldstein:2014aja,
      author         = "Goldstein, Gary. R. and Hernandez, J. Osvaldo Gonzalez
                        and Liuti, Simonetta",
      title          = "{Flavor dependence of chiral odd generalized parton
                        distributions and the tensor charge from the analysis of
                        combined $\pi^0$ and $\eta$ exclusive electroproduction
                        data}",
      year           = "2014",
      eprint         = "1401.0438",
      archivePrefix  = "arXiv",
      primaryClass   = "hep-ph",
      SLACcitation   = "%%CITATION = ARXIV:1401.0438;%%",
}

@article{Goldstein:2010gu,
      author         = "Goldstein, Gary R. and Hernandez, J. OsvaldoGonzalez and
                        Liuti, Simonetta",
      title          = "{Flexible Parametrization of Generalized Parton
                        Distributions from Deeply Virtual Compton Scattering
                        Observables}",
      journal        = "Phys.Rev.",
      volume         = "D84",
      pages          = "034007",
      doi            = "10.1103/PhysRevD.84.034007",
      year           = "2011",
      eprint         = "1012.3776",
      archivePrefix  = "arXiv",
      primaryClass   = "hep-ph",
      SLACcitation   = "%%CITATION = ARXIV:1012.3776;%%",
}

@article{GonzalezHernandez:2012jv,
      author         = "Gonzalez-Hernandez, J. Osvaldo and Liuti, Simonetta and
                        Goldstein, Gary R. and Kathuria, Kunal",
      title          = "{Interpretation of the Flavor Dependence of Nucleon Form
                        Factors in a Generalized Parton Distribution Model}",
      journal        = "Phys.Rev.",
      volume         = "C88",
      pages          = "065206",
      doi            = "10.1103/PhysRevC.88.065206",
      year           = "2013",
      eprint         = "1206.1876",
      archivePrefix  = "arXiv",
      primaryClass   = "hep-ph",
      SLACcitation   = "%%CITATION = ARXIV:1206.1876;%%",
}

@article{Meissner:2009ww,
    author = "Meissner, Stephan and Metz, Andreas and Schlegel, Marc",
    title = "{Generalized parton correlation functions for a spin-1/2 hadron}",
    eprint = "0906.5323",
    archivePrefix = "arXiv",
    primaryClass = "hep-ph",
    reportNumber = "JLAB-THY-09-1018",
    doi = "10.1088/1126-6708/2009/08/056",
    journal = "JHEP",
    volume = "08",
    pages = "056",
    year = "2009"
}

@article{Lorce:2013pza,
    author = "Lorc{\'e}, C. and Pasquini, B.",
    title = "{Structure analysis of the generalized correlator of quark and gluon for a spin-1/2 target}",
    eprint = "1307.4497",
    archivePrefix = "arXiv",
    primaryClass = "hep-ph",
    doi = "10.1007/JHEP09(2013)138",
    journal = "JHEP",
    volume = "09",
    pages = "138",
    year = "2013"
}

@article{Cates:2011pz,
    author = "Cates, G. D. and de Jager, C. W. and Riordan, S. and Wojtsekhowski, B.",
    title = "{Flavor decomposition of the elastic nucleon electromagnetic form factors}",
    eprint = "1103.1808",
    archivePrefix = "arXiv",
    primaryClass = "nucl-ex",
    reportNumber = "JLAB-PHY-11-1325",
    doi = "10.1103/PhysRevLett.106.252003",
    journal = "Phys. Rev. Lett.",
    volume = "106",
    pages = "252003",
    year = "2011"
}

@article{Qattan:2012zf,
    author = "Qattan, I. A. and Arrington, J.",
    title = "{Flavor decomposition of the nucleon electromagnetic form factors}",
    eprint = "1209.0683",
    archivePrefix = "arXiv",
    primaryClass = "nucl-ex",
    doi = "10.1103/PhysRevC.86.065210",
    journal = "Phys. Rev. C",
    volume = "86",
    pages = "065210",
    year = "2012"
}

@article{Qattan:2015qxa,
    author = "Qattan, I. A. and Arrington, J. and Alsaad, A.",
    title = "{Flavor decomposition of the nucleon electromagnetic form factors at low $Q^2$}",
    eprint = "1502.02872",
    archivePrefix = "arXiv",
    primaryClass = "nucl-ex",
    doi = "10.1103/PhysRevC.91.065203",
    journal = "Phys. Rev. C",
    volume = "91",
    number = "6",
    pages = "065203",
    year = "2015"
}

@article{Venkat:2010by,
    author = "Venkat, Siddharth and Arrington, John and Miller, Gerald A. and Zhan, Xiaohui",
    title = "{Realistic Transverse Images of the Proton Charge and Magnetic Densities}",
    eprint = "1010.3629",
    archivePrefix = "arXiv",
    primaryClass = "nucl-th",
    reportNumber = "NT@UW-10-15",
    doi = "10.1103/PhysRevC.83.015203",
    journal = "Phys. Rev. C",
    volume = "83",
    pages = "015203",
    year = "2011"
}

@article{Miller:2007uy,
    author = "Miller, Gerald A.",
    title = "{Charge Density of the Neutron}",
    eprint = "0705.2409",
    archivePrefix = "arXiv",
    primaryClass = "nucl-th",
    reportNumber = "NT@UW-07-07",
    doi = "10.1103/PhysRevLett.99.112001",
    journal = "Phys. Rev. Lett.",
    volume = "99",
    pages = "112001",
    year = "2007"
}

@article{Liuti:2004hd,
    author = "Liuti, S. and Taneja, S. K.",
    title = "{Generalized parton distributions and color transparency phenomena}",
    eprint = "hep-ph/0405014",
    archivePrefix = "arXiv",
    doi = "10.1103/PhysRevD.70.074019",
    journal = "Phys. Rev. D",
    volume = "70",
    pages = "074019",
    year = "2004"
}

@article{Brodsky:2022bum,
    author = "Brodsky, Stanley J. and de Teramond, Guy F.",
    title = "{Onset of Color Transparency in Holographic Light-Front QCD}",
    eprint = "2202.13283",
    archivePrefix = "arXiv",
    primaryClass = "hep-ph",
    reportNumber = "SLAC-PUB-17656",
    doi = "10.3390/physics4020042",
    journal = "MDPI Physics",
    volume = "4",
    number = "2",
    pages = "633--646",
    year = "2022"
}

@article{Soper:1976jc,
    author = "Soper, Davison E.",
    title = "{The Parton Model and the Bethe-Salpeter Wave Function}",
    reportNumber = "Print-76-0959 (PRINCETON)",
    doi = "10.1103/PhysRevD.15.1141",
    journal = "Phys. Rev. D",
    volume = "15",
    pages = "1141",
    year = "1977"
}

@article{Diehl:2002he,
    author = "Diehl, M.",
    title = "{Generalized parton distributions in impact parameter space}",
    eprint = "hep-ph/0205208",
    archivePrefix = "arXiv",
    doi = "10.1007/s10052-002-1016-9",
    journal = "Eur. Phys. J. C",
    volume = "25",
    pages = "223--232",
    year = "2002",
    note = "[Erratum: Eur.Phys.J.C 31, 277--278 (2003)]"
}

@article{Golec-Biernat:1998zbo,
    author = "Golec-Biernat, Krzysztof J. and Martin, Alan D.",
    title = "{Off diagonal parton distributions and their evolution}",
    eprint = "hep-ph/9807497",
    archivePrefix = "arXiv",
    reportNumber = "DTP-98-48",
    doi = "10.1103/PhysRevD.59.014029",
    journal = "Phys. Rev. D",
    volume = "59",
    pages = "014029",
    year = "1999"
}

@article{Jaffe:1989jz,
    author = "Jaffe, R. L. and Manohar, Aneesh",
    title = "{The $g_1$ Problem: Fact and Fantasy on the Spin of the Proton}",
    reportNumber = "MIT-CTP-1706-REV, MIT-CTP-1706",
    doi = "10.1016/0550-3213(90)90506-9",
    journal = "Nucl. Phys. B",
    volume = "337",
    pages = "509--546",
    year = "1990"
}

@article{Rajan:2017cpx,
    author = "Rajan, Abha and Engelhardt, Michael and Liuti, Simonetta",
    title = "{Lorentz Invariance and QCD Equation of Motion Relations for Generalized Parton Distributions and the Dynamical Origin of Proton Orbital Angular Momentum}",
    eprint = "1709.05770",
    archivePrefix = "arXiv",
    primaryClass = "hep-ph",
    doi = "10.1103/PhysRevD.98.074022",
    journal = "Phys. Rev. D",
    volume = "98",
    number = "7",
    pages = "074022",
    year = "2018"
}

@article{Kiptily:2002nx,
    author = "Kiptily, D. V. and Polyakov, M. V.",
    title = "{Genuine twist three contributions to the generalized parton distributions from instantons}",
    eprint = "hep-ph/0212372",
    archivePrefix = "arXiv",
    reportNumber = "RUB-TP2-15-02",
    doi = "10.1140/epjc/s2004-01957-3",
    journal = "Eur. Phys. J. C",
    volume = "37",
    pages = "105--114",
    year = "2004"
}

@article{Ji:2020hii,
    author = "Ji, Xiangdong and Yuan, Feng",
    title = "{Transverse spin sum rule of the proton}",
    eprint = "2008.04349",
    archivePrefix = "arXiv",
    primaryClass = "hep-ph",
    doi = "10.1016/j.physletb.2020.135786",
    journal = "Phys. Lett. B",
    volume = "810",
    pages = "135786",
    year = "2020"
}

@article{Bakker:2004ib,
    author = "Bakker, B. L. G. and Leader, E. and Trueman, T. L.",
    title = "{A Critique of the angular momentum sum rules and a new angular momentum sum rule}",
    eprint = "hep-ph/0406139",
    archivePrefix = "arXiv",
    reportNumber = "BNL-HET-04-6, IC-HEP-04-2",
    doi = "10.1103/PhysRevD.70.114001",
    journal = "Phys. Rev. D",
    volume = "70",
    pages = "114001",
    year = "2004"
}

@article{Leader:2011cr,
    author = "Leader, Elliot",
    title = "{New relation between transverse angular momentum and generalized parton distributions}",
    eprint = "1109.1230",
    archivePrefix = "arXiv",
    primaryClass = "hep-ph",
    doi = "10.1103/PhysRevD.85.051501",
    journal = "Phys. Rev. D",
    volume = "85",
    pages = "051501",
    year = "2012"
}

@article{Ji:2012vj,
    author = "Ji, Xiangdong and Xiong, Xiaonu and Yuan, Feng",
    title = "{Transverse Polarization of the Nucleon in Parton Picture}",
    eprint = "1209.3246",
    archivePrefix = "arXiv",
    primaryClass = "hep-ph",
    doi = "10.1016/j.physletb.2012.09.027",
    journal = "Phys. Lett. B",
    volume = "717",
    pages = "214--218",
    year = "2012"
}

@article{Hatta:2012jm,
    author = "Hatta, Yoshitaka and Tanaka, Kazuhiro and Yoshida, Shinsuke",
    title = "{Twist-three relations of gluonic correlators for the transversely polarized nucleon}",
    eprint = "1211.2918",
    archivePrefix = "arXiv",
    primaryClass = "hep-ph",
    doi = "10.1007/JHEP02(2013)003",
    journal = "JHEP",
    volume = "02",
    pages = "003",
    year = "2013"
}

@article{Leader:2012ar,
    author = "Leader, Elliot",
    title = "{A critical assessment of the angular momentum sum rules}",
    eprint = "1211.3957",
    archivePrefix = "arXiv",
    primaryClass = "hep-ph",
    doi = "10.1016/j.physletb.2013.01.050",
    journal = "Phys. Lett. B",
    volume = "720",
    pages = "120--124",
    year = "2013",
    note = "[Erratum: Phys.Lett.B 726, 927--927 (2013)]"
}

@article{Lorce:2018zpf,
    author = "Lorc{\'e}, C{\'e}dric",
    title = "{The relativistic center of mass in field theory with spin}",
    eprint = "1805.05284",
    archivePrefix = "arXiv",
    primaryClass = "hep-ph",
    doi = "10.1140/epjc/s10052-018-6249-3",
    journal = "Eur. Phys. J. C",
    volume = "78",
    number = "9",
    pages = "785",
    year = "2018"
}

@article{Taneja:2011sy,
    author = "Taneja, Swadhin K. and Kathuria, Kunal and Liuti, Simonetta and Goldstein, Gary R.",
    title = "{Angular momentum sum rule for spin one hadronic systems}",
    eprint = "1101.0581",
    archivePrefix = "arXiv",
    primaryClass = "hep-ph",
    doi = "10.1103/PhysRevD.86.036008",
    journal = "Phys. Rev. D",
    volume = "86",
    pages = "036008",
    year = "2012"
}

@article{Moiseeva:2008qd,
    author = "Moiseeva, Alena M. and Polyakov, Maxim V.",
    title = "{Dual parameterization and Abel transform tomography for twist-3 DVCS}",
    eprint = "0803.1777",
    archivePrefix = "arXiv",
    primaryClass = "hep-ph",
    doi = "10.1016/j.nuclphysb.2010.02.008",
    journal = "Nucl. Phys. B",
    volume = "832",
    pages = "241--250",
    year = "2010"
}

@article{Rajan:2018zzy,
    author = "Rajan, Abha and Gorda, Tyler and Liuti, Simonetta and Yagi, Kent",
    title = "{Bounds on the Equation of State of Neutron Stars from High Energy Deeply Virtual Exclusive Experiments}",
    eprint = "1812.01479",
    archivePrefix = "arXiv",
    primaryClass = "hep-ph",
    month = "12",
    year = "2018"
}

@article{Lorce:2020onh,
    author = "Lorc{\'e}, C{\'e}dric",
    title = "{Charge Distributions of Moving Nucleons}",
    eprint = "2007.05318",
    archivePrefix = "arXiv",
    primaryClass = "hep-ph",
    doi = "10.1103/PhysRevLett.125.232002",
    journal = "Phys. Rev. Lett.",
    volume = "125",
    number = "23",
    pages = "232002",
    year = "2020"
}

@article{Epelbaum:2022fjc,
    author = "Epelbaum, E. and Gegelia, J. and Lange, N. and Mei{\ss}ner, U. -G. and Polyakov, M. V.",
    title = "{Definition of Local Spatial Densities in Hadrons}",
    eprint = "2201.02565",
    archivePrefix = "arXiv",
    primaryClass = "hep-ph",
    doi = "10.1103/PhysRevLett.129.012001",
    journal = "Phys. Rev. Lett.",
    volume = "129",
    number = "1",
    pages = "012001",
    year = "2022"
}

@article{Jaffe:2020ebz,
    author = "Jaffe, Robert L.",
    title = "{Ambiguities in the definition of local spatial densities in light hadrons}",
    eprint = "2010.15887",
    archivePrefix = "arXiv",
    primaryClass = "hep-ph",
    reportNumber = "MIT-CTP/5253",
    doi = "10.1103/PhysRevD.103.016017",
    journal = "Phys. Rev. D",
    volume = "103",
    number = "1",
    pages = "016017",
    year = "2021"
}

@article{Liu:2026pbf,
    author = "Liu, Keh-Fei",
    title = "{Pressure-energy equations of state of the nucleon}",
    eprint = "2605.04163",
    archivePrefix = "arXiv",
    primaryClass = "hep-ph",
    doi = "10.1103/1hsy-xvqx",
    journal = "Phys. Rev. D",
    volume = "114",
    number = "3",
    pages = "034042",
    year = "2026"
}

@article{Mantysaari:2016jaz,
    author = {M{\"a}ntysaari, Heikki and Schenke, Bj{\"o}rn},
    title = "{Revealing proton shape fluctuations with incoherent diffraction at high energy}",
    eprint = "1607.01711",
    archivePrefix = "arXiv",
    primaryClass = "hep-ph",
    doi = "10.1103/PhysRevD.94.034042",
    journal = "Phys. Rev. D",
    volume = "94",
    number = "3",
    pages = "034042",
    year = "2016"
}

@article{Mantysaari:2016ykx,
    author = {M{\"a}ntysaari, Heikki and Schenke, Bj{\"o}rn},
    title = "{Evidence of strong proton shape fluctuations from incoherent diffraction}",
    eprint = "1603.04349",
    archivePrefix = "arXiv",
    primaryClass = "hep-ph",
    doi = "10.1103/PhysRevLett.117.052301",
    journal = "Phys. Rev. Lett.",
    volume = "117",
    number = "5",
    pages = "052301",
    year = "2016"
}

@article{Rossi:1977cy,
    author = "Rossi, G. C. and Veneziano, G.",
    title = "{A Possible Description of Baryon Dynamics in Dual and Gauge Theories}",
    reportNumber = "CERN-TH-2287",
    doi = "10.1016/0550-3213(77)90178-X",
    journal = "Nucl. Phys. B",
    volume = "123",
    pages = "507--545",
    year = "1977"
}

@article{Kharzeev:1996sq,
    author = "Kharzeev, D.",
    title = "{Can gluons trace baryon number?}",
    eprint = "nucl-th/9602027",
    archivePrefix = "arXiv",
    reportNumber = "CERN-TH-95-343, BI-TP-95-42",
    doi = "10.1016/0370-2693(96)00435-2",
    journal = "Phys. Lett. B",
    volume = "378",
    pages = "238--246",
    year = "1996"
}

@article{STAR:2024lvy,
    author = "Aboona, B. E. and others",
    collaboration = "STAR",
    title = "{Tracking the baryon number with nuclear collisions}",
    eprint = "2408.15441",
    archivePrefix = "arXiv",
    primaryClass = "nucl-ex",
    doi = "10.1126/science.ads5962",
    journal = "Science",
    volume = "393",
    number = "6812",
    pages = "ads5962",
    year = "2026"
}

@article{Panjsheeri:2025zrm,
    author = "Panjsheeri, Zaki and Pandey, Saraswati and Semp, Brannon and Liuti, Simonetta",
    title = "{Connected and disconnected contributions to nucleon form factors and parton distributions}",
    eprint = "2512.20853",
    archivePrefix = "arXiv",
    primaryClass = "hep-ph",
    month = "12",
    year = "2025"
}

@article{Liu:2012ch,
    author = "Liu, Keh-Fei and Chang, Wen-Chen and Cheng, Hai-Yang and Peng, Jen-Chieh",
    title = "{Connected-Sea Partons}",
    eprint = "1206.4339",
    archivePrefix = "arXiv",
    primaryClass = "hep-ph",
    reportNumber = "UK-12-06",
    doi = "10.1103/PhysRevLett.109.252002",
    journal = "Phys. Rev. Lett.",
    volume = "109",
    pages = "252002",
    year = "2012"
}

@article{Hou:2022ajg,
    author = "Hou, Tie-Jiun and Yan, Mengshi and Liang, Jian and Liu, Keh-Fei and Yuan, C. -P.",
    title = "{Connected and disconnected sea partons from the CT18 parametrization of PDFs}",
    eprint = "2206.02431",
    archivePrefix = "arXiv",
    primaryClass = "hep-ph",
    doi = "10.1103/PhysRevD.106.096008",
    journal = "Phys. Rev. D",
    volume = "106",
    number = "9",
    pages = "096008",
    year = "2022"
}

@article{Panjsheeri:2024ysh,
    author = "Panjsheeri, Zaki and Bautista, Joshua and Liuti, Simonetta",
    title = "{The Correlated Spatial Structure of the Proton: Two-body densities as a framework for dynamical imaging}",
    eprint = "2405.05842",
    archivePrefix = "arXiv",
    primaryClass = "hep-ph",
    doi = "10.22323/1.456.0021",
    journal = "PoS",
    volume = "SPIN2023",
    pages = "021",
    year = "2024"
}

@article{Panjsheeri:2024gmw,
    author = "Panjsheeri, Zaki and Bautista, Joshua and Liuti, Simonetta",
    title = "{Two-body densities as a framework for dynamical imaging and their connection to ultra-peripheral collisions}",
    doi = "10.17161/r71z8v22",
    journal = "Phys. Proc. UPC",
    volume = "1",
    pages = "12",
    year = "2024"
}

@article{Miller:2025zte,
    author = "Miller, Gerald A.",
    title = "{Impossibility of obtaining time-independent, three-dimensional, spherically symmetric densities of confined systems of relativistically moving constituents}",
    eprint = "2507.14388",
    archivePrefix = "arXiv",
    primaryClass = "hep-ph",
    reportNumber = "NT@UW-25-8",
    doi = "10.1103/jkn7-4fzj",
    journal = "Phys. Rev. C",
    volume = "112",
    number = "4",
    pages = "045204",
    year = "2025"
}

@article{Hatta:2022bxn,
    author = "Hatta, Yoshitaka and Zhou, Jian",
    title = "{Small-$x$ evolution of the gluon GPD $E_g$}",
    eprint = "2207.03378",
    archivePrefix = "arXiv",
    primaryClass = "hep-ph",
    doi = "10.1103/PhysRevLett.129.252002",
    journal = "Phys. Rev. Lett.",
    volume = "129",
    number = "25",
    pages = "252002",
    year = "2022"
}

@article{Landshoff:1970ff,
    author = "Landshoff, P. V. and Polkinghorne, J. C. and Short, R. D.",
    title = "{a Nonperturbative parton model of current interactions}",
    doi = "10.1016/0550-3213(71)90375-0",
    journal = "Nucl. Phys. B",
    volume = "28",
    pages = "225--239",
    year = "1971"
}

@article{Zhang:2026dzi,
    author = "Zhang, Ziqi and Mondal, Chandan and Xu, Siqi and Zhao, Xingbo and Vary, James P.",
    collaboration = "BLFQ",
    title = "{Dynamical gluon effects in twist-3 generalized parton distributions of the proton}",
    eprint = "2601.07590",
    archivePrefix = "arXiv",
    primaryClass = "hep-ph",
    doi = "10.1103/1yjm-sqck",
    journal = "Phys. Rev. D",
    volume = "113",
    number = "9",
    pages = "094031",
    year = "2026"
}

@article{Zhang:2025nll,
    author = "Zhang, Pengxiang and Liu, Yiping and Xu, Siqi and Mondal, Chandan and Zhao, Xingbo and Vary, James P.",
    collaboration = "BLFQ",
    title = "{Gluon skewed generalized parton distributions of proton from a light-front Hamiltonian approach}",
    eprint = "2501.10119",
    archivePrefix = "arXiv",
    primaryClass = "hep-ph",
    doi = "10.1016/j.physletb.2025.139584",
    journal = "Phys. Lett. B",
    volume = "866",
    pages = "139584",
    year = "2025"
}

@article{Chakrabarti:2024hwx,
    author = "Chakrabarti, Dipankar and Choudhary, Poonam and Gurjar, Bheemsehan and Maji, Tanmay and Mondal, Chandan and Mukherjee, Asmita",
    title = "{Gluon generalized parton distributions of the proton at nonzero skewness}",
    eprint = "2402.16503",
    archivePrefix = "arXiv",
    primaryClass = "hep-ph",
    doi = "10.1103/PhysRevD.109.114040",
    journal = "Phys. Rev. D",
    volume = "109",
    number = "11",
    pages = "114040",
    year = "2024"
}

@article{Chakrabarti:2015ama,
    author = "Chakrabarti, Dipankar and Mondal, Chandan",
    title = "{Chiral-odd generalized parton distributions for proton in a light-front quark-diquark model}",
    eprint = "1509.00598",
    archivePrefix = "arXiv",
    primaryClass = "hep-ph",
    doi = "10.1103/PhysRevD.92.074012",
    journal = "Phys. Rev. D",
    volume = "92",
    number = "7",
    pages = "074012",
    year = "2015"
}

@article{Chakrabarti:2005zm,
    author = "Chakrabarti, D. and Mukherjee, A.",
    title = "{Generalized parton distributions in the impact parameter space with non-zero skewedness}",
    eprint = "hep-ph/0506006",
    archivePrefix = "arXiv",
    doi = "10.1103/PhysRevD.72.034013",
    journal = "Phys. Rev. D",
    volume = "72",
    pages = "034013",
    year = "2005"
}

@article{Diehl:2004cx,
    author = "Diehl, M. and Feldmann, Th. and Jakob, R. and Kroll, P.",
    title = "{Generalized parton distributions from nucleon form-factor data}",
    eprint = "hep-ph/0408173",
    archivePrefix = "arXiv",
    reportNumber = "DESY-04-146, CERN-PH-04-154, WUB-04-08, CERN-PH-TH-04-154",
    doi = "10.1140/epjc/s2004-02063-4",
    journal = "Eur. Phys. J. C",
    volume = "39",
    pages = "1--39",
    year = "2005"
}

@article{Diehl:2000xz,
    author = "Diehl, M. and Feldmann, T. and Jakob, R. and Kroll, P.",
    title = "{The overlap representation of skewed quark and gluon distributions}",
    eprint = "hep-ph/0009255",
    archivePrefix = "arXiv",
    reportNumber = "SLAC-PUB-8613, WUB-00-10, PITHA-00-21",
    doi = "10.1016/S0550-3213(00)00684-2",
    journal = "Nucl. Phys. B",
    volume = "596",
    pages = "33--65",
    year = "2001",
    note = "[Erratum: Nucl.Phys.B 605, 647--647 (2001)]"
}

@article{Diehl:1998kh,
    author = "Diehl, M. and Feldmann, T. and Jakob, R. and Kroll, P.",
    title = "{Linking parton distributions to form-factors and Compton scattering}",
    eprint = "hep-ph/9811253",
    archivePrefix = "arXiv",
    reportNumber = "DESY-98-172, WUB-98-37, FNT-T-98-10",
    doi = "10.1007/s100529901100",
    journal = "Eur. Phys. J. C",
    volume = "8",
    pages = "409--434",
    year = "1999"
}

@article{Rajan:2016tlg,
    author = "Rajan, Abha and Courtoy, Aurore and Engelhardt, Michael and Liuti, Simonetta",
    title = "{Parton Transverse Momentum and Orbital Angular Momentum Distributions}",
    eprint = "1601.06117",
    archivePrefix = "arXiv",
    primaryClass = "hep-ph",
    doi = "10.1103/PhysRevD.94.034041",
    journal = "Phys. Rev. D",
    volume = "94",
    number = "3",
    pages = "034041",
    year = "2016"
}

@article{Bhattacharya:2023yvo,
    author = "Bhattacharya, Shohini and Zheng, Duxin and Zhou, Jian",
    title = "{Accessing the gluon GTMD F1,4 in exclusive {\ensuremath{\pi}}0 production in ep collisions}",
    eprint = "2304.05784",
    archivePrefix = "arXiv",
    primaryClass = "hep-ph",
    doi = "10.1103/PhysRevD.109.096029",
    journal = "Phys. Rev. D",
    volume = "109",
    number = "9",
    pages = "096029",
    year = "2024"
}

@article{Liu:2024umn,
    author = "Liu, Yiping and Xu, Siqi and Mondal, Chandan and Hu, Zhi and Zhao, Xingbo and Vary, James P.",
    collaboration = "BLFQ",
    title = "{Skewed generalized parton distributions of proton from basis light-front quantization}",
    eprint = "2403.05922",
    archivePrefix = "arXiv",
    primaryClass = "hep-ph",
    doi = "10.1016/j.physletb.2024.138809",
    journal = "Phys. Lett. B",
    volume = "855",
    pages = "138809",
    year = "2024"
}

@article{Echevarria:2022ztg,
    author = "Echevarria, Miguel G. and Gutierrez Garcia, Patricia A. and Scimemi, Ignazio",
    title = "{GTMDs and the factorization of exclusive double Drell-Yan}",
    eprint = "2208.00021",
    archivePrefix = "arXiv",
    primaryClass = "hep-ph",
    reportNumber = "IPARCOS-UCM-23-023",
    doi = "10.1016/j.physletb.2023.137881",
    journal = "Phys. Lett. B",
    volume = "840",
    pages = "137881",
    year = "2023"
}

@article{Burkardt:2012sd,
    author = "Burkardt, Matthias",
    title = "{Parton Orbital Angular Momentum and Final State Interactions}",
    eprint = "1205.2916",
    archivePrefix = "arXiv",
    primaryClass = "hep-ph",
    doi = "10.1103/PhysRevD.88.014014",
    journal = "Phys. Rev. D",
    volume = "88",
    number = "1",
    pages = "014014",
    year = "2013"
}

@article{Kriesten:2020wcx,
    author = "Kriesten, Brandon and Liuti, Simonetta",
    title = "{Theory of deeply virtual Compton scattering off the unpolarized proton}",
    eprint = "2004.08890",
    archivePrefix = "arXiv",
    primaryClass = "hep-ph",
    doi = "10.1103/PhysRevD.105.016015",
    journal = "Phys. Rev. D",
    volume = "105",
    number = "1",
    pages = "016015",
    year = "2022"
}

@article{Brodsky:1973hm,
    author = "Brodsky, Stanley J. and Close, Francis E. and Gunion, J. F.",
    title = "{A GAUGE - INVARIANT SCALING MODEL OF CURRENT INTERACTIONS WITH REGGE BEHAVIOR AND FINITE FIXED POLE SUM RULES}",
    reportNumber = "SLAC-PUB-1243",
    doi = "10.1103/PhysRevD.8.3678",
    journal = "Phys. Rev. D",
    volume = "8",
    pages = "3678",
    year = "1973"
}

@article{Radyushkin:1998rt,
    author = "Radyushkin, A. V.",
    title = "{Nonforward parton densities and soft mechanism for form-factors and wide angle Compton scattering in QCD}",
    eprint = "hep-ph/9803316",
    archivePrefix = "arXiv",
    reportNumber = "JLAB-THY-98-10",
    doi = "10.1103/PhysRevD.58.114008",
    journal = "Phys. Rev. D",
    volume = "58",
    pages = "114008",
    year = "1998"
}

@article{Kovchegov:2025yyl,
    author = "Kovchegov, Yuri V. and Santiago, M. Gabriel and Sun, Huachen",
    title = "{Unpolarized GPDs at small x and non-zero skewness}",
    eprint = "2512.10086",
    archivePrefix = "arXiv",
    primaryClass = "hep-ph",
    doi = "10.1016/j.physletb.2026.140470",
    journal = "Phys. Lett. B",
    volume = "877",
    pages = "140470",
    year = "2026"
}

@article{Hatta:2016aoc,
    author = "Hatta, Yoshitaka and Nakagawa, Yuya and Yuan, Feng and Zhao, Yong and Xiao, Bowen",
    title = "{Gluon orbital angular momentum at small-$x$}",
    eprint = "1612.02445",
    archivePrefix = "arXiv",
    primaryClass = "hep-ph",
    reportNumber = "YITP-16-133",
    doi = "10.1103/PhysRevD.95.114032",
    journal = "Phys. Rev. D",
    volume = "95",
    number = "11",
    pages = "114032",
    year = "2017"
}

@article{HERMES:2001bob,
    author = "Airapetian, A. and others",
    collaboration = "HERMES",
    title = "{Measurement of the beam spin azimuthal asymmetry associated with deeply virtual Compton scattering}",
    eprint = "hep-ex/0106068",
    archivePrefix = "arXiv",
    reportNumber = "DESY-01-091",
    doi = "10.1103/PhysRevLett.87.182001",
    journal = "Phys. Rev. Lett.",
    volume = "87",
    pages = "182001",
    year = "2001"
}

@article{H1:2001nez,
    author = "Adloff, C. and others",
    collaboration = "H1",
    title = "{Measurement of deeply virtual Compton scattering at HERA}",
    eprint = "hep-ex/0107005",
    archivePrefix = "arXiv",
    reportNumber = "DESY-01-093",
    doi = "10.1016/S0370-2693(01)00939-X",
    journal = "Phys. Lett. B",
    volume = "517",
    pages = "47--58",
    year = "2001"
}

@article{CLAS:2001wjj,
    author = "Stepanyan, S. and others",
    collaboration = "CLAS",
    title = "{Observation of exclusive deeply virtual Compton scattering in polarized electron beam asymmetry measurements}",
    eprint = "hep-ex/0107043",
    archivePrefix = "arXiv",
    reportNumber = "JLAB-PHY-01-84",
    doi = "10.1103/PhysRevLett.87.182002",
    journal = "Phys. Rev. Lett.",
    volume = "87",
    pages = "182002",
    year = "2001"
}

@article{ZEUS:2003pwh,
    author = "Chekanov, S. and others",
    collaboration = "ZEUS",
    title = "{Measurement of deeply virtual Compton scattering at HERA}",
    eprint = "hep-ex/0305028",
    archivePrefix = "arXiv",
    reportNumber = "DESY-03-059",
    doi = "10.1016/j.physletb.2003.08.048",
    journal = "Phys. Lett. B",
    volume = "573",
    pages = "46--62",
    year = "2003"
}

@article{H1:2005gdw,
    author = "Aktas, A. and others",
    collaboration = "H1",
    title = "{Measurement of deeply virtual compton scattering at HERA}",
    eprint = "hep-ex/0505061",
    archivePrefix = "arXiv",
    reportNumber = "DESY-05-065",
    doi = "10.1140/epjc/s2005-02345-3",
    journal = "Eur. Phys. J. C",
    volume = "44",
    pages = "1--11",
    year = "2005"
}

@article{HERMES:2006pre,
    author = "Airapetian, A. and others",
    collaboration = "HERMES",
    title = "{The Beam-charge azimuthal asymmetry and deeply virtual compton scattering}",
    eprint = "hep-ex/0605108",
    archivePrefix = "arXiv",
    reportNumber = "DESY-06-078",
    doi = "10.1103/PhysRevD.75.011103",
    journal = "Phys. Rev. D",
    volume = "75",
    pages = "011103",
    year = "2007"
}

@article{JeffersonLabHallA:2007jdm,
    author = "Mazouz, M. and others",
    collaboration = "Jefferson Lab Hall A",
    title = "{Deeply virtual compton scattering off the neutron}",
    eprint = "0709.0450",
    archivePrefix = "arXiv",
    primaryClass = "nucl-ex",
    reportNumber = "JLAB-PHY-07-707",
    doi = "10.1103/PhysRevLett.99.242501",
    journal = "Phys. Rev. Lett.",
    volume = "99",
    pages = "242501",
    year = "2007"
}

@article{H1:2007vrx,
    author = "Aaron, F. D. and others",
    collaboration = "H1",
    title = "{Measurement of deeply virtual Compton scattering and its t-dependence at HERA}",
    eprint = "0709.4114",
    archivePrefix = "arXiv",
    primaryClass = "hep-ex",
    reportNumber = "DESY-07-142",
    doi = "10.1016/j.physletb.2007.11.093",
    journal = "Phys. Lett. B",
    volume = "659",
    pages = "796--806",
    year = "2008"
}

@article{CLAS:2007clm,
    author = "Girod, F. X. and others",
    collaboration = "CLAS",
    title = "{Measurement of Deeply virtual Compton scattering beam-spin asymmetries}",
    eprint = "0711.4805",
    archivePrefix = "arXiv",
    primaryClass = "hep-ex",
    reportNumber = "IRFU-08-24, JLAB-PHY-07-767",
    doi = "10.1103/PhysRevLett.100.162002",
    journal = "Phys. Rev. Lett.",
    volume = "100",
    pages = "162002",
    year = "2008"
}

@article{ZEUS:2008hcd,
    author = "Chekanov, S. and others",
    collaboration = "ZEUS",
    title = "{A Measurement of the Q**2, W and t dependences of deeply virtual Compton scattering at HERA}",
    eprint = "0812.2517",
    archivePrefix = "arXiv",
    primaryClass = "hep-ex",
    reportNumber = "DESY-08-132",
    doi = "10.1088/1126-6708/2009/05/108",
    journal = "JHEP",
    volume = "05",
    pages = "108",
    year = "2009"
}

@article{CLAS:2008ahu,
    author = "Gavalian, G. and others",
    collaboration = "CLAS",
    title = "{Beam spin asymmetries in deeply virtual Compton scattering (DVCS) with CLAS at 4.8 GeV}",
    eprint = "0812.2950",
    archivePrefix = "arXiv",
    primaryClass = "hep-ex",
    reportNumber = "JLAB-PHY-08-930",
    doi = "10.1103/PhysRevC.80.035206",
    journal = "Phys. Rev. C",
    volume = "80",
    pages = "035206",
    year = "2009"
}

@article{H1:2009wnw,
    author = "Aaron, F. D. and others",
    collaboration = "H1",
    title = "{Deeply Virtual Compton Scattering and its Beam Charge Asymmetry in e+- Collisions at HERA}",
    eprint = "0907.5289",
    archivePrefix = "arXiv",
    primaryClass = "hep-ex",
    reportNumber = "DESY-09-109, DESY09-109",
    doi = "10.1016/j.physletb.2009.10.035",
    journal = "Phys. Lett. B",
    volume = "681",
    pages = "391--399",
    year = "2009"
}

@article{HERMES:2009cqe,
    author = "Airapetian, A. and others",
    collaboration = "HERMES",
    title = "{Separation of contributions from deeply virtual Compton scattering and its interference with the Bethe-Heitler process in measurements on a hydrogen target}",
    eprint = "0909.3587",
    archivePrefix = "arXiv",
    primaryClass = "hep-ex",
    reportNumber = "DESY-09-143",
    doi = "10.1088/1126-6708/2009/11/083",
    journal = "JHEP",
    volume = "11",
    pages = "083",
    year = "2009"
}

@article{HERMES:2010hnl,
    author = "Airapetian, A. and others",
    collaboration = "HERMES",
    title = "{Measurement of azimuthal asymmetries associated with deeply virtual Compton scattering on a longitudinally polarized deuterium target}",
    eprint = "1008.3996",
    archivePrefix = "arXiv",
    primaryClass = "hep-ex",
    reportNumber = "DESY-10-136",
    doi = "10.1016/j.nuclphysb.2010.09.010",
    journal = "Nucl. Phys. B",
    volume = "842",
    pages = "265--298",
    year = "2011"
}

@article{JeffersonLabHallA:2015dwe,
    author = "Defurne, M. and others",
    collaboration = "Jefferson Lab Hall A",
    title = "{E00-110 experiment at Jefferson Lab Hall A: Deeply virtual Compton scattering off the proton at 6 GeV}",
    eprint = "1504.05453",
    archivePrefix = "arXiv",
    primaryClass = "nucl-ex",
    reportNumber = "IRFU-15-12, JLAB-PHY-15-2038",
    doi = "10.1103/PhysRevC.92.055202",
    journal = "Phys. Rev. C",
    volume = "92",
    number = "5",
    pages = "055202",
    year = "2015"
}

@article{CLAS:2015uuo,
    author = "Jo, H. S. and others",
    collaboration = "CLAS",
    title = "{Cross sections for the exclusive photon electroproduction on the proton and Generalized Parton Distributions}",
    eprint = "1504.02009",
    archivePrefix = "arXiv",
    primaryClass = "hep-ex",
    reportNumber = "JLAB-PHY-15-2037",
    doi = "10.1103/PhysRevLett.115.212003",
    journal = "Phys. Rev. Lett.",
    volume = "115",
    number = "21",
    pages = "212003",
    year = "2015"
}

@article{CLAS:2015bqi,
    author = "Pisano, S. and others",
    collaboration = "CLAS",
    title = "{Single and double spin asymmetries for deeply virtual Compton scattering measured with CLAS and a longitudinally polarized proton target}",
    eprint = "1501.07052",
    archivePrefix = "arXiv",
    primaryClass = "hep-ex",
    reportNumber = "JLAB-PHY-15-2005",
    doi = "10.1103/PhysRevD.91.052014",
    journal = "Phys. Rev. D",
    volume = "91",
    number = "5",
    pages = "052014",
    year = "2015"
}

@article{HERMES:2012gbh,
    author = "Airapetian, A. and others",
    collaboration = "HERMES",
    title = "{Beam-helicity and beam-charge asymmetries associated with deeply virtual Compton scattering on the unpolarised proton}",
    eprint = "1203.6287",
    archivePrefix = "arXiv",
    primaryClass = "hep-ex",
    reportNumber = "DESY-12-040",
    doi = "10.1007/JHEP07(2012)032",
    journal = "JHEP",
    volume = "07",
    pages = "032",
    year = "2012"
}

@article{HERMES:2011bou,
    author = "Airapetian, A. and others",
    collaboration = "HERMES",
    title = "{Measurement of double-spin asymmetries associated with deeply virtual Compton scattering on a transversely polarized hydrogen target}",
    eprint = "1106.2990",
    archivePrefix = "arXiv",
    primaryClass = "hep-ex",
    reportNumber = "DESY-11-100",
    doi = "10.1016/j.physletb.2011.08.067",
    journal = "Phys. Lett. B",
    volume = "704",
    pages = "15--23",
    year = "2011"
}

@article{HERMES:2010dsx,
    author = "Airapetian, A. and others",
    collaboration = "HERMES",
    title = "{Exclusive Leptoproduction of Real Photons on a Longitudinally Polarised Hydrogen Target}",
    eprint = "1004.0177",
    archivePrefix = "arXiv",
    primaryClass = "hep-ex",
    reportNumber = "DESY-10-046",
    doi = "10.1007/JHEP06(2010)019",
    journal = "JHEP",
    volume = "06",
    pages = "019",
    year = "2010"
}

@article{Belitsky:2008bz,
    author = "Belitsky, Andrei V. and Mueller, Dieter",
    title = "{Refined analysis of photon leptoproduction off spinless target}",
    eprint = "0809.2890",
    archivePrefix = "arXiv",
    primaryClass = "hep-ph",
    doi = "10.1103/PhysRevD.79.014017",
    journal = "Phys. Rev. D",
    volume = "79",
    pages = "014017",
    year = "2009"
}

@article{Belitsky:2012ch,
    author = {Belitsky, Andrei V. and M{\"u}ller, Dieter and Ji, Yao},
    title = "{Compton scattering: from deeply virtual to quasi-real}",
    eprint = "1212.6674",
    archivePrefix = "arXiv",
    primaryClass = "hep-ph",
    doi = "10.1016/j.nuclphysb.2013.11.014",
    journal = "Nucl. Phys. B",
    volume = "878",
    pages = "214--268",
    year = "2014"
}

@article{Anikin:2000em,
    author = "Anikin, I. V. and Pire, B. and Teryaev, O. V.",
    title = "{On the gauge invariance of the DVCS amplitude}",
    eprint = "hep-ph/0003203",
    archivePrefix = "arXiv",
    doi = "10.1103/PhysRevD.62.071501",
    journal = "Phys. Rev. D",
    volume = "62",
    pages = "071501",
    year = "2000"
}

@article{Radyushkin:2000ap,
    author = "Radyushkin, A. V. and Weiss, C.",
    title = "{DVCS amplitude at tree level: Transversality, twist - three, and factorization}",
    eprint = "hep-ph/0010296",
    archivePrefix = "arXiv",
    reportNumber = "JLAB-THY-00-37, RUB-TPII-17-00",
    doi = "10.1103/PhysRevD.63.114012",
    journal = "Phys. Rev. D",
    volume = "63",
    pages = "114012",
    year = "2001"
}

@article{Kivel:2000fg,
    author = "Kivel, N. and Polyakov, Maxim V. and Vanderhaeghen, M.",
    title = "{DVCS on the nucleon: Study of the twist - three effects}",
    eprint = "hep-ph/0012136",
    archivePrefix = "arXiv",
    reportNumber = "TPR-00-22, RUB-TP2-23-00, RUB-TPII-23-00, MKPH-T-00-25",
    doi = "10.1103/PhysRevD.63.114014",
    journal = "Phys. Rev. D",
    volume = "63",
    pages = "114014",
    year = "2001"
}

@article{Belitsky:2000vx,
    author = "Belitsky, Andrei V. and Mueller, Dieter",
    title = "{Twist- three effects in two photon processes}",
    eprint = "hep-ph/0007031",
    archivePrefix = "arXiv",
    doi = "10.1016/S0550-3213(00)00542-3",
    journal = "Nucl. Phys. B",
    volume = "589",
    pages = "611--630",
    year = "2000"
}

@article{Braun:2012bg,
    author = "Braun, V. M. and Manashov, A. N. and Pirnay, B.",
    title = "{Finite-t and target mass corrections to DVCS on a scalar target}",
    eprint = "1205.3332",
    archivePrefix = "arXiv",
    primaryClass = "hep-ph",
    reportNumber = "IPHT-T12-038",
    doi = "10.1103/PhysRevD.86.014003",
    journal = "Phys. Rev. D",
    volume = "86",
    pages = "014003",
    year = "2012"
}

@article{Braun:2012hq,
    author = "Braun, V. M. and Manashov, A. N. and Pirnay, B.",
    title = "{Finite-t and target mass corrections to deeply virtual Compton scattering}",
    eprint = "1209.2559",
    archivePrefix = "arXiv",
    primaryClass = "hep-ph",
    doi = "10.1103/PhysRevLett.109.242001",
    journal = "Phys. Rev. Lett.",
    volume = "109",
    pages = "242001",
    year = "2012"
}

@article{Ji:1997nk,
    author = "Ji, Xiang-Dong and Osborne, Jonathan",
    title = "{One loop QCD corrections to deeply virtual Compton scattering: The Parton helicity independent case}",
    eprint = "hep-ph/9707254",
    archivePrefix = "arXiv",
    reportNumber = "UMD-PP-97-001, DOE-ER-40762-124",
    doi = "10.1103/PhysRevD.57.R1337",
    journal = "Phys. Rev. D",
    volume = "57",
    pages = "1337--1340",
    year = "1998"
}

@article{Belitsky:1997rh,
    author = "Belitsky, Andrei V. and Mueller, Dieter",
    title = "{Predictions from conformal algebra for the deeply virtual Compton scattering}",
    eprint = "hep-ph/9709379",
    archivePrefix = "arXiv",
    reportNumber = "NTZ-23-97",
    doi = "10.1016/S0370-2693(97)01390-7",
    journal = "Phys. Lett. B",
    volume = "417",
    pages = "129--140",
    year = "1998"
}

@article{Mankiewicz:1997bk,
    author = "Mankiewicz, L. and Piller, G. and Stein, E. and Vanttinen, M. and Weigl, T.",
    title = "{NLO corrections to deeply virtual Compton scattering}",
    eprint = "hep-ph/9712251",
    archivePrefix = "arXiv",
    reportNumber = "TUM-T39-97-31, DFTT-73-97",
    doi = "10.1016/S0370-2693(98)00190-7",
    journal = "Phys. Lett. B",
    volume = "425",
    pages = "186--192",
    year = "1998",
    note = "[Erratum: Phys.Lett.B 461, 423--423 (1999)]"
}

@article{Ji:1998xh,
    author = "Ji, Xiang-Dong and Osborne, Jonathan",
    title = "{One loop corrections and all order factorization in deeply virtual Compton scattering}",
    eprint = "hep-ph/9801260",
    archivePrefix = "arXiv",
    reportNumber = "UMD-PP-98-074, DOE-ER-40762-139",
    doi = "10.1103/PhysRevD.58.094018",
    journal = "Phys. Rev. D",
    volume = "58",
    pages = "094018",
    year = "1998"
}

@article{Belitsky:1999sg,
    author = "Belitsky, Andrei V. and Mueller, Dieter and Niedermeier, L. and Schafer, A.",
    title = "{Deeply virtual Compton scattering in next-to-leading order}",
    eprint = "hep-ph/9908337",
    archivePrefix = "arXiv",
    doi = "10.1016/S0370-2693(99)01283-6",
    journal = "Phys. Lett. B",
    volume = "474",
    pages = "163--169",
    year = "2000"
}

@article{Freund:2001hm,
    author = "Freund, A. and McDermott, M. F.",
    title = "{A Next-to-leading order analysis of deeply virtual Compton scattering}",
    eprint = "hep-ph/0106124",
    archivePrefix = "arXiv",
    reportNumber = "LTH-507",
    doi = "10.1103/PhysRevD.65.091901",
    journal = "Phys. Rev. D",
    volume = "65",
    pages = "091901",
    year = "2002"
}

@article{Freund:2001rk,
    author = "Freund, Andreas and McDermott, Martin F.",
    title = "{A Next-to-leading order QCD analysis of deeply virtual Compton scattering amplitudes}",
    eprint = "hep-ph/0106319",
    archivePrefix = "arXiv",
    doi = "10.1103/PhysRevD.65.074008",
    journal = "Phys. Rev. D",
    volume = "65",
    pages = "074008",
    year = "2002"
}

@article{Freund:2001hd,
    author = "Freund, Andreas and McDermott, Martin",
    title = "{A Detailed next-to-leading order QCD analysis of deeply virtual Compton scattering observables}",
    eprint = "hep-ph/0111472",
    archivePrefix = "arXiv",
    doi = "10.1007/s100520200928",
    journal = "Eur. Phys. J. C",
    volume = "23",
    pages = "651--674",
    year = "2002"
}

@article{Pire:2011st,
    author = "Pire, B. and Szymanowski, L. and Wagner, J.",
    title = "{NLO corrections to timelike, spacelike and double deeply virtual Compton scattering}",
    eprint = "1101.0555",
    archivePrefix = "arXiv",
    primaryClass = "hep-ph",
    doi = "10.1103/PhysRevD.83.034009",
    journal = "Phys. Rev. D",
    volume = "83",
    pages = "034009",
    year = "2011"
}

@article{Moutarde:2013qs,
    author = "Moutarde, H. and Pire, B. and Sabatie, F. and Szymanowski, L. and Wagner, J.",
    title = "{Timelike and spacelike deeply virtual Compton scattering at next-to-leading order}",
    eprint = "1301.3819",
    archivePrefix = "arXiv",
    primaryClass = "hep-ph",
    reportNumber = "CPHT-RR002.0113, IRFU-13-01",
    doi = "10.1103/PhysRevD.87.054029",
    journal = "Phys. Rev. D",
    volume = "87",
    number = "5",
    pages = "054029",
    year = "2013"
}

@article{Berger:2001xd,
    author = "Berger, Edgar R. and Diehl, M. and Pire, B.",
    title = "{Time - like Compton scattering: Exclusive photoproduction of lepton pairs}",
    eprint = "hep-ph/0110062",
    archivePrefix = "arXiv",
    reportNumber = "CPHT-S010-0201, DESY-01-119",
    doi = "10.1007/s100520200917",
    journal = "Eur. Phys. J. C",
    volume = "23",
    pages = "675--689",
    year = "2002"
}

@article{Boer:2015hma,
    author = {Bo{\"e}r, Marie and Guidal, Michel and Vanderhaeghen, Marc},
    title = "{Single and double polarization observables in timelike Compton scattering off proton}",
    eprint = "1501.00270",
    archivePrefix = "arXiv",
    primaryClass = "hep-ph",
    month = "1",
    year = "2015"
}

@article{Guidal:2002kt,
    author = "Guidal, M. and Vanderhaeghen, M.",
    title = "{Double deeply virtual Compton scattering off the nucleon}",
    eprint = "hep-ph/0208275",
    archivePrefix = "arXiv",
    doi = "10.1103/PhysRevLett.90.012001",
    journal = "Phys. Rev. Lett.",
    volume = "90",
    pages = "012001",
    year = "2003"
}

@article{Belitsky:2002tf,
    author = "Belitsky, Andrei V. and Mueller, Dieter",
    title = "{Exclusive electroproduction of lepton pairs as a probe of nucleon structure}",
    eprint = "hep-ph/0210313",
    archivePrefix = "arXiv",
    reportNumber = "DOE-ER-40762-270, UMD-PP-03-024",
    doi = "10.1103/PhysRevLett.90.022001",
    journal = "Phys. Rev. Lett.",
    volume = "90",
    pages = "022001",
    year = "2003"
}

@article{Belitsky:2003fj,
    author = "Belitsky, Andrei V. and Mueller, Dieter",
    title = "{Probing generalized parton distributions with electroproduction of lepton pairs off the nucleon}",
    eprint = "hep-ph/0307369",
    archivePrefix = "arXiv",
    reportNumber = "DOE-ER-40762-264, UMD-PP-03-055",
    doi = "10.1103/PhysRevD.68.116005",
    journal = "Phys. Rev. D",
    volume = "68",
    pages = "116005",
    year = "2003"
}

@article{Kumericki:2011zc,
    author = "Kumericki, Kresimir and Lautenschlager, Tobias and Mueller, Dieter and Passek-Kumericki, Kornelija and Schaefer, Andreas and Meskauskas, Mantas",
    title = "{Accessing GPDs from Experiment --- Potential of A High-Luminosity EIC ---}",
    eprint = "1105.0899",
    archivePrefix = "arXiv",
    primaryClass = "hep-ph",
    month = "5",
    year = "2011"
}

@article{Altinoluk:2012fb,
    author = "Altinoluk, T. and Pire, B. and Szymanowski, L. and Wallon, S.",
    title = "{Soft-collinear resummation in deeply virtual Compton scattering}",
    eprint = "1206.3115",
    archivePrefix = "arXiv",
    primaryClass = "hep-ph",
    reportNumber = "LPT-12-56",
    month = "6",
    year = "2012"
}

@article{Altinoluk:2012nt,
    author = "Altinoluk, T. and Pire, B. and Szymanowski, L. and Wallon, S.",
    title = "{Resumming soft and collinear contributions in deeply virtual Compton scattering}",
    eprint = "1207.4609",
    archivePrefix = "arXiv",
    primaryClass = "hep-ph",
    reportNumber = "CPHT-RR066.0712, LPT-ORSAY-12-82",
    doi = "10.1007/JHEP10(2012)049",
    journal = "JHEP",
    volume = "10",
    pages = "049",
    year = "2012"
}

@article{Collins:1998be,
    author = "Collins, John C. and Freund, Andreas",
    title = "{Proof of factorization for deeply virtual Compton scattering in QCD}",
    eprint = "hep-ph/9801262",
    archivePrefix = "arXiv",
    reportNumber = "PSU-TH-192",
    doi = "10.1103/PhysRevD.59.074009",
    journal = "Phys. Rev. D",
    volume = "59",
    pages = "074009",
    year = "1999"
}

@article{Belitsky:1999hf,
    author = "Belitsky, Andrei V. and Freund, A. and Mueller, Dieter",
    title = "{Evolution kernels of skewed parton distributions: Method and two loop results}",
    eprint = "hep-ph/9912379",
    archivePrefix = "arXiv",
    doi = "10.1016/S0550-3213(00)00012-2",
    journal = "Nucl. Phys. B",
    volume = "574",
    pages = "347--406",
    year = "2000"
}

@article{Braun:2011zr,
    author = "Braun, V. M. and Manashov, A. N.",
    title = "{Kinematic power corrections in off-forward hard reactions}",
    eprint = "1108.2394",
    archivePrefix = "arXiv",
    primaryClass = "hep-ph",
    doi = "10.1103/PhysRevLett.107.202001",
    journal = "Phys. Rev. Lett.",
    volume = "107",
    pages = "202001",
    year = "2011"
}

@article{Braun:2014sta,
    author = {Braun, Vladimir M. and Manashov, Alexander N. and M{\"u}ller, Dieter and Pirnay, Bjoern M.},
    title = "{Deeply Virtual Compton Scattering to the twist-four accuracy: Impact of finite-$t$ and target mass corrections}",
    eprint = "1401.7621",
    archivePrefix = "arXiv",
    primaryClass = "hep-ph",
    doi = "10.1103/PhysRevD.89.074022",
    journal = "Phys. Rev. D",
    volume = "89",
    number = "7",
    pages = "074022",
    year = "2014"
}

@article{Berthou:2015oaw,
    author = "Berthou, B. and others",
    title = "{PARTONS: PARtonic Tomography Of Nucleon Software}: {A computing framework for the phenomenology of Generalized Parton Distributions}",
    eprint = "1512.06174",
    archivePrefix = "arXiv",
    primaryClass = "hep-ph",
    doi = "10.1140/epjc/s10052-018-5948-0",
    journal = "Eur. Phys. J. C",
    volume = "78",
    number = "6",
    pages = "478",
    year = "2018"
}

@article{Qiu:2023mrm,
    author = "Qiu, Jian-Wei and Yu, Zhite",
    title = "{Extraction of the Parton Momentum-Fraction Dependence of Generalized Parton Distributions from Exclusive Photoproduction}",
    eprint = "2305.15397",
    archivePrefix = "arXiv",
    primaryClass = "hep-ph",
    reportNumber = "JLAB-THY-23-3828, JLAB-THY-23-3828, MSUHEP-23-015",
    doi = "10.1103/PhysRevLett.131.161902",
    journal = "Phys. Rev. Lett.",
    volume = "131",
    number = "16",
    pages = "161902",
    year = "2023"
}

@article{Moutarde:2018kwr,
    author = "Moutarde, H. and Sznajder, P. and Wagner, J.",
    title = "{Border and skewness functions from a leading order fit to DVCS data}",
    eprint = "1807.07620",
    archivePrefix = "arXiv",
    primaryClass = "hep-ph",
    doi = "10.1140/epjc/s10052-018-6359-y",
    journal = "Eur. Phys. J. C",
    volume = "78",
    number = "11",
    pages = "890",
    year = "2018"
}

@article{Engelhardt:2017miy,
    author = "Engelhardt, M.",
    title = "{Quark orbital dynamics in the proton from Lattice QCD -- from Ji to Jaffe-Manohar orbital angular momentum}",
    eprint = "1701.01536",
    archivePrefix = "arXiv",
    primaryClass = "hep-lat",
    doi = "10.1103/PhysRevD.95.094505",
    journal = "Phys. Rev. D",
    volume = "95",
    number = "9",
    pages = "094505",
    year = "2017"
}

@article{Engelhardt:2024kcf,
    author = "Engelhardt, M. and Hasan, N. and Krieg, S. and Liuti, S. and Meinel, S. and Negele, J. and Pochinsky, A. and Rodekamp, M. and Syritsyn, S.",
    title = "{Quark orbital angular momentum in the proton from a twist-3 generalized parton distribution}",
    doi = "10.22323/1.456.0075",
    journal = "PoS",
    volume = "SPIN2023",
    pages = "075",
    year = "2024"
}

@article{Green:2015wqa,
    author = "Green, Jeremy and Meinel, Stefan and Engelhardt, Michael and Krieg, Stefan and Laeuchli, Jesse and Negele, John and Orginos, Kostas and Pochinsky, Andrew and Syritsyn, Sergey",
    title = "{High-precision calculation of the strange nucleon electromagnetic form factors}",
    eprint = "1505.01803",
    archivePrefix = "arXiv",
    primaryClass = "hep-lat",
    reportNumber = "JLAB-THY-15-2043, RBRC-1140",
    doi = "10.1103/PhysRevD.92.031501",
    journal = "Phys. Rev. D",
    volume = "92",
    number = "3",
    pages = "031501",
    year = "2015"
}

@article{Sufian:2020coz,
    author = {Sufian, Raza Sabbir and Liu, Tianbo and Alexandru, Andrei and Brodsky, Stanley J. and de T{\'e}ramond, Guy F. and Dosch, Hans G{\"u}nter and Draper, Terrence and Liu, Keh-Fei and Yang, Yi-Bo},
    title = "{Constraints on charm-anticharm asymmetry in the nucleon from lattice QCD}",
    eprint = "2003.01078",
    archivePrefix = "arXiv",
    primaryClass = "hep-lat",
    reportNumber = "JLAB-THY-20-3155, SLAC-PUB-17515",
    doi = "10.1016/j.physletb.2020.135633",
    journal = "Phys. Lett. B",
    volume = "808",
    pages = "135633",
    year = "2020"
}

@article{Diehl:2007uc,
    author = "Diehl, M. and Feldmann, Th. and Kroll, P.",
    title = "{Form factors and other measures of strangeness in the nucleon}",
    eprint = "0711.4304",
    archivePrefix = "arXiv",
    primaryClass = "hep-ph",
    reportNumber = "DESY-07-209, SI-HEP-2007-18, WUB-07-11, DESY 07-209, SI-HEP-2007-18, WUB 07-11",
    doi = "10.1103/PhysRevD.77.033006",
    journal = "Phys. Rev. D",
    volume = "77",
    pages = "033006",
    year = "2008"
}

@article{Hackett:2023rif,
    author = "Hackett, Daniel C. and Pefkou, Dimitra A. and Shanahan, Phiala E.",
    title = "{Gravitational Form Factors of the Proton from Lattice QCD}",
    eprint = "2310.08484",
    archivePrefix = "arXiv",
    primaryClass = "hep-lat",
    reportNumber = "MIT-CTP/5630, FERMILAB-PUB-23-592-T",
    doi = "10.1103/PhysRevLett.132.251904",
    journal = "Phys. Rev. Lett.",
    volume = "132",
    number = "25",
    pages = "251904",
    year = "2024"
}

@article{Panjsheeri:2025vpa,
    author = "Panjsheeri, Zaki and Adams, Douglas Q. and Khawaja, Adil and Pandey, Saraswati and Tezgin, Kemal and Liuti, Simonetta",
    title = "{Updated flexible global parametrization of generalized parton distributions from elastic and deep inelastic inclusive scattering data}",
    eprint = "2511.03065",
    archivePrefix = "arXiv",
    primaryClass = "hep-ph",
    month = "11",
    year = "2025"
}

@article{Shanahan:2018pib,
    author = "Shanahan, P. E. and Detmold, W.",
    title = "{Gluon gravitational form factors of the nucleon and the pion from lattice QCD}",
    eprint = "1810.04626",
    archivePrefix = "arXiv",
    primaryClass = "hep-lat",
    reportNumber = "MIT-CTP/5069",
    doi = "10.1103/PhysRevD.99.014511",
    journal = "Phys. Rev. D",
    volume = "99",
    number = "1",
    pages = "014511",
    year = "2019"
}

@article{Musatov:1997pu,
    author = "Musatov, I. V. and Radyushkin, A. V.",
    title = "{Transverse momentum and Sudakov effects in exclusive QCD processes: Gamma* gamma pi0 form-factor}",
    eprint = "hep-ph/9702443",
    archivePrefix = "arXiv",
    reportNumber = "JLAB-THY-97-07",
    doi = "10.1103/PhysRevD.56.2713",
    journal = "Phys. Rev. D",
    volume = "56",
    pages = "2713--2735",
    year = "1997"
}

@article{Bertone:2023jeh,
    author = "Bertone, Valerio and del Castillo, Rafael F. and Echevarria, Miguel G. and del R{\'\i}o, {\'O}scar and Rodini, Simone",
    title = "{One-loop evolution of twist-2 generalized parton distributions}",
    eprint = "2311.13941",
    archivePrefix = "arXiv",
    primaryClass = "hep-ph",
    reportNumber = "DESY-23-188, IPARCOS-UCM-23-128",
    doi = "10.1103/PhysRevD.109.034023",
    journal = "Phys. Rev. D",
    volume = "109",
    number = "3",
    pages = "034023",
    year = "2024"
}

@article{Hatta:2025obw,
    author = "Hatta, Yoshitaka and Schoenleber, Jakob",
    title = "{Probing quantum entanglement with generalized parton distributions at the Electron-Ion Collider}",
    eprint = "2511.04537",
    archivePrefix = "arXiv",
    primaryClass = "hep-ph",
    doi = "10.1103/qdb2-k2nh",
    journal = "Phys. Rev. D",
    volume = "113",
    number = "9",
    pages = "094016",
    year = "2026"
}

@article{Scopetta:2004kj,
    author = "Scopetta, Sergio",
    title = "{Generalized parton distributions of He-3}",
    eprint = "nucl-th/0404014",
    archivePrefix = "arXiv",
    doi = "10.1103/PhysRevC.70.015205",
    journal = "Phys. Rev. C",
    volume = "70",
    pages = "015205",
    year = "2004"
}

@article{Guzey:2008fe,
    author = "Guzey, V. and Thomas, A. W. and Tsushima, K.",
    title = "{Medium modifications of the bound nucleon GPDs and incoherent DVCS on nuclear targets}",
    eprint = "0806.3288",
    archivePrefix = "arXiv",
    primaryClass = "hep-ph",
    reportNumber = "JLAB-THY-08-837",
    doi = "10.1016/j.physletb.2009.01.064",
    journal = "Phys. Lett. B",
    volume = "673",
    pages = "9--14",
    year = "2009"
}

@article{Chang:2025pgi,
    author = "Chang, Wan and Aschenauer, Elke-Caroline and Jentsch, Alexander and Kumar, Arjun and Tu, Zhoudunming and Yin, Zhongbao",
    title = "{Opportunities for imaging light nuclei with a second interaction region at the Electron-Ion Collider}",
    eprint = "2511.05638",
    archivePrefix = "arXiv",
    primaryClass = "nucl-ex",
    doi = "10.1103/y4yv-y9dn",
    journal = "Phys. Rev. D",
    volume = "113",
    number = "3",
    pages = "032018",
    year = "2026"
}

@article{Fucini:2018gso,
    author = "Fucini, Sara and Scopetta, Sergio and Viviani, Michele",
    title = "{Coherent deeply virtual Compton scattering off $^4$He}",
    eprint = "1805.05877",
    archivePrefix = "arXiv",
    primaryClass = "nucl-th",
    doi = "10.1103/PhysRevC.98.015203",
    journal = "Phys. Rev. C",
    volume = "98",
    number = "1",
    pages = "015203",
    year = "2018"
}

@article{Goeke:2009tu,
    author = "Goeke, K. and Guzey, V. and Siddikov, M.",
    title = "{Leading twist nuclear shadowing, nuclear generalized parton distributions and nuclear DVCS at small x}",
    eprint = "0901.4711",
    archivePrefix = "arXiv",
    primaryClass = "hep-ph",
    reportNumber = "USM-TH-243, JLAB-THY-09-941",
    doi = "10.1103/PhysRevC.79.035210",
    journal = "Phys. Rev. C",
    volume = "79",
    pages = "035210",
    year = "2009"
}

@article{Liuti:2005qj,
    author = "Liuti, S. and Taneja, S. K.",
    title = "{Nuclear medium modifications of hadrons from generalized parton distributions}",
    eprint = "hep-ph/0504027",
    archivePrefix = "arXiv",
    doi = "10.1103/PhysRevC.72.034902",
    journal = "Phys. Rev. C",
    volume = "72",
    pages = "034902",
    year = "2005"
}

@article{Kirchner:2003wt,
    author = "Kirchner, A. and Mueller, Dieter",
    title = "{Deeply virtual Compton scattering off nuclei}",
    eprint = "hep-ph/0302007",
    archivePrefix = "arXiv",
    doi = "10.1140/epjc/s2003-01415-x",
    journal = "Eur. Phys. J. C",
    volume = "32",
    pages = "347--375",
    year = "2003"
}

@article{Liuti:2005gi,
    author = "Liuti, S. and Taneja, S. K.",
    title = "{Microscopic description of deeply virtual Compton scattering off spin-0 nuclei}",
    eprint = "hep-ph/0505123",
    archivePrefix = "arXiv",
    doi = "10.1103/PhysRevC.72.032201",
    journal = "Phys. Rev. C",
    volume = "72",
    pages = "032201",
    year = "2005"
}

@article{Duplancic:2023kwe,
    author = "Duplan{\v{c}}i{\'c}, Goran and Nabeebaccus, Saad and Passek-Kumeri{\v{c}}ki, Kornelija and Pire, Bernard and Szymanowski, Lech and Wallon, Samuel",
    title = "{Probing chiral-even and chiral-odd leading twist quark generalized parton distributions through the exclusive photoproduction of a {\ensuremath{\gamma}}{\ensuremath{\rho}} pair}",
    eprint = "2302.12026",
    archivePrefix = "arXiv",
    primaryClass = "hep-ph",
    doi = "10.1103/PhysRevD.107.094023",
    journal = "Phys. Rev. D",
    volume = "107",
    number = "9",
    pages = "094023",
    year = "2023"
}

@article{ALICE:2023jgu,
    author = "Acharya, Shreyasi and others",
    collaboration = "ALICE",
    title = "{Energy dependence of coherent photonuclear production of J/{\ensuremath{\psi}} mesons in ultra-peripheral Pb-Pb collisions at $ \sqrt{{\textrm{s}}_{\textrm{NN}}} $ = 5.02 TeV}",
    eprint = "2305.19060",
    archivePrefix = "arXiv",
    primaryClass = "nucl-ex",
    reportNumber = "CERN-EP-2023-100",
    doi = "10.1007/JHEP10(2023)119",
    journal = "JHEP",
    volume = "10",
    pages = "119",
    year = "2023"
}

@article{Cosyn:2020kfe,
    author = "Cosyn, W. and Pire, B. and Szymanowski, L.",
    title = "{Diffractive two-meson electroproduction with a nucleon and deuteron target}",
    eprint = "2007.01923",
    archivePrefix = "arXiv",
    primaryClass = "hep-ph",
    reportNumber = "CPHT-RR040.062020",
    doi = "10.1103/PhysRevD.102.054003",
    journal = "Phys. Rev. D",
    volume = "102",
    number = "5",
    pages = "054003",
    year = "2020"
}

@article{Cepila:2017nef,
    author = "Cepila, Jan and Contreras, Jesus Guillermo and Krelina, Michal",
    title = "{Coherent and incoherent $\mathrm{J/}\psi$ photonuclear production in an energy-dependent hot-spot model}",
    eprint = "1711.01855",
    archivePrefix = "arXiv",
    primaryClass = "hep-ph",
    doi = "10.1103/PhysRevC.97.024901",
    journal = "Phys. Rev. C",
    volume = "97",
    number = "2",
    pages = "024901",
    year = "2018"
}

@article{Mantysaari:2017dwh,
    author = {M{\"a}ntysaari, Heikki and Schenke, Bj{\"o}rn},
    title = "{Probing subnucleon scale fluctuations in ultraperipheral heavy ion collisions}",
    eprint = "1703.09256",
    archivePrefix = "arXiv",
    primaryClass = "hep-ph",
    doi = "10.1016/j.physletb.2017.07.063",
    journal = "Phys. Lett. B",
    volume = "772",
    pages = "832--838",
    year = "2017"
}

@article{ALICE:2023gcs,
    author = "Acharya, Shreyasi and others",
    collaboration = "ALICE",
    title = "{First Measurement of the |t| Dependence of Incoherent J/{\ensuremath{\psi}} Photonuclear Production}",
    eprint = "2305.06169",
    archivePrefix = "arXiv",
    primaryClass = "nucl-ex",
    reportNumber = "CERN-EP-2023-080",
    doi = "10.1103/PhysRevLett.132.162302",
    journal = "Phys. Rev. Lett.",
    volume = "132",
    number = "16",
    pages = "162302",
    year = "2024"
}

@article{ALICE:2021gpt,
    author = "Acharya, Shreyasi and others",
    collaboration = "ALICE",
    title = "{Coherent $J/\psi$ and $\psi'$ photoproduction at midrapidity in ultra-peripheral Pb-Pb collisions at $\sqrt{s_{\mathrm{NN}}}~=~5.02$ TeV}",
    eprint = "2101.04577",
    archivePrefix = "arXiv",
    primaryClass = "nucl-ex",
    reportNumber = "CERN-EP-2021-002",
    doi = "10.1140/epjc/s10052-021-09437-6",
    journal = "Eur. Phys. J. C",
    volume = "81",
    number = "8",
    pages = "712",
    year = "2021"
}

@article{Armstrong:2017zqr,
    author = "Armstrong, Whitney and others",
    title = "{Tagged EMC Measurements on Light Nuclei}",
    eprint = "1708.00891",
    archivePrefix = "arXiv",
    primaryClass = "nucl-ex",
    month = "8",
    year = "2017"
}

@article{Moreno:2026mqk,
    author = "Moreno, Eric A. and Bright-Thonney, Samuel and Novak, Andrzej and Garcia, Dolores and Harris, Philip",
    title = "{AI Agents Can Already Autonomously Perform Experimental High Energy Physics}",
    eprint = "2603.20179",
    archivePrefix = "arXiv",
    primaryClass = "hep-ex",
    month = "3",
    year = "2026"
}

@article{Bakshi:2025fgx,
    author = "Bakshi, S. D. and others",
    title = "{ArgoLOOM: agentic AI for fundamental physics from quarks to cosmos}",
    eprint = "2510.02426",
    archivePrefix = "arXiv",
    primaryClass = "hep-ph",
    reportNumber = "ANL-199516",
    month = "10",
    year = "2025"
}

@article{Dutta:2012ii,
    author = "Dutta, D. and Hafidi, K. and Strikman, M.",
    title = "{Color Transparency: past, present and future}",
    eprint = "1211.2826",
    archivePrefix = "arXiv",
    primaryClass = "nucl-th",
    doi = "10.1016/j.ppnp.2012.11.001",
    journal = "Prog. Part. Nucl. Phys.",
    volume = "69",
    pages = "1--27",
    year = "2013"
}

@article{Jain:1995dd,
    author = "Jain, Pankaj and Pire, Bernard and Ralston, John P.",
    title = "{Quantum color transparency and nuclear filtering}",
    eprint = "hep-ph/9511333",
    archivePrefix = "arXiv",
    reportNumber = "KANSAS-95-11-14",
    doi = "10.1016/0370-1573(95)00071-2",
    journal = "Phys. Rept.",
    volume = "271",
    pages = "67--179",
    year = "1996"
}

@article{Aschenauer:2013hhw,
    author = "Aschenauer, Elke-Caroline and Fazio, Salvatore and Kumericki, Kresimir and Mueller, Dieter",
    title = "{Deeply Virtual Compton Scattering at a Proposed High-Luminosity Electron-Ion Collider}",
    eprint = "1304.0077",
    archivePrefix = "arXiv",
    primaryClass = "hep-ph",
    doi = "10.1007/JHEP09(2013)093",
    journal = "JHEP",
    volume = "09",
    pages = "093",
    year = "2013"
}

@article{Aschenauer:2022aeb,
    author = "Aschenauer, E. C. and Batozskaya, V. and Fazio, S. and Gates, K. and Moutarde, H. and Sokhan, D. and Spiesberger, H. and Sznajder, P. and Tezgin, K.",
    title = "{EpIC: novel Monte Carlo generator for exclusive processes}",
    eprint = "2205.01762",
    archivePrefix = "arXiv",
    primaryClass = "hep-ph",
    doi = "10.1140/epjc/s10052-022-10651-z",
    journal = "Eur. Phys. J. C",
    volume = "82",
    number = "9",
    pages = "819",
    year = "2022"
}

@article{Aschenauer:2025cdq,
    author = "Aschenauer, E. C. and others",
    title = "{Study of deeply virtual Compton scattering at the future electron-ion collider}",
    eprint = "2503.05908",
    archivePrefix = "arXiv",
    primaryClass = "hep-ph",
    doi = "10.1103/fy8y-bjc9",
    journal = "Phys. Rev. D",
    volume = "112",
    number = "3",
    pages = "036010",
    year = "2025"
}

@article{Lin:2020rxa,
    author = "Lin, Huey-Wen",
    title = "{Nucleon Tomography and Generalized Parton Distribution at Physical Pion Mass from Lattice QCD}",
    eprint = "2008.12474",
    archivePrefix = "arXiv",
    primaryClass = "hep-ph",
    reportNumber = "MSUHEP-20-014, MSUHEP-20-014",
    doi = "10.1103/PhysRevLett.127.182001",
    journal = "Phys. Rev. Lett.",
    volume = "127",
    number = "18",
    pages = "182001",
    year = "2021"
}

@article{Grocholski:2019pqj,
    author = "Grocholski, O. and Moutarde, H. and Pire, B. and Sznajder, P. and Wagner, J.",
    title = "{Data-driven study of timelike Compton scattering}",
    eprint = "1912.09853",
    archivePrefix = "arXiv",
    primaryClass = "hep-ph",
    doi = "10.1140/epjc/s10052-020-7700-9",
    journal = "Eur. Phys. J. C",
    volume = "80",
    number = "2",
    pages = "171",
    year = "2020"
}

@article{Mueller:2005ed,
    author = "Mueller, Dieter and Schafer, A.",
    title = "{Complex conformal spin partial wave expansion of generalized parton distributions and distribution amplitudes}",
    eprint = "hep-ph/0509204",
    archivePrefix = "arXiv",
    doi = "10.1016/j.nuclphysb.2006.01.019",
    journal = "Nucl. Phys. B",
    volume = "739",
    pages = "1--59",
    year = "2006"
}

@article{Mamo:2024vjh,
    author = "Mamo, Kiminad A. and Zahed, Ismail",
    title = "{String-based parametrization of nucleon GPDs at any skewness: A comparison to lattice QCD}",
    eprint = "2404.13245",
    archivePrefix = "arXiv",
    primaryClass = "hep-ph",
    doi = "10.1103/PhysRevD.110.114016",
    journal = "Phys. Rev. D",
    volume = "110",
    number = "11",
    pages = "114016",
    year = "2024"
}

@article{Mamo:2024jwp,
    author = "Mamo, Kiminad A. and Zahed, Ismail",
    title = "{Parametrization of Generalized Parton Distributions from t-Channel String Exchange in AdS Spaces}",
    eprint = "2411.04162",
    archivePrefix = "arXiv",
    primaryClass = "hep-ph",
    doi = "10.1103/PhysRevLett.133.241901",
    journal = "Phys. Rev. Lett.",
    volume = "133",
    number = "24",
    pages = "241901",
    year = "2024"
}

@article{Brodsky:2008qu,
    author = "Brodsky, Stanley J. and Llanes-Estrada, Felipe J. and Szczepaniak, Adam P.",
    title = "{Local Two-Photon Couplings and the J=0 Fixed Pole in Real and Virtual Compton Scattering}",
    eprint = "0812.0395",
    archivePrefix = "arXiv",
    primaryClass = "hep-ph",
    reportNumber = "SLAC-PUB-13478",
    doi = "10.1103/PhysRevD.79.033012",
    journal = "Phys. Rev. D",
    volume = "79",
    pages = "033012",
    year = "2009"
}

@article{Ahmad:2006gn,
    author = "Ahmad, Saeed and Honkanen, Heli and Liuti, Simonetta and Taneja, Swadhin K.",
    title = "{Generalized Parton Distributions from Hadronic Observables: Zero Skewness}",
    eprint = "hep-ph/0611046",
    archivePrefix = "arXiv",
    doi = "10.1103/PhysRevD.75.094003",
    journal = "Phys. Rev. D",
    volume = "75",
    pages = "094003",
    year = "2007"
}

@article{Ahmad:2009fvg,
    author = "Ahmad, Saeed and Honkanen, Heli and Liuti, Simonetta and Taneja, Swadhin K.",
    title = "{Generalized Parton Distributions from Hadronic Observables: Non-Zero Skewness}",
    eprint = "0708.0268",
    archivePrefix = "arXiv",
    primaryClass = "hep-ph",
    doi = "10.1140/epjc/s10052-009-1073-4",
    journal = "Eur. Phys. J. C",
    volume = "63",
    pages = "407--421",
    year = "2009"
}

@article{Liu:2023cse,
    author = "Liu, Keh-Fei",
    title = "{Hadrons, superconductor vortices, and cosmological constant}",
    eprint = "2302.11600",
    archivePrefix = "arXiv",
    primaryClass = "hep-ph",
    doi = "10.1016/j.physletb.2023.138418",
    journal = "Phys. Lett. B",
    volume = "849",
    pages = "138418",
    year = "2024"
}

@article{Cotogno:2019xcl,
    author = "Cotogno, Sabrina and Lorc{\'e}, C{\'e}dric and Lowdon, Peter",
    title = "{Poincar{\'e} constraints on the gravitational form factors for massive states with arbitrary spin}",
    eprint = "1905.11969",
    archivePrefix = "arXiv",
    primaryClass = "hep-th",
    doi = "10.1103/PhysRevD.100.045003",
    journal = "Phys. Rev. D",
    volume = "100",
    number = "4",
    pages = "045003",
    year = "2019"
}

@article{Alkasassbeh:2024aws,
    author = "Alkasassbeh, Osamah and Rajan, Abha and Engelhardt, Michael and Liuti, Simonetta",
    title = "{Transverse orbital angular momentum in the proton}",
    eprint = "2410.21604",
    archivePrefix = "arXiv",
    primaryClass = "hep-ph",
    doi = "10.1016/j.physletb.2026.140373",
    journal = "Phys. Lett. B",
    volume = "876",
    pages = "140373",
    year = "2026"
}

@article{Schwartz:2026ekw,
    author = "Schwartz, Matthew D.",
    title = "{Resummation of the C-Parameter Sudakov Shoulder Using Effective Field Theory}",
    eprint = "2601.02484",
    archivePrefix = "arXiv",
    primaryClass = "hep-ph",
    month = "1",
    year = "2026"
}

@article{Martinez-Fernandez:2026web,
    author = "Mart{\'\i}nez-Fern{\'a}ndez, V{\'\i}ctor and Pire, B. and Sznajder, P. and Wagner, J.",
    title = "{Coherent deeply virtual Compton scattering on helium-4 beyond leading power}",
    eprint = "2604.25677",
    archivePrefix = "arXiv",
    primaryClass = "hep-ph",
    month = "4",
    year = "2026"
}

@article{Martinez-Fernandez:2026zog,
    author = "Mart{\'\i}nez-Fern{\'a}ndez, V. and Pire, B. and Sznajder, P. and Wagner, J.",
    title = "{Quark and gluon tomography of the helium-4 nucleus}",
    eprint = "2605.18519",
    archivePrefix = "arXiv",
    primaryClass = "hep-ph",
    month = "5",
    year = "2026"
}

@article{Cosyn:2026gyy,
    author = "Cosyn, Wim and Freese, Adam and Sosa, Alan",
    title = "{Quantum stress and torsion distributions in the deuteron}",
    eprint = "2602.18298",
    archivePrefix = "arXiv",
    primaryClass = "nucl-th",
    reportNumber = "JLAB-THY-26-4579",
    doi = "10.1103/kpl8-9nyx",
    journal = "Phys. Rev. C",
    volume = "113",
    number = "5",
    pages = "055208",
    year = "2026"
}

@article{Guo:2025jiz,
    author = "Guo, Yuxun and Yuan, Feng and Zhao, Wenbin",
    title = "{Bayesian Inferring Nucleon Gravitational Form Factors via Near-Threshold J/{\ensuremath{\psi}} Photoproduction}",
    eprint = "2501.10532",
    archivePrefix = "arXiv",
    primaryClass = "hep-ph",
    doi = "10.1103/3x7r-ythq",
    journal = "Phys. Rev. Lett.",
    volume = "135",
    number = "11",
    pages = "111902",
    year = "2025"
}

@article{Hechenberger:2025wnz,
    author = "Hechenberger, Florian and Mamo, Kiminad A. and Zahed, Ismail",
    title = "{String-based axial and helicity-flip GPDs: A comparison to lattice QCD}",
    eprint = "2508.00817",
    archivePrefix = "arXiv",
    primaryClass = "hep-ph",
    doi = "10.1103/7lxw-g6tq",
    journal = "Phys. Rev. D",
    volume = "112",
    number = "7",
    pages = "074018",
    year = "2025"
}

@article{Alexandrou:2020zbe,
    author = "Alexandrou, Constantia and Cichy, Krzysztof and Constantinou, Martha and Hadjiyiannakou, Kyriakos and Jansen, Karl and Scapellato, Aurora and Steffens, Fernanda",
    title = "{Unpolarized and helicity generalized parton distributions of the proton within lattice QCD}",
    eprint = "2008.10573",
    archivePrefix = "arXiv",
    primaryClass = "hep-lat",
    reportNumber = "DESY-20-150",
    doi = "10.1103/PhysRevLett.125.262001",
    journal = "Phys. Rev. Lett.",
    volume = "125",
    number = "26",
    pages = "262001",
    year = "2020"
}

@article{Holligan:2023jqh,
    author = "Holligan, Jack and Lin, Huey-Wen",
    title = "{Systematic improvement of x-dependent unpolarized nucleon generalized parton distributions in lattice-QCD calculation}",
    eprint = "2312.10829",
    archivePrefix = "arXiv",
    primaryClass = "hep-lat",
    reportNumber = "MSUHEP-23-033",
    doi = "10.1103/PhysRevD.110.034503",
    journal = "Phys. Rev. D",
    volume = "110",
    number = "3",
    pages = "034503",
    year = "2024"
}

@article{Panteleeva:2021iip,
    author = "Panteleeva, Julia Yu. and Polyakov, Maxim V.",
    title = "{Forces inside the nucleon on the light front from 3D Breit frame force distributions: Abel tomography case}",
    eprint = "2102.10902",
    archivePrefix = "arXiv",
    primaryClass = "hep-ph",
    doi = "10.1103/PhysRevD.104.014008",
    journal = "Phys. Rev. D",
    volume = "104",
    number = "1",
    pages = "014008",
    year = "2021"
}

\end{document}